\documentclass[iop, apj, twocolappendix, numberedappendix, appendixfloats]{emulateapj}
\usepackage{graphics}
\usepackage{subfigure}
\usepackage{graphicx}
\usepackage{threeparttable}
\usepackage{multirow}
\usepackage{epsf}
\usepackage{bm}
\usepackage{color}
\usepackage{float}
\usepackage{times}
\usepackage{rotating}
\usepackage{CJK}
\usepackage[titletoc]{appendix}

\newcommand\nature{{Nature}}

\shortauthors{zhang et al. 2021}
\shorttitle{A SYSTEMATIC SEARCH FOR DUAL AGNs IN MERGING GALAXIES}

\begin{document}

\title{ \textbf{A} \textbf{S}YS\textbf{T}EMATIC SEA\textbf{R}CH F\textbf{O}R \textbf{D}UAL \textbf{A}GNs IN ME\textbf{R}G\textbf{IN}G \textbf{G}ALAXIES \textbf{(ASTRO-DARING)}:\\ III: results from the SDSS spectroscopic surveys }

   \author{Yang-Wei Zhang\altaffilmark{1,3},
   Yang Huang       \altaffilmark{2,5},
   Jin-Ming Bai     \altaffilmark{1,4,5},
   Xiao-Wei Liu     \altaffilmark{2,5},
   Jian-guo Wang     \altaffilmark{1},
   Xiao-bo Dong     \altaffilmark{1}
   }
\altaffiltext{1}{Yunnan Observatories, Chinese Academy of Sciences, Kunming, Yunnan 650011, China; zhangyangwei@ynao.ac.cn; baijinming@ynao.ac.cn}
\altaffiltext{2}{South-Western Institute for Astronomy Research, Yunnan University, Kunming 650500, China; yanghuang@ynu.edu.cn}
\altaffiltext{3}{University of Chinese Academy of Sciences, Beijing 100049, China}
\altaffiltext{4}{Key Laboratory for the Structure and Evolution of Celestial Objects, Chinese Academy of Sciences, Kunming 650011, China}
\altaffiltext{5}{Corresponding authors}

\begin{abstract}

As the third installment in a series systematically searching dual active galactic nuclei (AGN) amongst merging galaxies, we present the results of 20 dual AGNs found by using the SDSS fiber spectra. 
To reduce the flux contamination from both the fiber aperture and seeing effects, the angular separation of two cores in our merging galaxy pairs sample is restricted at least larger than 3\arcsec. 
By careful analysis of the emission lines, 20 dual AGNs are identified from 61 merging galaxies with their two cores both observed by the SDSS spectroscopic surveys. 
15 of them are identified for the first time. 
The identification efficiency is about 32.79$\%$ (20/61), comparable to our former results (16 dual AGNs identified from 41 merging galaxies) based on the long-slit spectroscopy. 
Interestingly, two of the 20 dual AGNs show two prominent cores in radio images and their radio powers show they as the radio-excess AGNs.
So far, 31 dual AGNs are found by our project and this is the current largest dual AGN sample, ever constructed with a consistent approach.
This sample, together with more candidates from ongoing observations, is of vital importance to study the AGN physics and the coevolution between the supermassive black holes and their host galaxies.

\end{abstract}


\section{Introduction}

Every massive galaxy is believed to host a supermassive black hole (SMBH) in the central region (e.g., \citealt{Kormendy1995}) and grows up through accretion and merger in the standard $\Lambda$ cold dark matter cosmology (e.g., \citealt{Kelly2013}). 
During the merger of two galaxies, the SMBH pairs are expected to exist and may be triggered as AGN by accreting a large number of gas into the central SMBH (e.g., \citealt{Begelman1980, Hopkins2008}).
The kpc scale dual AGNs become natural products when the SMBH pairs are triggered as AGNs simultaneously.
Moreover, dual AGNs are precursors of the pc-scale binary AGNs that will finally coalesce within Hubble time (e.g., \citealt{Rodriguez2006}) and could be the prominent gravitational wave signals in the Universe (e.g., \citealt{Arzoumanian2016, Goulding2019}).

Unlike the AGN pair (with separations greater than 15$-$20 kpc) representing the earliest stage of the binary SMBH evolution, the dual AGNs are much closer (with separations between few hundred pc and 10$-$15 kpc) and thus serve as a vital observational tool to probe the galaxy evolution and AGN physics, especially the coevolution between the central BHs and their host galaxies (e.g., \citealt{Volonteri2003, Colpi and Dotti2011, Yu2011, Kormendy2013}).
However, identifying dual AGNs is not an easy task and the currently confirmed number (around 70, see a compilation in Huang et al. 2021 -- hereafter Paper I; Zhang et al. 2021a -- Paper II) is largely smaller than that of the theoretical prediction (e.g., \citealt{Komossa_Zensus_review2016}).
In prior to this project, the most ambitious dual AGN systematically searching method is based on the follow-up observations of double-peaked AGNs (DPAGNs) found either from the SDSS survey (e.g., \citealt{wang2009, Liu2010a, Smith2010}) or from the DEEP survey (e.g., \citealt{Gerke2007, Comerford2009a}).
However, only 2$\%$ to 5$\%$ of those selected DPAGNs, as revealed by long-slit or integral field-unit spectroscopy follow-up observations, are confirmed as dual AGNs (e.g., \citealt{Shen2011, Comerford2011, Comerford2012, Fu2012}).
The double-peaked emission-line profiles are largely contributed by the complicate gas kinematics around single AGN rather than dual AGN (e.g., \citealt{Shen2011, Fu2012, Nevin2016}).
Recently, other systematically searching methods are also proposed based on hard X-ray imaging (e.g., \citealt{Komossa2003, Koss2011, Koss2016, Liu2013, Comerford2015}), radio imaging (e.g., \citealt{Fu2011b, Fu2015a, Muller-Sanchez2015}), and infrared colors (e.g., \citealt{Satyapal2014, Satyapal2017}).
But the number of identified dual AGNs is still quite limited.

As stated in our Paper I and II, an innovative method of systematically searching of dual AGNs in merging galaxies has been developed and 222 candidates are selected. A brief description of the candidate selection is presented in Section\,2.
In our Paper II, 41 of the 222 candidates have been observed by the long-slit spectroscopy using the YFOSC mounted on the Lijiang 2.4\,m telescope (LJT) of Yunnan observatories. By careful analysis, 16 of the 41 observed merging galaxies are identified as dual AGN and the identification efficiency is about 40 $\%$.


\begin{table*}[!htp]
\scriptsize
\centering
\caption{SDSS Observational information for the 20 dual AGNs}
\begin{threeparttable}
\begin{tabular}{cccccccccccc}
\hline
\hline
Name & Core   & R.A.(J2000)  & Decl.(J2000)   & Seeing    & Separation   & $g$ band & Flux$_{\rm contam}$ & Observing date  & PA\\
   
       &         &    &                &  (\arcsec)  & (\arcsec /kpc) &   (mag) &         &         (UT)       & (\arcdeg)     \\   
   (1)    &  (2)   & (3) & (4)  &  (5)  & (6) & (7) & (8) & (9)  & (10)   \\   
\hline
\hline

 \multirow{2}{*}{J080544.13+113040.30}& J0805+1130EN & 121.43468 & 11.51171  & 1.50 & \multirow{2}{*}{3.3/10.8} & 19.60 
 &  1.4 \%       &        20110211 & \multirow{2}{*}{56.0}\\
                                        & J0805+1130WS & 121.43391 & 11.51120  & 2.33 &  &  18.13   &   1.2 \%         &20060223\\ 
\hline

 \multirow{2}{*}{J084809.69+351532.12}& J0848+3515EN & 132.04216 & 35.25953   & 1.47 & \multirow{2}{*}{5.7/6.3} & 16.16  &  0.0 \%  &20030202 & \multirow{2}{*}{67.3}\\
                                      & J0848+3515WS & 132.04038  & 35.25892  & 1.91 &                          & 16.18   &  0.0 \%   &20110312 & \\ 
\hline 

 \multirow{2}{*}{J090714.61+520350.61}& J0907+5203EN & 136.81088 & 52.06406    & 2.06 & \multirow{2}{*}{7.4/8.5} & 17.83 &  0.0 \%  &20010324 & \multirow{2}{*}{11.9}\\
                                      & J0907+5203WS & 136.81020 & 52.06206    & 2.34 &                          & 16.93 &  0.0 \%  &20010329 & \\ 
\hline   

 \multirow{2}{*}{J091448.94+085324.45}& J0914+0853WN & 138.70394 & 8.89013  & 1.83 & \multirow{2}{*}{3.8/9.4} & 18.35 & 0.2 \%  &20111230 & \multirow{2}{*}{153.4}\\
                                      & J0914+0853ES & 138.70440 & 8.88920  & 1.58 &                          & 18.83 & 0.1 \%  &20031201 & \\ 
\hline 

 \multirow{2}{*}{J095559.35+395438.87}& J0955+3954EN & 148.99952 & 39.91302  & 1.85 & \multirow{2}{*}{10.1/9.7} &  16.52 & 0.0 \%   &20030506 & \multirow{2}{*}{37.3} \\
                                      & J0955+3954WS & 148.99731 & 39.91079  & 2.31  &                    &  16.61 & 0.0 \%  &20120119 & \\ 
\hline 

 \multirow{2}{*}{J100602.50+071131.80}& J1006+0711EN & 151.51044 & 7.19217  & 2.28 & \multirow{2}{*}{5.6/12.5} &  16.88 & 0.0 \%  &20030102 & \multirow{2}{*}{81.8} \\
                                      & J1006+0711WS & 151.50889 & 7.19195  & 4.06 &                           &  16.98 & 2.1 \%   &20030422 & \\ 
\hline

 \multirow{2}{*}{J110639.56+433620.64}& J1106+4336W & 166.66297 & 43.60582  & 1.71 & \multirow{2}{*}{4.9/10.5} & 18.04 & 0.0 \%   &20040226 & \multirow{2}{*}{93.4} \\
                                      & J1106+4336E & 166.66485 & 43.60574  & 1.67 &                           & 17.33 & 0.0 \%   &20040218 & \\ 
\hline 

 \multirow{2}{*}{J111519.98+542316.75}& J1115+5423EN & 168.83326 & 54.38799  & 1.94 & \multirow{2}{*}{8.8/11.8} & 16.33 & 0.0 \%  &20030309 & \multirow{2}{*}{48.6}\\
                                      & J1115+5423WS & 168.83016 & 54.38639  & 1.97 &                           & 17.64 & 0.0 \%  &20030102 & \\ 
\hline 

 \multirow{2}{*}{J114411.74+102202.40}& J1144+1022EN & 176.04894 & 10.36735  & 2.17 & \multirow{2}{*}{6.5/14.7} &  17.69 & 0.0 \%  &20030405 & \multirow{2}{*}{12.1}\\
                                      & J1144+1022WS & 176.04855 & 10.36558  & 2.08 &                           &  18.48 & 0.0 \%  &20120223 & \\ 
\hline 

 \multirow{2}{*}{J121418.25+293146.70}& J1214+2931EN & 183.57607  & 29.52964  & 1.45 & \multirow{2}{*}{7.9/9.3} & 16.08 & 0.0 \%  &20060324 & \multirow{2}{*}{60.8}\\
                                      & J1214+2931WS & 183.57419  & 29.52873  & 1.50 &                          & 16.87 & 0.0 \%  &20130306 & \\ 
\hline  

 \multirow{2}{*}{J122217.85-000743.70}& J1222$-$0007EN & 185.57438 & $-$0.12881  & 	3.33 & \multirow{2}{*}{4.6/13.5} & 18.27 & 1.9 \%  &20010322  & \multirow{2}{*}{80.5} \\
                                      & J1222$-$0007WS & 185.57308 & $-$0.12903  & 	2.29 &  & 18.73 & 0.2 \%  &20010401 & \\ 
\hline 

 \multirow{2}{*}{J133031.98-003613.80}& J1330$-$0036WN & 202.63229 & $-$0.60333  & 3.47 & \multirow{2}{*}{4.2/4.4} &  15.48 & 1.7 \%    & 20000428 & \multirow{2}{*}{116.9} \\
                                      & J1330$-$0036ES & 202.63329 & $-$0.60383  & 2.64  &                 &  16.93 & 5.0 \%    &	20010219 & \\ 
\hline 

 \multirow{2}{*}{J133817.27+481632.20}& J1338+4816EN & 204.57407 & 48.27806  & 2.35 & \multirow{2}{*}{10.2/5.6} & 14.48 & 0.0 \%  & 	20030503 & \multirow{2}{*}{29.2} \\
                                      & J1338+4816WS & 204.57200 & 48.27561  & 2.10  &    &   14.43 & 0.0 \%   &	20090114 & \\ 
\hline 

 \multirow{2}{*}{J150134.72+544734.07}& J1501+5447EN & 225.39469 & 54.79280  & 1.87 & \multirow{2}{*}{3.8/10.7} & 18.62 & 0.3 \%  & 20020319 & \multirow{2}{*}{28.7} \\
                                      & J1501+5447WS & 225.39373 & 54.79179  & 1.78  &                    &  18.23 & 0.2 \%  & 20020401 & \\
\hline 

\multirow{2}{*}{J151751.77+252353.38}& J1517+2523E & 229.46573 & 25.39816  & 1.75 & \multirow{2}{*}{5.5/7.5} &  15.63 & 0.0 \%  & 20110412 & \multirow{2}{*}{96.8} \\
                                      & J1517+2523W & 229.46406 & 25.39834  & 2.29 &  &  19.15 & 0.2 \%   & 20070522 & \\ 
\hline

 \multirow{2}{*}{J155344.31+302508.50}& J1553+3025EN & 238.43466 & 30.41904  & 1.13 & \multirow{2}{*}{3.5/13.1} &  18.38 & 0.0 \%  &20050413 & \multirow{2}{*}{40.4}\\
                                      & J1553+3025WS & 238.43393 & 30.41830  & 1.82 &  &  21.12 & 4.1 \%   &20110705 & \\ 
\hline

 \multirow{2}{*}{J155850.44+272323.93}  & J1558+2723WN & 239.70974 & 27.39106  & 1.85 & \multirow{2}{*}{4.1/7.2} & 19.22 & 1.6 \%  & 20110527 & \multirow{2}{*}{160.0}\\
                                        & J1558+2723ES & 239.71019 & 27.38998  & 2.43 &  &  16.11 & 0.1 \%  & 20030702 & \\ 
\hline

 \multirow{2}{*}{J164507.91+205759.43}& J1645+2057WN & 251.28297 & 20.96651  & 1.50 & \multirow{2}{*}{4.2/9.7} & 18.11 & 0.0 \%  & 20040611 & \multirow{2}{*}{157.2} \\
                                      & J1645+2057ES & 251.28348 & 20.96543  & 1.45 &    & 17.99 & 0.0 \% &	20040524 & \\ 
\hline 

 \multirow{2}{*}{J171322.58+325627.90}& J1713+3256E & 258.34552 & 32.94121  & 1.99 & \multirow{2}{*}{4.3/8.0} & 20.26 & 1.5 \%  &20080505 & \multirow{2}{*}{84.6} \\
                                      & J1713+3256W & 258.34412 & 32.94110  & 1.95 &  &   17.23 & 0.0 \%   &20020517\\ 
\hline 

 \multirow{2}{*}{J220635.08+000323.16}& J2206+0003WN & 331.64572 & 0.05766  & 1.75 & \multirow{2}{*}{4.7/4.3} & 15.97 & 0.0 \%  &20030920 & \multirow{2}{*}{159.6} \\
                                      & J2206+0003ES & 331.64618 & 0.05643  & 1.64 &    &  15.97 & 0.0 \%   &20031020\\ 
                                        
    
\hline

\end{tabular}
\label{obs_log} 
\begin{flushleft}
Notes: Column (1): Object name; Column (2): The two optical cores of the system; Column (3) \& (4): The right ascension (R.A.) and declination (Decl.) of each core; Column (5): The 80th-percentile of seeing during exposure for each core in arcsecond; Column (6): The separation of two cores in the unit of arcsecond and kpc; Column (7): The SDSS $g$ band magnitude of each core; Column (8): The flux contamination of each core, the values are derived from a Monte Carlo approach (see Section\,2 for details); Column (9): The observing date of each spectrum; Column (10): The position angle (PA) of two cores on the sky, in degrees east of north.
\end{flushleft}
\end{threeparttable}
\end{table*}




\begin{table*}[!htp]
\scriptsize
\centering
\caption{BPT classifications}
\begin{threeparttable}
\begin{tabular}{ccccccccc}
\hline
\hline

Name &  log([O III]/${\mathrm{H}\beta}) $  & log([N II]/${\mathrm{H}\alpha})$ & log([S II]/${\mathrm{H}\alpha})$  & log([O I]/${\mathrm{H}\alpha})$  &  BPT$_{\rm [N\,II]}$ & BPT$_{\rm [S\,II]}$ & BPT$_{\rm [O\,I]}$ & Classification    \\

\hline  
\multicolumn{7}{c}{Identified dual AGNs (20 sources)}\\
      
 \hline 
 
J0805+1130EN  & $ $--$ $ & $ $--$ $ & $ $--$ $ & $ $--$ $ & $ $--$ $ & $ $--$ $ & $ $--$ $     & Type I AGN  \\         
J0805+1130WS  & $ $--$ $ & $ $--$ $ & $ $--$ $ & $ $--$ $ & $ $--$ $ & $ $--$ $ & $ $--$ $ & Type I AGN  \\   
         
 \hline 
 
 J0848+3515EN  & $ 0.49 \pm 0.02 $ & $ -0.20 \pm 0.05 $ & $ -0.44 \pm 0.02 $ & $ -1.08 \pm 0.03 $ & AGN & Seyfert & Seyfert & Seyfert  \\    
 J0848+3515WS  & $ $--$ $ & $ $--$ $ & $ $--$ $ & $ $--$ $ & $ $--$ $ & $ $--$ $ & $ $--$ $ & Type I AGN  \\    
\hline 

J0907+5203EN & $ 0.29 \pm 0.00 $ & $ -0.37 \pm 0.01 $ & $ -0.32 \pm 0.00 $ & $ -0.94 \pm 0.01 $ & Comp & Seyfert  & Seyfert  &  Ambiguous AGN  \\ 
J0907+5203WS & $ 0.71 \pm 0.02 $ & $ -0.33 \pm 0.01 $ & $ -0.29 \pm 0.01 $ & $ -0.85 \pm 0.03 $ & AGN  & Seyfert  & Seyfert  &  Seyfert  \\

  \hline 
 
J0914+0853WN  & $ 0.35 \pm 0.10 $ & $ 0.10 \pm 0.01 $ & $ -0.04 \pm 0.02 $ & $ -0.75 \pm 0.04 $ & AGN & LINER &  LINER &LINER \\
J0914+0853ES  & $ $--$ $ & $ $--$ $ & $ $--$ $ & $ $--$ $ & $ $--$ $ & $ $--$ $ & $ $--$ $ &  Type I AGN  \\         

   \hline 

J0955+3954EN & $ -0.05 \pm 0.05 $ & $ -0.34 \pm 0.02 $ & $ -0.30 \pm 0.02 $ & $ -1.01 \pm 0.02 $ & Comp & HII  & LINER  &  Ambiguous AGN  \\ 
J0955+3954WS & $ 0.26 \pm 0.01 $ & $ -0.12 \pm 0.01 $ & $ -0.02 \pm 0.01 $ & $ -0.69 \pm 0.02 $ & AGN & LINER  & LINER  &  LINER  \\ 
  \hline

J1006+0711EN  & $ $--$ $ & $ $--$ $ & $ $--$ $ & $ $--$ $ & $ $--$ $ & $ $--$ $ & $ $--$ $ & Type I AGN  \\         
J1006+0711WS  & $ $--$ $ & $ $--$ $ & $ $--$ $ & $ $--$ $ & $ $--$ $ & $ $--$ $ & $ $--$ $ &  Type I AGN  \\         

 \hline  
 
J1106+4336W  & $ 0.56 \pm 0.03 $ & $ -0.28 \pm 0.04 $ & $ -0.30 \pm 0.05 $ & $ -0.94 \pm 0.18 $ & AGN & Seyfert & Seyfert &  Seyfert  \\
J1106+4336E  & $ $--$ $ & $ $--$ $ & $ $--$ $ & $ $--$ $ & $ $--$ $ & $ $--$ $ & $ $--$ $ & Type I AGN  \\         
  \hline

J1115+5423EN  & $ 0.84 \pm 0.04 $ & $ -0.18 \pm 0.01 $ & $ -0.38 \pm 0.01 $ & $ -1.10 \pm 0.02 $ & AGN & Seyfert & Seyfert &  Seyfert  \\
J1115+5423WS  & $ 0.88 \pm 0.01 $ & $ -0.33 \pm 0.01 $ & $ -0.41 \pm 0.02 $ & $ -1.25 \pm 0.07 $ & AGN & Seyfert & Seyfert &  Seyfert  \\ 
     
 \hline      
J1144+1022EN  & $ $--$ $ & $ $--$ $ & $ $--$ $ & $ $--$ $ & $ $--$ $ & $ $--$ $ & $ $--$ $ &  Type I AGN  \\
J1144+1022WS  & $ 0.31 \pm 0.06 $ & $ 0.00 \pm 0.01 $ & $ -0.01 \pm 0.01 $ & $ -0.75 \pm 0.05 $ & AGN & LINER & LINER & LINER  \\ 

\hline 

 J1214+2931EN  & $ 0.84 \pm 0.05 $ & $ 0.16 \pm 0.01 $ & $ -0.03 \pm 0.01 $ & $ -0.90 \pm 0.02 $ & AGN & Seyfert & Seyfert & Seyfert  \\ 
 J1214+2931WS  & $ $--$ $ & $ $--$ $ & $ $--$ $ & $ $--$ $ & $ $--$ $ & $ $--$ $ & $ $--$ $ & Type I AGN  \\
  
 \hline 
    
J1222-0007EN  & $ $--$ $ & $ $--$ $ & $ $--$ $ & $ $--$ $ & $ $--$ $ & $ $--$ $ & $ $--$ $ &  Type I AGN  \\
J1222-0007WS  & $ 0.78 \pm 0.01 $ & $ -0.60 \pm 0.03 $ & $ -0.72 \pm 0.05 $ & $ -1.62 \pm 0.16 $ & AGN & Seyfert & Seyfert &  Seyfert  \\
 
\hline 

J1330-0036WN & $ 0.20 \pm 0.01 $ & $ -0.33 \pm 0.01 $ & $ -0.44 \pm 0.01 $ & $ -1.34 \pm 0.06 $ & Comp & HII  & HII      & Comp  \\ 
J1330-0036ES & $ 0.35 \pm 0.02 $ & $ -0.17 \pm 0.02 $ & $ -0.51 \pm 0.01 $ & $ -1.30 \pm 0.02 $ & AGN  & HII  & Seyfert  & Ambiguous AGN  \\  
 
\hline

J1338+4816EN & $ $--$ $ & $ $--$ $ & $ $--$ $ & $ $--$ $ & $ $--$ $ & $ $--$ $ & $ $--$ $ & Type I AGN  \\ 
J1338+4816WS & $ 0.40  \pm 0.02 $ & $ -0.26 \pm 0.02 $ & $ -0.44 \pm 0.01 $ & $ -1.30 \pm 0.06 $ & AGN  & Seyfert & Seyfert &  Seyfert  \\  
\hline

J1501+5447EN & $ $--$ $ & $ $--$ $ & $ $--$ $ & $ $--$ $ & $ $--$ $ & $ $--$ $ & $ $--$ $ & Type I AGN  \\
J1501+5447WS & $ 0.11 \pm 0.03 $ & $ -0.10 \pm 0.01 $ & $ -0.38 \pm 0.03 $ & $ -1.29 \pm 0.06 $ & Comp & HII  & HII      &  Comp  \\  

\hline 
    
J1517+2523E  & $ 0.61 \pm 0.06 $ & $ 0.09 \pm 0.04 $ & $ -0.05 \pm 0.03 $ & $ -0.67 \pm 0.04 $ & AGN & LINER & Seyfert &  Ambiguous AGN  \\
J1517+2523W  & $ 0.36 \pm 0.03 $ & $ 0.13 \pm 0.02 $ & $ -0.30 \pm 0.05 $ & $ -0.85 \pm 0.05 $ & AGN & Seyfert & Seyfert & Seyfert  \\ 
 
    \hline 
J1553+3025EN  & $ $--$ $ & $ $--$ $ & $ $--$ $ & $ $--$ $ & $ $--$ $ & $ $--$ $ & $ $--$ $ & Type I AGN  \\
J1553+3025WS  & $ 0.17 \pm 0.03 $ & $ -0.25 \pm 0.02 $ & $ -0.30 \pm 0.05 $ & $ -0.84 \pm 0.05 $ &  Comp  & LINER & LINER & Ambiguous AGN  \\

\hline 

J1558+2723WN  & $ $--$ $ & $ $--$ $ & $ $--$ $ & $ $--$ $ & $ $--$ $ & $ $--$ $ & $ $--$ $ &  Type I AGN  \\
J1558+2723ES  & $ $--$ $ & $ $--$ $ & $ $--$ $ & $ $--$ $ & $ $--$ $ & $ $--$ $ & $ $--$ $ &  Type I AGN  \\

\hline

J1645+2057WN & $ 0.43 \pm 0.02 $ & $ -0.22 \pm 0.02 $ & $ -0.45 \pm 0.02 $ & $ -1.28 \pm 0.03 $ & AGN  & Seyfert  & Seyfert  &  Seyfert  \\
J1645+2057ES & $ 0.33 \pm 0.01 $ & $ -0.28 \pm 0.01 $ & $ -0.51 \pm 0.01 $ & $ -1.36 \pm 0.02 $ & Comp & HII      & HII      & Comp  \\
  
 \hline 
     
J1713+3256E  & $ 0.84 \pm 0.09 $ & $ 0.10 \pm 0.05 $ & $ -0.27 \pm 0.09 $ & $ -0.96 \pm 0.11 $ & AGN & Seyfert & Seyfert & Seyfert  \\
J1713+3256W  & $ $--$ $ & $ $--$ $ & $ $--$ $ & $ $--$ $ & $ $--$ $ & $ $--$ $ & $ $--$ $ &Type I AGN  \\

 \hline 
    
J2206+0003WN  & $ 0.01 \pm 0.02 $ & $ -0.07 \pm 0.01 $ & $ -0.17 \pm 0.01 $ & $ -0.89 \pm 0.01 $ & Comp & LINER & LINER & Ambiguous AGN  \\
J2206+0003ES  & $ 0.28 \pm 0.01 $ & $ 0.15 \pm 0.02 $ & $ 0.08 \pm 0.02 $ & $ -0.46 \pm 0.02 $ & AGN & LINER & LINER &  LINER  \\

 \hline 

\end{tabular}
\label{BPT_classify} 
\begin{flushleft}
   
   Notes: BPT$_{\rm [N II]}$, BPT$_{\rm [S II]}$, BPT$_{\rm [O I]}$: Result of BPT classifications in the log ([O III]/${\mathrm{H}\beta})$ -- log ([N II]/${\mathrm{H}\alpha})$, log ([O III]/${\mathrm{H}\beta})$ -- log ([S II]/${\mathrm{H}\alpha})$
   and log ([O III]/${\mathrm{H}\beta})$ -- log ([O I]/${\mathrm{H}\alpha})$ planes, respectively.  
   Seyfert: Seyfert galaxy.
   Comp: AGN/Starforming composite galaxy.
   H\,{\sc ii}: Star--forming region.
  LINER: Low ionization nuclear emission line region.
  Ambiguous AGN: AGN of ambiguous classification, i.e. those are classified as one type of AGN in one or two of the BPT diagram(s) and classified as another type of AGN in the remaining diagram(s).
Classification: Classification based on BPT diagrams with the SDSS data.

\end{flushleft}
\end{threeparttable}
\end{table*} 


\begin{figure*}[ht]
\centering
  \includegraphics[width=5.20cm,height=5.0cm]{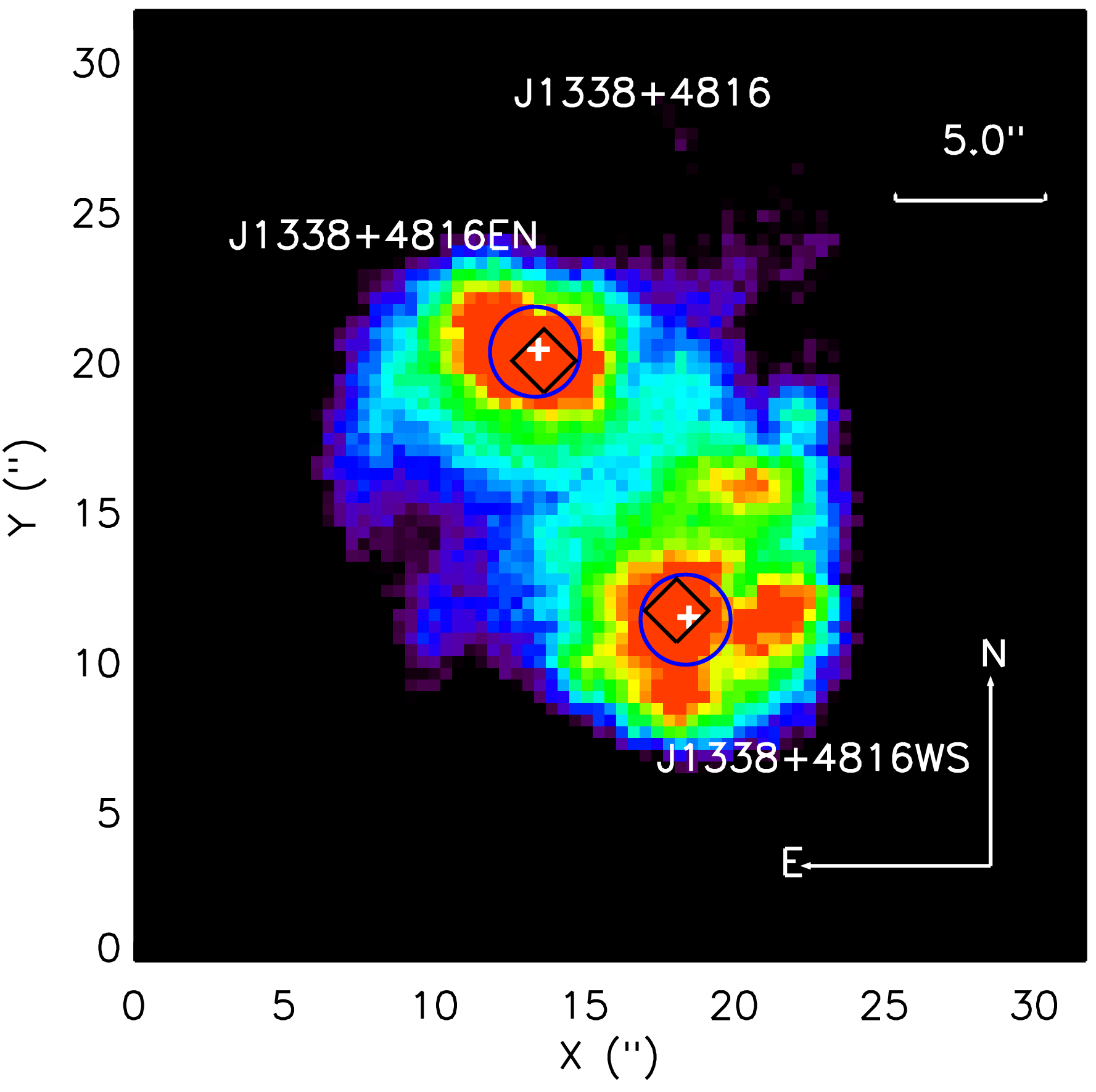}
  \includegraphics[width=12.2cm,height=5.2cm]{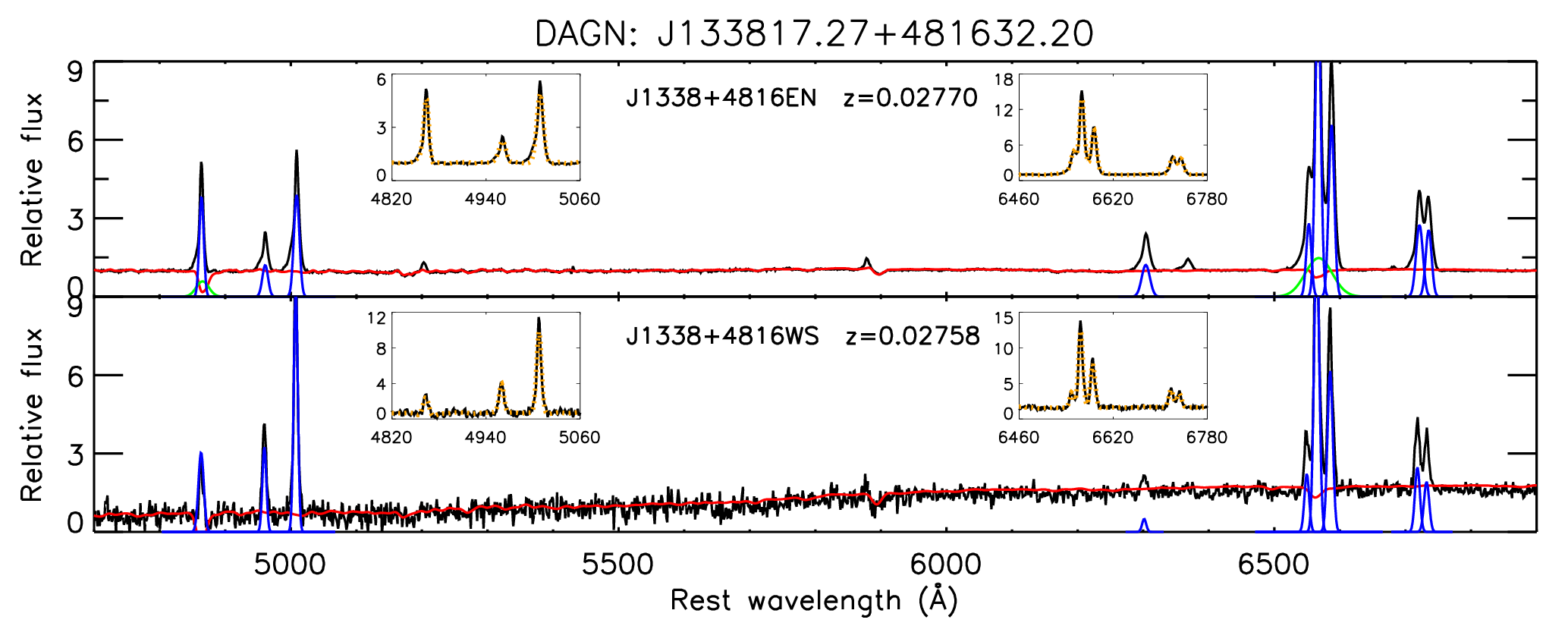}
  
\caption{Left: Pseudo color image of J1338+4816 in SDSS $g$ band. The position of the radio detection given in the FIRST catalog is marked by the black diamond. The fiber position of the SDSS spectrum is marked by the blue circle.
Right: The spectra of dual AGN: J133817.27+481632.20 respectively for J1338+4816EN (bottom) and J1338+4816WS (top) at rest-frame optical wavelengths. We used pPXF to subtract continuous spectrum and got emission lines of AGN. The red lines represent the continuum component, the blue lines represent the narrow emission lines component and the green lines represent the broad emission lines component.} 
\label{The spectra fitting of J1338+4816}
\end{figure*}

\begin{figure*}[ht]
\centering
\includegraphics[scale=0.68]{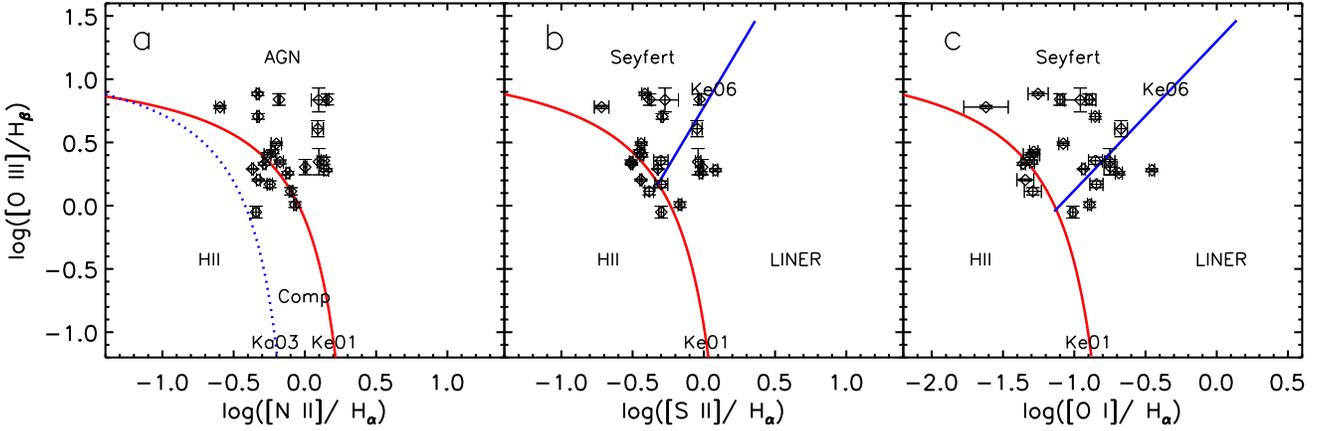}  
\caption{BPT diagram of 23 narrow line AGNs from 20 dual AGNs in the BPT-[N\,{\sc ii}], BPT-[S\,{\sc ii}] and BPT-[O\,{\sc i}] panels. In panel a, galaxy above the red solid line (Ke01; \citealt{Kewley2001}) is classified as AGN, below the blue dashed line is purely starforming galaxy (Ka03; \citealt{Kauffmann2003}), between the red solid line (Ke01) and the blue dashed line (Ka03) is AGN/starforming composite galaxy (Comp). In the AGN region, Seyferts and LINER are, respectively, located above and below the blue solid branch (Ke06; \citealt{Kewley2006}), as shown in panels b and c. The detail information is presented in Table\,\ref{BPT_classify}. } 
\label{BPT_DAGN}
\end{figure*}



\begin{table*}[!htp]
\scriptsize
\centering
\caption{Identified dual AGNs}
\begin{threeparttable}
\begin{tabular}{ccccccccccc}
\hline
\hline

DAGN & AGN &  $Z_{\rm sdss}$  & FWHM$_{\rm NLR}$ & FWHM$_{\rm H\beta}$ & FWHM$_{\rm H\alpha}$  & $V_{\rm offset}$   & Sep  & $W_{1}$-$W_{2}$&  Radio  &Classification   \\
    &           &      & (km $\mathrm{s}^{-1}$) &    (km $\mathrm{s}^{-1}$)  &    (km $\mathrm{s}^{-1}$)  &    (km $\mathrm{s}^{-1}$)  & kpc & mag  &    &       \\

\hline  
\hline

\multirow{2}{*}{J080544.13+113040.30}&  J0805+1130EN   &$0.19738  $  &$492 \pm 23 $  & $1635 \pm 34 $ & $3275 \pm 37 $ & \multirow{2}{*}{$ 540 \pm 15 $}  & \multirow{2}{*}{ 10.8 }  & \multirow{2}{*}{ 0.692 }    & \multirow{2}{*}{ Y } & Type I AGN    \\

                                       &  J0805+1130WS   &$0.19918  $  &$931 \pm 12 $  & $2612 \pm 126 $ & $5388 \pm 17 $ &                                 &           &                &     & Type I AGN     \\ 
                                       
\hline

\multirow{2}{*}{J084809.69+351532.12} &   J0848+3515EN    &$0.05727 $  &$333 \pm 30 $  &$489 \pm 22$   &$492 \pm 13 $  & \multirow{2}{*}{$90 \pm 15 $}   & \multirow{2}{*}{ 6.2 } & \multirow{2}{*}{ 0.632 }    &   \multirow{2}{*}{ Y }   &  Seyfert  \\
                                      &   J0848+3515WS    &$0.05697 $  &$677 \pm 14 $  &$1788\pm 65$   &$2664 \pm 15 $ &                                 &              &                &       &  Type I AGN\\
\hline

\multirow{2}{*}{J090714.61+520350.61}&  J0907+5203EN   &$0.06024  $  &$462\pm 13 $  & $453 \pm 23 $ & $440 \pm 13 $ & 
\multirow{2}{*}{$ 207 \pm 4 $}  & \multirow{2}{*}{ 8.5 }  & 0.429     &     Y    & Ambiguous AGN  \\
                                     &  J0907+5203WS   &$0.05955  $  &$406 \pm 16 $  & $373 \pm 28 $ & $407 \pm 14 $ &                                 &               &   0.368     &   Y   & Seyfert  \\ 

\hline

\multirow{2}{*}{J091448.94+085324.45}&  J0914+0853WN   &$0.13997  $  &$763 \pm 23 $  & $460 \pm 43 $ & $575 \pm 15 $ & \multirow{2}{*}{$ 18 \pm 11 $}  & \multirow{2}{*}{ 9.4 }  &  0.184    &    \multirow{2}{*}{ Y }   & LINER    \\

                                       &  J0914+0853ES   &$0.13991  $  &$359 \pm 29 $  & $1404 \pm 23 $ & $2071 \pm 13 $ &                                 &                 &   0.792     &      & Type I AGN     \\ 
                                       
\hline

\multirow{2}{*}{J095559.35+395438.87}&  J0955+3954EN   &$0.04876  $  &$291 \pm 20 $  & $305 \pm 35 $ & $280 \pm 21 $ & 
\multirow{2}{*}{$ 159 \pm 4 $}  & \multirow{2}{*}{ 9.7 }  & \multirow{2}{*}{ 0.481 }     &     Y    & Ambiguous AGN  \\
                                     &  J0955+3954WS   &$0.04929  $  &$520 \pm 11 $  & $516 \pm 20 $ & $519 \pm 11 $ &                                 &               &               &   Y   & LINER  \\ 

\hline

\multirow{2}{*}{J100602.50+071131.80}&  J1006+0711EN   &$0.12051  $  &$785 \pm 12 $  & $970 \pm 40  $ & $2121 \pm 15 $ & \multirow{2}{*}{$ 399 \pm 8 $}  & \multirow{2}{*}{ 12.5 }  & \multirow{2}{*}{ 1.194 }    &  \multirow{2}{*}{ Y }     & Type I AGN    \\

                                       &  J1006+0711WS   &$0.12184 $  &$987 \pm 15 $  & $1000 \pm 41 $ & $2547 \pm 23 $ &                                 &               &               &      & Type I AGN     \\ 
                                       
\hline

\multirow{2}{*}{J110639.56+433620.64}&  J1106+4336W   &$0.12473  $  &$460 \pm 21 $  & $530 \pm 22  $ & $560 \pm 18 $ & \multirow{2}{*}{$ 141 \pm 11 $}  & \multirow{2}{*}{ 10.5 }  & \multirow{2}{*}{ 0.782 }    &   \multirow{2}{*}{ Y }    & Seyfert    \\

                                       &  J1106+4336E   &$0.12426  $  &$595 \pm 17 $  & $1335 \pm 41 $ & $3943 \pm 12 $ &                                 &             &             &        & Type I AGN     \\ 
                                       
\hline

\multirow{2}{*}{J111519.98+542316.75}&  J1115+5423EN   &$0.07043  $  &$526 \pm 18 $  & $549 \pm 29  $ & $516 \pm 11 $ & \multirow{2}{*}{$ 264 \pm 4 $}  & \multirow{2}{*}{ 11.8 }  & \multirow{2}{*}{ 0.787 }    &    \multirow{2}{*}{ Y }   & Seyfert   \\

                                       &  J1115+5423WS   &$0.07131  $  &$361 \pm 27 $  & $375 \pm 28 $ & $320 \pm 18 $ &                                 &                 &           &       & Seyfert   \\ 
                                       
\hline

\multirow{2}{*}{J114411.74+102202.40}&  J1144+1022EN   &$0.12594  $  &$496 \pm 20 $  & $3440 \pm 13  $ & $3401 \pm 13 $ & \multirow{2}{*}{$ 189 \pm 11 $}  & \multirow{2}{*}{ 14.7 }  & 1.003    &   \multirow{2}{*}{ Y }    & Type I AGN  \\

                                       &  J1144+1022WS   &$0.12531 $  &$479 \pm 20 $  & $371 \pm 36 $ & $486 \pm 11 $ &                                 &                    &   0.128     &       & LINER  \\

\hline	

\multirow{2}{*}{J121418.25+293146.70}  &   J1214+2931EN     &$0.06326   $  &$618 \pm 17 $  &$571 \pm 25   $  &$517 \pm 11$ & \multirow{2}{*}{$ 72 \pm 9 $}   & \multirow{2}{*}{ 9.3 } & 0.059    &     \multirow{2}{*}{ Y }    &  Seyfert    \\
                                       &   J1214+2931WS     &$0.06350   $  &$444 \pm 22 $ &$1410 \pm 167 $  &$2405 \pm 16$&               &       &  1.274    &      &  Type I AGN    \\
\hline

\multirow{2}{*}{J122217.85-000743.70}&  J1222-0007EN   &$0.17290  $  &$792 \pm 12 $  & $1295 \pm 50  $ & $2000 \pm 22 $ & \multirow{2}{*}{$ 246 \pm 13 $}  & \multirow{2}{*}{ 13.5 } & \multirow{2}{*}{ 1.296 }     &   \multirow{2}{*}{ Y }    & Type I AGN  \\

                                       &  J1222-0007WS   &$0.17208  $  &$380 \pm 26 $  & $480 \pm 22 $ & $348 \pm 16 $ &                                 &                      &       &      & Seyfert   \\ 
                                       
\hline

\multirow{2}{*}{J133031.98-003613.80}&  J1330-0036WN   &$0.05424  $  &$280 \pm 23 $  & $324 \pm 32 $ & $297 \pm 19 $ & 
\multirow{2}{*}{$ 24 \pm 4 $}  & \multirow{2}{*}{ 4.4 }  & \multirow{2}{*}{ 0.598 }     &  \multirow{2}{*}{ Y }     & Comp   \\
                                     &  J1330-0036ES   &$0.05416  $  &$521 \pm 19 $  & $353 \pm 29 $ & $348 \pm 16 $ &                                 &               &               &      &  Ambiguous AGN  \\ 

\hline

\multirow{2}{*}{J133817.27+481632.20}&  J1338+4816EN   &$0.02770  $  &$566 \pm 15 $  & $1819 \pm 34 $ & $2214 \pm 18 $ & 
\multirow{2}{*}{$ 36 \pm 9 $}  & \multirow{2}{*}{ 5.6 }  & \multirow{2}{*}{ 0.590 }     &   Y      & Type I AGN   \\
                                     &  J1338+4816WS   &$0.02758  $  &$419 \pm 16 $  & $554 \pm 19 $ & $423 \pm 13 $ &                                 &               &               &   Y   &  Seyfert  \\ 

\hline

\multirow{2}{*}{J150134.72+544734.07}&  J1501+5447EN   &$0.16421 $  &$560 \pm 17 $  & $1435 \pm 149 $ & $2502 \pm 18 $ & 
\multirow{2}{*}{$ 168 \pm 8 $}  & \multirow{2}{*}{ 10.7 }  & \multirow{2}{*}{ 0.897 }     &   \multirow{2}{*}{ Y }   & Type I AGN   \\
                                     &  J1501+5447WS   &$0.16365 $  &$452 \pm 23 $  & $426 \pm 25 $ & $410 \pm 14 $ &                                 &               &               &      &  Comp  \\

\hline

\multirow{2}{*}{J151751.77+252353.38}&  J1517+2523E   &$0.07125 $  &$575 \pm 17 $  & $619 \pm 22  $ & $885 \pm 13 $ & \multirow{2}{*}{$ 225 \pm 13 $}  & \multirow{2}{*}{ 7.5 }  & \multirow{2}{*}{ 0.256 }    &   \multirow{2}{*}{ Y }    & Ambiguous AGN   \\

                                       &  J1517+2523W   &$0.07050  $  &$473 \pm 21 $  & $343 \pm 30 $ & $399 \pm 14 $ &                                 &                        &     &       & Seyfert   \\ 
                                       
\hline

\multirow{2}{*}{J155344.31+302508.50}&  J1553+3025EN   &$0.23699  $  &$936 \pm 10 $  & $3091 \pm 35  $ & $3616 \pm 12 $ & \multirow{2}{*}{$ 222 \pm 8 $}  & \multirow{2}{*}{ 13.1 }  & \multirow{2}{*}{ 1.195 }    &    \multirow{2}{*}{ Y }    & Type I AGN   \\

                                       &  J1553+3025WS   &$0.23773  $  &$667 \pm 15 $  & $477 \pm 22 $ & $523 \pm 11 $ &                                 &                        &     &       & Ambiguous AGN  \\ 
                                       
\hline

\multirow{2}{*}{J155850.44+272323.93}&  J1558+2723WN   &$0.09517  $  &$636 \pm 11 $  & $1283 \pm 16  $ & $1936 \pm 18 $ & \multirow{2}{*}{$ 492 \pm 15 $}  & \multirow{2}{*}{ 7.2 }  & \multirow{2}{*}{ 0.658 }    &    \multirow{2}{*}{ Y }    & Type I AGN   \\

                                     &  J1558+2723ES   &$0.09353  $  &$663 \pm 13 $  & $1037 \pm 55 $ & $1890 \pm 36 $ &                                 &                        &     &       & Type I AGN  \\ 
                                       
\hline

\multirow{2}{*}{J164507.91+205759.43}&  J1645+2057WN   &$0.13004  $  &$428 \pm 23 $  & $526 \pm 20 $ & $511 \pm 12 $ & 
\multirow{2}{*}{$ 252 \pm 7 $}  & \multirow{2}{*}{ 9.7 }  &  0.569     &    \multirow{2}{*}{ Y }    & Seyfert  \\
                                     &  J1645+2057ES   &$0.13088 $  &$333 \pm 30 $  & $364 \pm 29 $ & $346 \pm 17 $ &                                 &               &      0.569        &       &  Comp  \\

\hline

\multirow{2}{*}{J171322.58+325627.90}&  J1713+3256E   &$0.10141  $  &$358 \pm 28 $  & $540 \pm 38  $ & $419 \pm 19 $ & \multirow{2}{*}{$ 51 \pm 15 $}  & \multirow{2}{*}{ 8.0 } & \multirow{2}{*}{ 0.631 }     &   \multirow{2}{*}{ Y }     & Seyfert   \\

                                       &  J1713+3256W   &$0.10158  $  &$458 \pm 21 $  & $5008 \pm 19 $ & $5166 \pm 17 $ &                                 &                      &       &       & Type I AGN  \\ 
                                       
\hline

\multirow{2}{*}{J220635.08+000323.16}&  J2206+0003WN   &$0.04657  $  &$392 \pm 25 $  & $320 \pm 32  $ & $329 \pm 17 $ & \multirow{2}{*}{$ 129 \pm 7 $}  & \multirow{2}{*}{ 4.3 }  & \multirow{2}{*}{ 0.107 }     &   \multirow{2}{*}{ Y }    & Ambiguous AGN  \\

                                       &  J2206+0003ES   &$0.04614  $  &$615 \pm 16 $  & $511 \pm 20 $ & $590 \pm 12 $ &                                 &               &              &       & LINER  \\ 
                                       
\hline	


\end{tabular}
\label{finally_DAGN} 
\begin{flushleft}

   Notes: $Z_{\rm m}$: The redshift is from the median of main strong emission lines that having higher S/N.
          FWHM$_{\rm NLR}$: The FWHM of narrow line region is mesured from [O\,{\sc iii}] $\lambda$5007 line.
          FWHM$_{\rm BLR}$: The FWHM of broad line region is mesured from H${\alpha}$ broad line component.
          $V_{\rm offset}$: The velocity offset of two AGN cores.
          Sep : The separation of two AGN optical cores.
          $W_{1}-W_{2}$: The color between $W_{1}$ and $W_{2}$ bands for the resolved or unresolved core(s) from WISE.
          Classification: The classification of two AGN optical cores. For Type II AGN, we classify them based on the BPT diagrams (see Fig.\,2 and Table\,2) and for Type I AGN, we identify them by the value of FWHM of H$\alpha$ or H${\beta}$.

\end{flushleft}
\end{threeparttable}
\end{table*}

In this work, we continue our systematic searching of dual AGNs by the fiber spectra from the SDSS archived data.
Doing so, 61 merging galaxies from our 222 candidates are found with both cores observed by the SDSS fiber spectra under good seeing conditions. By analysis of emission lines,  20 systems are identified as dual AGNs and the efficiency is similar to that in our Paper II, implying the success of our searching approach. 
The paper is structured as follows. 
The data and reduction is described in Section~\ref{Data_and_Reduction}. 
The main results are presented in Section~\ref{Result} and a discussion is given in Section~\ref{Applications}. 
Finally, we summary in Section~\ref{Conclusions}. 
Cosmological parameters $H_{0}$ = 70 km\,s$^{-1}$ Mpc$^{-1}$, $\Omega_{\rm m}$ = 0.3, and $\Omega_{\rm \Lambda}=0.7$ are adopted throughout the paper and all wavelengths are in vacuum units.

\section{Data \& Reduction}
\label{Data_and_Reduction}

\subsection{The SDSS spectroscopic surveys}
\label{image_spectra}

The SDSS survey has been running for almost 20 years to create a most detailed map of the Universe (through several phases, SDSS-I: 2000-2005; e.g., \citealt{York2000, Abazajian2003}; SDSS-II: 2005-2008; \citealt{Frieman2008}; SDSS-III: 2008-2014; \citealt{Eisenstein2011} and SDSS-IV: 2014-2020; \citealt{Blanton2017}).
So far, multicolor images have been taken almost one third of the sky and more than three million spectra for astronomical objects have been obtained (including almost a million spectra for galaxies)\footnote{https://www.sdss.org}.
The selected merging galaxies with two optical cores larger than several arcseconds can be spatially resolved by the high quality SDSS images that taken under dark photometric nights with good seeing in five SDSS broad bands ($u, g, r, i, z$; \citealt{Fukugita1996}).
The optical spectra data we adopted in the current work were collected from SDSS I to IV and all released in SDSS DR15 (\citealt{Aguado2019}).
The spectra were taken with the double-armed SDSS/BOSS spectrographs with 3\arcsec/2\arcsec fiber aperture covering the whole optical wavelength range. 
The typical spectral resolving power $R$ is around 2000, corresponding to a native SDSS instrument broadening $\sigma$ $\sim$ 60 -- 70 km $\mathrm{s}^{-1}$ (\citealt{Smee2013}). 

\subsection{Sample Selection}
\label{Sample Selection}

As mentioned in our Paper I and II, we have developed a high efficiency method of systematically searching and identifying dual AGNs from kpc scale merging galaxies and a total of 222 candidates are selected.  
To select those candidates, we first select merging galaxies that are spatially resolved from the SDSS photometric catalog: galaxies exhibiting two optical cores separated by less than 8$\arcsec$ and at least one opctical core with SDSS fiber spectra (\citealt{Paris2018, Lyke2020}).
Secondly, the selected merging galaxies are at least with one radio detection from the FIRST radio survey (\citealt{Helfand2015}).
Typically, enhanced radio power are expected from merge-driven star formations or starbursts (e.g., \citealt{Bell2006, Jogee2009, Robaina2009}). 
To a certain extent, the requirement of at least one radio detection can help remove the physically unrelated pairs from the merging galaxies sample.
Finally, only merging galaxies with at least one core previously identified as an AGN by analyzing SDSS fiber spectra (\citealt{Paris2018}) are selected as dual AGN candidates.
In this manner, a total of 222 candidates are selected from the merging galaxies sample.
The redshifts of our selected candidates are smaller than 0.25.
For candidate with higher redshift, two cores of the kpc merging galaxy may fall in the 2\arcsec/3\arcsec fiber (corresponding to 7.5 or 11.7\,kpc at a redshift of 0.25).

In this work, we continue to search and identify dual AGNs based on the SDSS fiber spectra.
Doing so, the selected 222 candidates were cross-matched to the database of SDSS DR15 (\citealt{Aguado2019}) and 61 candidates have been found with both cores observed by the SDSS survey. 

For the merging galaxy, the spectra of two cores are not taken at the same time as two fibers on the same plate can't be placed closer than 55\arcsec and 62\arcsec due to fiber collisions (e.g., \citealt{Dawson2013}), respectively, for the original SDSS and the BOSS spectrographs.
To explore the nature of activity (e.g., star formation or accretion disk) of each core in the merging system, the contamination from another core should be controlled to a certain low fraction (i.e., 5$\%$ here).
To do so, we perform a simple simulation to obtain the flux contamination to each core, by considering the effects of fiber positioning uncertainty and the seeing\footnote{The 80th-percentile of seeing is adopted here.}.
In the simulation, the flux of each core is given by $g$-band magnitude and is assumed to follow a Gaussian distribution spatially with sigma given by the measured seeing.
The separation between the two cores is measured from the SDSS $g$-band images.
The aperture size is set to 3\arcsec and 2\arcsec if spectra observed by the original SDSS and the BOSS spectrographs, respectively.
In this manner, 61 merging galaxies with both cores observed by the SDSS surveys and flux contamination less than 5$\%$ are selected as our candidates.

\subsection{Subtraction of continuum }

The SDSS spectra have already been wavelength and flux calibrated. 
Before deriving the fluxes of emission lines, the continuum (including absorption features) of the host galaxy should be subtracted properly.
To do so, the penalised PiXel Fitting software (pPXF, originally described in \citealt{Cappellari2004} and upgraded in \citealt{Cappellari2017}) was adopted to fit the continuum and absorption features.
We operated the pPXF with stellar template of MILES Library (\citealt{Sanchez2016}).
This library contains 985 flux well-calibrated stars with wavelength coverage from 3525 \AA\, to 7500 \AA, at a spectral resolution of FWHM $\sim$ 2.51 \AA, $\sigma$ $\sim$ 64 km $\mathrm{s}^{-1}$, slightly higher than the SDSS's spectral resolution.
These templates are properly convolved in pPXF to the resolution of SDSS spectra before fitting. 
All prominent emission lines were masked when subtracting the continuum and absorption features.
As an example, we show the performance of this subtraction in Fig.\,1 for J1338+4816 system (as well as in the Appendix\,A for more examples). 

\subsection{The fitting of emission lines}
By properly subtracting the continuum and absorption features of host galaxy, we now perform emission line measurements using a IDL code developed in the current work.
In total, 9 strong optical emission lines (i.e., H${\beta}$, [O\,{\sc iii}] $\lambda\lambda$4959, 5007, [O\,{\sc i}] $\lambda$6300, H${\alpha}$, [N\,{\sc ii}] $\lambda\lambda$6549, 6583, [S\,{\sc ii}] $\lambda\lambda$6717, 6731) are fitted, and their properties, i.e., FWHMs, central wavelength, flux,  are measured.
We note that the derived FWHMs of those emission lines have been subtracted the instrumental broadening (FWHM $\sim 2.76$\AA).
The fluxes of emission lines have been corrected for the foreground Galactic extinction using the extinction map of \cite{Schlegel1998} and then the local reddening using the measured Balmer decrement\footnote{The intrinsic, dust-free Balmer decrement of H$\alpha$/H$\beta$ ratio of 3.1 is adopted here (\citealt{Osterbrock1989}).}.

When fitting the emission lines, for example, we set one Gaussian for [N\,{\sc ii}] $\lambda$6549 line, one Gaussian for [N\,{\sc ii}] $\lambda$6583 line, two Gaussians with the same central wavelength respectively for H${\alpha}$ broad and narrow line components. 
When H${\alpha}$ emission line does not have broad line component, we then set only one Gaussian for it. 
The fitting process of other emission lines are similar.
In Fig.\,1 (as well as in the Appendix\,A ), examples of the fitting are shown.

\begin{figure*}[ht]
  \centering
\includegraphics[width=5.8cm,height=5.8cm]{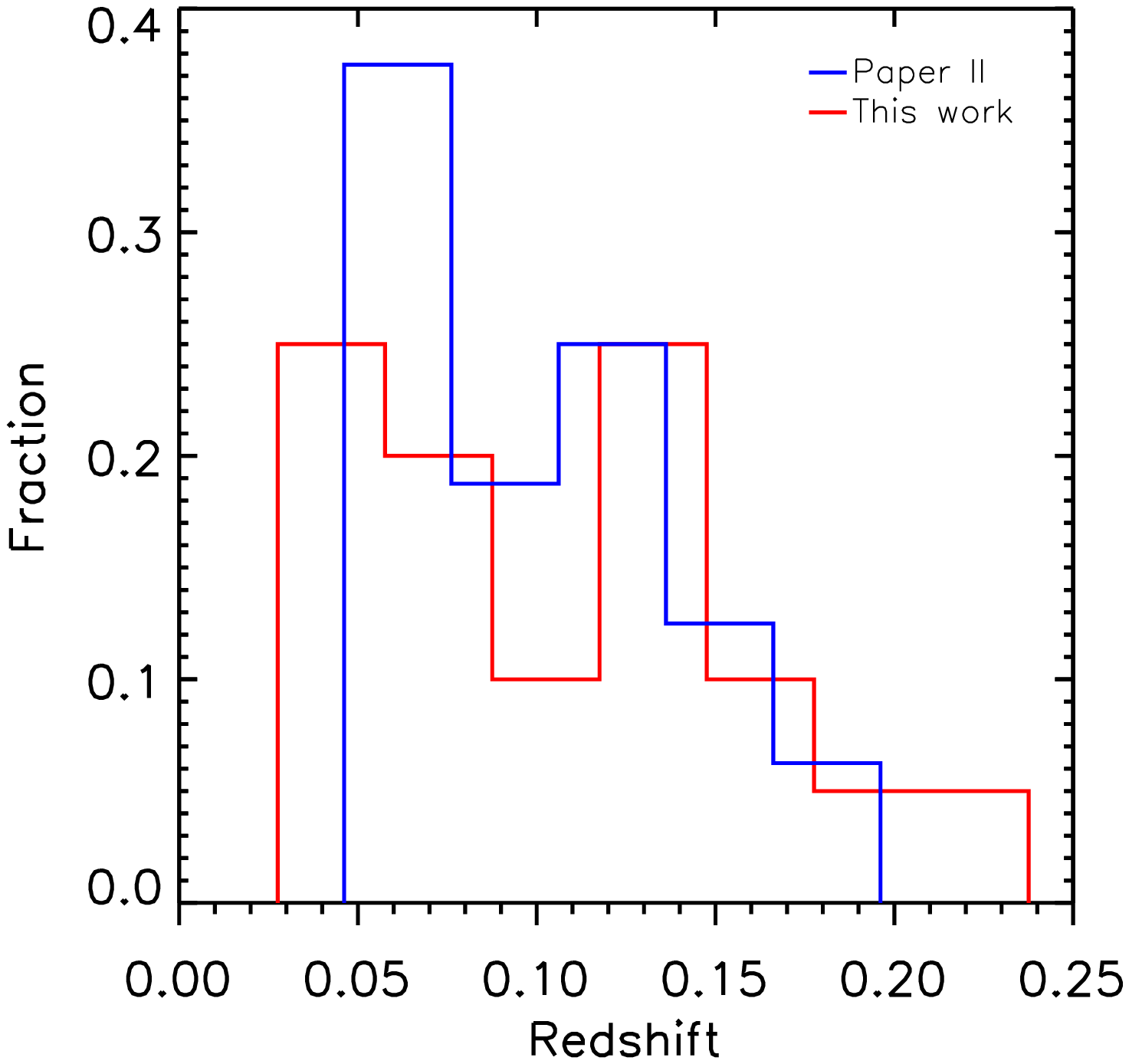}
\includegraphics[width=5.8cm,height=5.8cm]{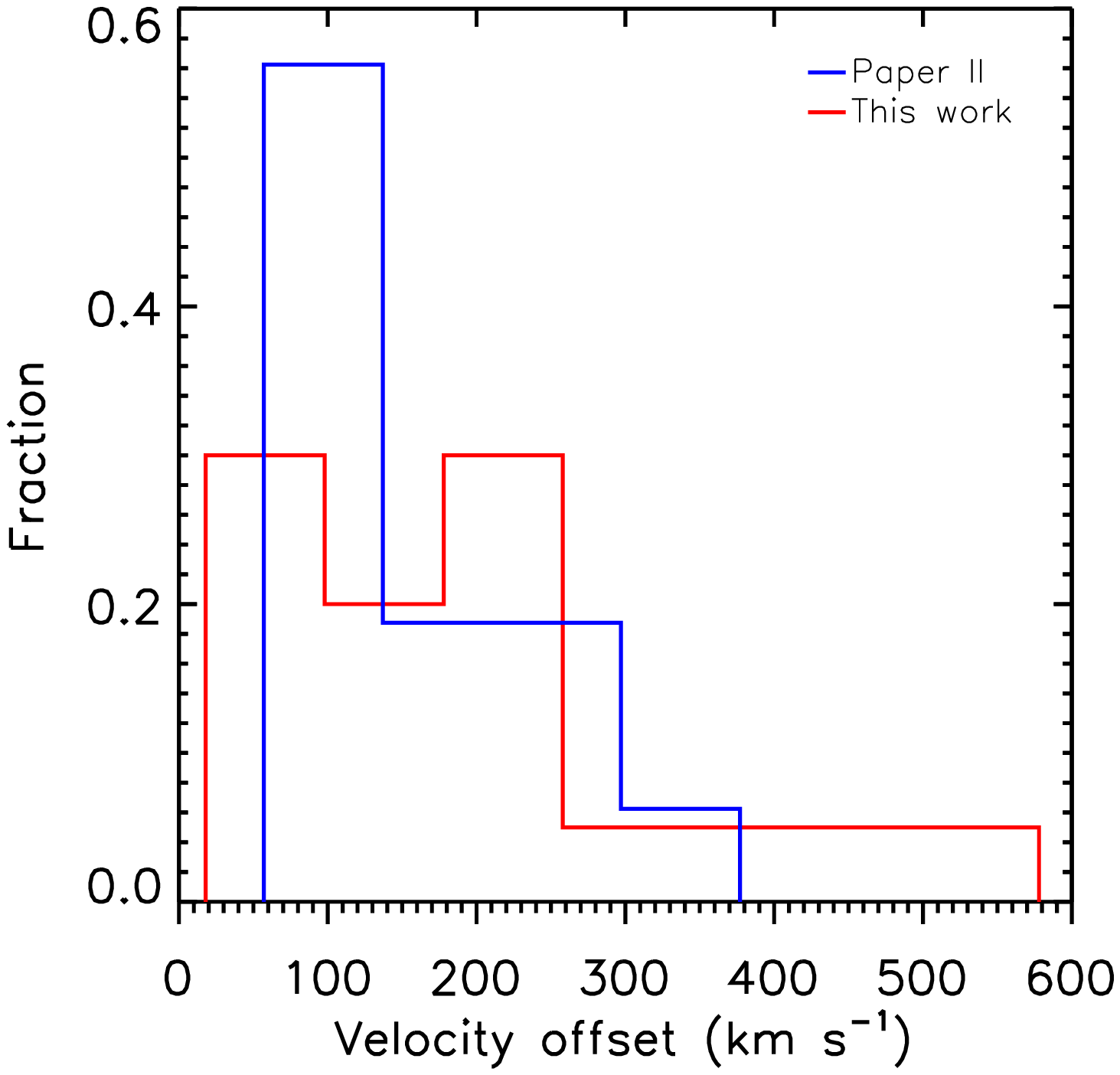}
\includegraphics[width=5.8cm,height=5.8cm]{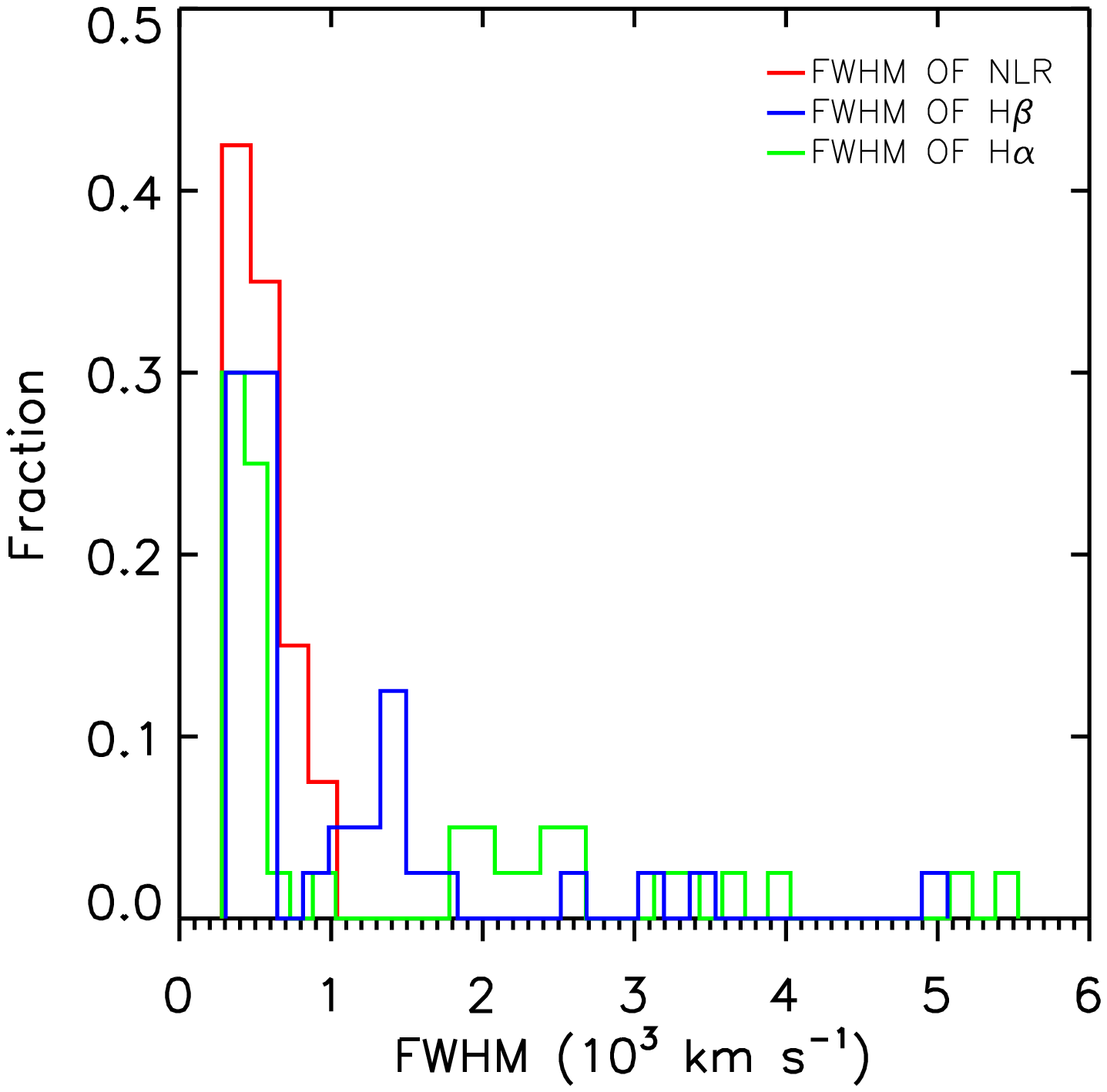}
\caption{Left: redshift distribution of dual AGNs; Middle: the distribution of velocity offset of dual AGNs; Right: FWHM distribution of dual AGNs.}
\label{distribution of result}
\end{figure*}


\section{Results \& Discussions}
\label{Result}

\subsection{Identify AGN}
\label{Identify_AGN}

In this paper, we adopt the same criteria to identify Type I and II AGNs as described in our Paper\,II.
Type I AGN is usually distinguished by the FWHM of the H${\alpha}$ broad line component: 
1) FWHM of broad H${\alpha}$ $>$ 2200 km $\mathrm{s}^{-1}$; or 2) FWHM of broad H${\alpha}$ $>$ 1200 km $\mathrm{s}^{-1}$ and h(broad H${\alpha}$)/h(narrow H${\alpha}$) $>$ 0.1, where h(broad H${\alpha}$) and h(narrow H${\alpha}$) are the heights of H${\alpha}$ broad and narrow line components, respectively. 
The height of H${\alpha}$ emission line is derived from the peak of the Gaussian fits.
The detail criteria of Type I AGN has been described in the work of \cite{Hao2005}.

For the Type II AGNs, we use the Baldwin-Phillips-Terlevich (BPT; \citealt{Baldwin1981}, \citealt{Kewley2001}, \citealt{Kauffmann2003}, \citealt{Kewley2006}) diagrams to identify them, using emission line flux ratios, 
i.e., [O\,{\sc iii}] $\lambda$5007/H$\beta$, [N\,{\sc ii}] $\lambda$6583/H$\alpha$, [S\,{\sc ii}] $\lambda\lambda$6717,6731/H$\alpha$, and [O\,{\sc i}] $\lambda$6300/H$\alpha$.
With the fluxes of emission lines measured in Section\,2.4, we show the positions of our final dual AGNs in the three BPT panels (BPT-[N\,{\sc ii}], BPT-[S\,{\sc ii}] and BPT-[O\,{\sc i}], see Fig.\,2).
The fluxes used here have been corrected for reddening and extinction effects.
Based on the classifications on BPT diagrams (see \citealt{Kewley2006} for more details), the Type II AGN can be further divided into the AGN/starforming composite galaxies (Comp), the Seyfert galaxies (Seyfert), the ambiguous galaxies (ambiguous AGN) and the Low ionization nuclear emission line region (LINER).
Ambiguous galaxies/AGNs are those that classified as one subtype of object in one or two of the BPT panel(s) but classified as another subtype of object in the remaining BPT panel(s).
Finally, we remark that the diagnostics purely based on optical line ratios can not provide secure classification of the nature of the systems with narrow line emission lines (especially, those systems classified as LINERs\footnote{The origin of LINER is still on hot debate. Most of the LINERs are likely to be AGNs (e.g., \citealt{Heckman2014, Netzer2015}) while the contribution from evolved post-AGB stars and low-mass X-ray binaries can not be ruled out (e.g., \citealt{Ho2008, Sarzi2010, Cid_Fernandes2011, Capetti_Baldi2011, Yan_Blanton2012}).}, composite galaxies and  ambiguous galaxies/AGNs).
Radio/X-ray follow-up observations of those dual AGN candidates are further required to provide more stringent constraints on their natures of the nuclei activities.

  
\begin{figure*}[htp]
\centering

\subfigure{\includegraphics[scale=0.61]{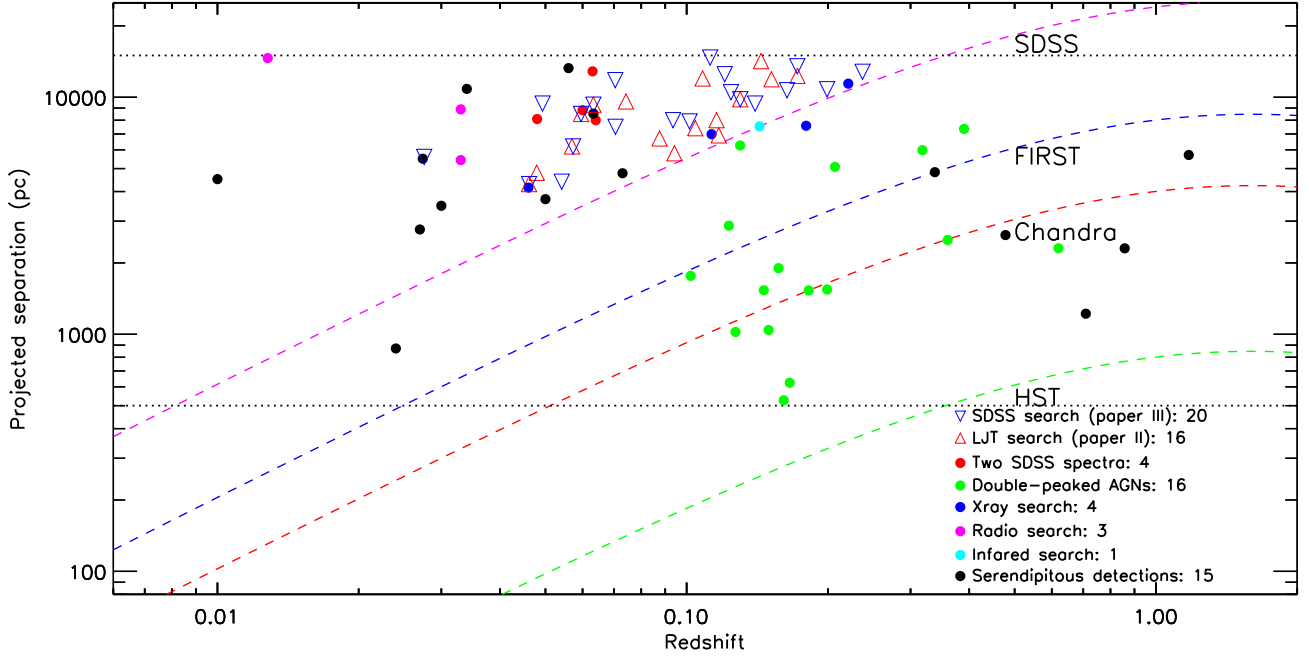}}

\caption{Identified dual AGNs in this work comparing with the known confirmed dual AGNs in paper II and other works. In this work, we have found 20 dual AGNs (15 dual AGNs are newly identified and 5 dual AGNs are already confirmed in paper II). The magenta, blue, red, green dashed lines, respectively, represent the critical resolutions of SDSS spectroscopy, FIRST, Chandra, HST catalogs, i.e., 3.0\arcsec, 1.0\arcsec, 0.5\arcsec, 0.1\arcsec, respectively. The two horizontal dashed lines are the typical separations, i.e., 0.5 and 15\,kpc, of dual AGNs, respectively.
The 43 known dual AGNs are found by the following methods: (1) Two SDSS spectra, (2) Double-peaked AGNs, (3) X-ray search, (4) Radio search, (5) Infaraed search and (6) Serendipitous detections. These methods here simply indicate how the dual AGNs were originally identified.
References for the 43 known dual AGNs: (1) \citealt{Junkkarinen2001}; (2) \citealt{Komossa2003}; (3) \citealt{Ballo2004}; (4) \citealt{Guainazzi2005}; (5) \citealt{Gerke2007}; (6) \citealt{Barth2008}; (7) \citealt{Bianchi2008}; (8) \citealt{Comerford2009a}; (9) \citealt{Comerford2009b}; (10) \citealt{Piconcelli2010}; (11) \citealt{Comerford2011}; (12) \citealt{Fu2011b}; (13) \citealt{Koss2011}; (14) \citealt{Liu2011}; (15) \citealt{McGurk2011}; (16) \citealt{Shen2011}; (17) \citealt{Barrows2012}; (18) \citealt{Fu2012}; (19) \citealt{Frey2012}; (20) \citealt{Koss2012}; (21) \citealt{Teng2012}; (22) \citealt{Mazzarella2012}; (23) \citealt{Liu2013}; (24) \citealt{Huang2014}; (25) \citealt{Comerford2015}; (26) \citealt{Fu2015a}; (27) \citealt{Muller-Sanchez2015}; (28) \citealt{Ellison2017}; (29) \citealt{Secrest2017}; (30) \citealt{Liu2018}.} 

\label{The_known_DAGNs}
\end{figure*}


\begin{figure}[ht]
\centering
\subfigure{\includegraphics[width=8.5cm,height=8.6cm]{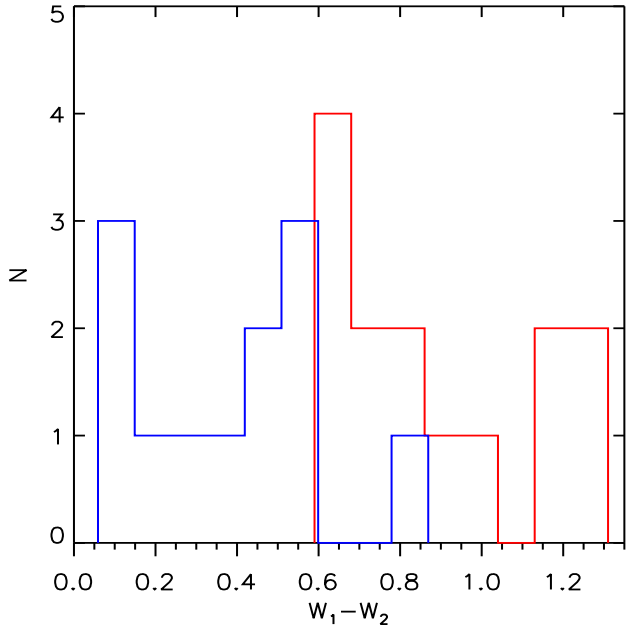}}
\caption{Distribution of $ W_{1}$-$W_{2}$ values of dual AGNs, the red histogram represents Type I AGN or the dual AGNs including at least one Type I AGN and the blue one represents Type II AGN or the dual AGNs without Type I AGN.}

\label{w1_w2_compare}
\end{figure} 

\begin{figure}[htp]
\centering

\subfigure{\includegraphics[width=8.5cm,height=8.2cm]{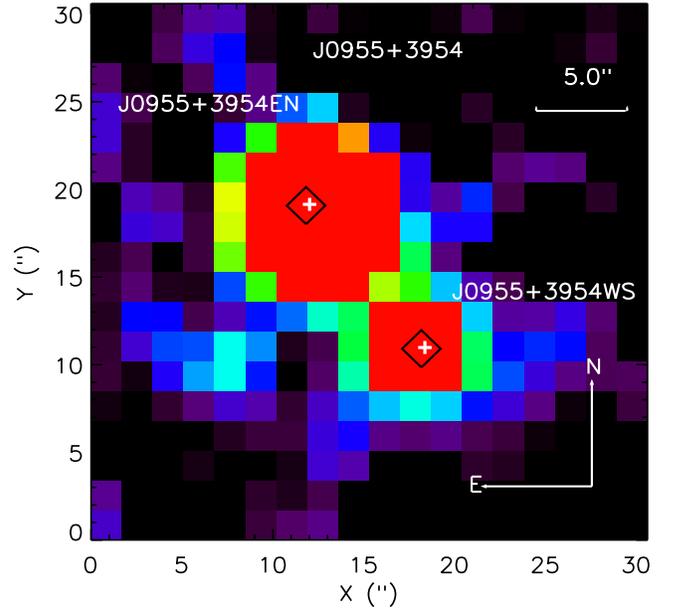}}
\caption{ FIRST 1.4 GHz pseudo-color image of J$0955+3954$ on a logarithmic scale.
The two nuclei, i.e., J$0955+3954$EN and J$0955+3954$WS, are clearly resolved, with the central positions marked by the black diamonds.
The white pluses here mark the optical positions from the SDSS images.
North is up and east is to the left. Spatial scale is also shown in the top-right corner.} 
\label{Radio image of J0955+3954}
\end{figure}

\begin{figure}[htp]
\centering

\subfigure{\includegraphics[width=8.5cm,height=8.2cm]{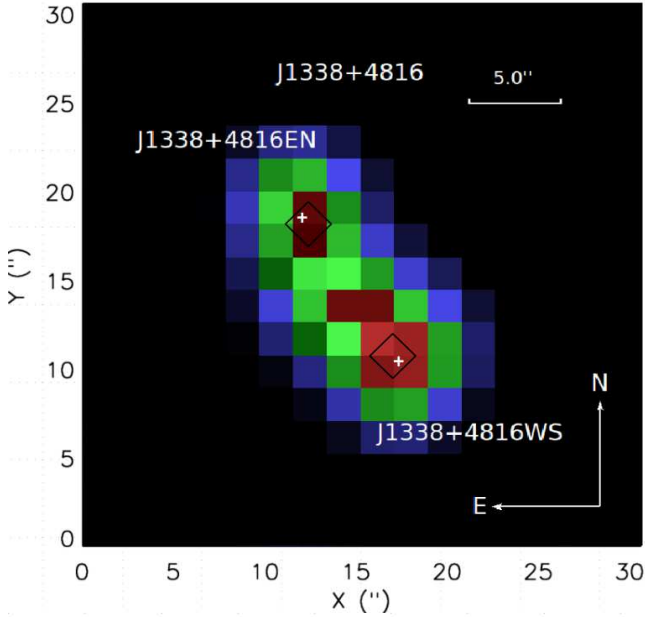}}
\caption{ Similar to Fig.\,6 but for J$1338+4861$.} 
\label{Radio image of J1338+4816}
\end{figure}

\begin{figure}[htp]
\centering
\subfigure{\includegraphics[scale=0.36,angle=0]{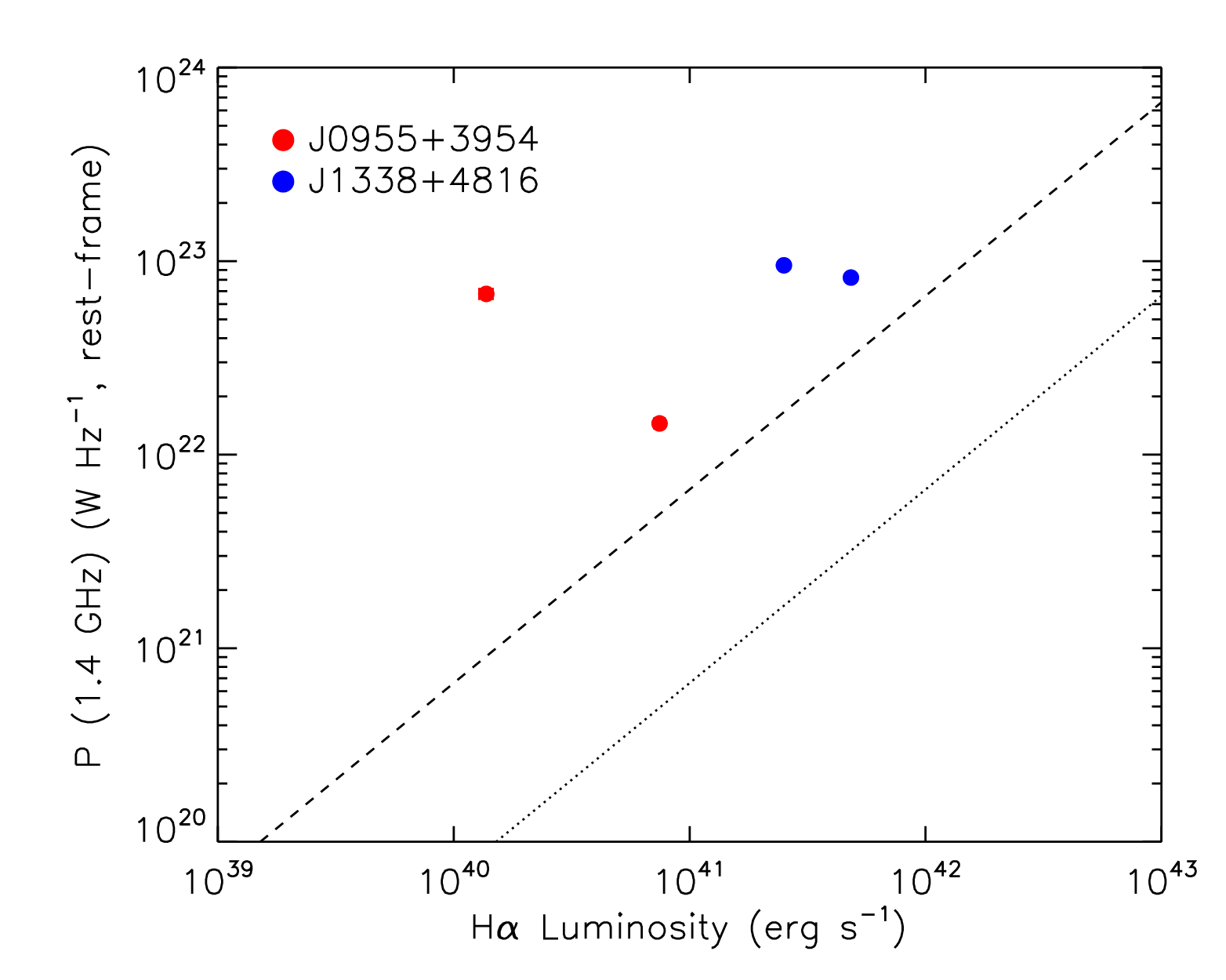}}
\caption{Rest-fram 1.4\,GHz radio power versus H$\alpha$ luminosity for J$0955+3954$ (red dots) and J$1338+4816$ (blue dots) systems.
Dotted line indicate radio-derived star formation rates (SFRs) equal to the H$\alpha$-derived SFRs (using the relations from \citealt{Hopkins2001}), while the dashed line mark the former 10 times larger than the latter.}
\label{haradio}
\end{figure}

  
\subsection{The identified dual AGNs sample}
\label{Dual AGN sample}
Using the criteria mentioned above, 20 dual AGNs are finally identified based on the fiber spectra from the SDSS survey.
The efficiency of this systematic searching is about 32.79$\%$ (20/61).
If excluding those systems containing LINERs, composite galaxies and  ambiguous galaxies/AGNs (their activity natures requiring further observations), the systematic searching efficiency is still as high as 16.39$\%$ (10/61).
The observation information of these 20 dual AGNs is shown in Table\,\ref{obs_log}. 
We note that 5 of 20 dual AGNs (J0848+3515, J0907+5203, J1214+2931, J1645+2057, J2206+0003) have been confirmed with the long-slit spectroscopy by using YFOSC mounted on LJT of Yunnan observatories (Paper II) separately. 
The properties (e.g., FWHM, velocity offset, emission line ratios and activity nature) of the five dual AGN revealed by the SDSS fiber spectra here are consistent with those measured by the long-slit spectra in our Paper II very well.

In the 20 dual AGNs, about 40.0$\%$ (16/40) are Type I AGNs, 
about 7.5$\%$ (3/40) are classified as Comps,
about 27.5$\%$ (11/40) are classified as Seyferts, 
about 10.0$\%$ (4/40) are LINERs,
and about 15.0$\%$ (6/40) are ambiguous galaxies/AGNs. 
The detail classifications of our dual AGNs are presented in Table\,\ref{BPT_classify}.

The redshift of those 20 dual AGN candidates ranges from 0.02758 to 0.23699 with a median value of 0.10158 shown in the left of Fig.\,\ref{distribution of result}. 
Compared to the dual AGN sample constructed by the long-slit spectroscopy by the LJT (Paper II), the redshift distribution of the current work is more flat along the redshift.
In the middle panel of Fig.\,\ref{distribution of result}, the velocity offset are ranging from 18 km $\mathrm{s}^{-1}$ to  540 km $\mathrm{s}^{-1}$ with a median value of 168 km $\mathrm{s}^{-1}$. 
The distribution of velocity offset is consistent with that of previous work (e.g., \citealt{Liu2011}, \citealt{Comerford2013}).

In the right panel of Fig.\,\ref{distribution of result}, the FWHM from [O\,{\sc iii}] $\lambda$5007 line (red line)  for NLRs and from H$\alpha$ (blue line) and  H$\beta$ (green line) lines for BLRs are shown.
The detail values can be checked from Table\,\ref{finally_DAGN}.
The current dual AGNs, together with those from our Paper II and previous literature, are shown in the redshift versus projected separation plane in Fig.\,\ref{The_known_DAGNs}.
Similar to our Paper II, the current dual AGNs are just right to fill the gap and largely improve completeness of the redshift distribution of dual AGNs.

Finally, the infrared color $W_1 - W_2$ distribution is shown in Fig.\,\ref{w1_w2_compare} for the dual AGNs.
Similar to the result of our Paper II, Type I and II AGNs show quite distinct distribution in  $W_1 - W_2$ color, with the Type I AGN more redder while Type II more bluer.

\section{Remarks on individual interesting objects}
\label{Applications}
In the current work, 20 dual AGNs were identified by the analysis of the SDSS fiber spectra.
For these dual AGNs, more solid evidences are required to confirm their nuclei natures, e.g., from the high angular resolution radio or X-ray imaging observations (e.g., \citealt{Fu2011b, Liu2013}).
Although no new follow-up observations in radio/X-ray bands, we check the images of these dual AGNs from the existing archival data, e.g., the FIRST survey (\citealt{Helfand2015}) and the Chandra Data Archive\footnote{\url{https://cxc.harvard.edu/cda/}}.

 Interestingly, two of our dual AGNs, i.e., J$0955+3954$ and J$1338+4816$, are found to show two prominent radio detections in FIRST images (see Figs.\,6 and 7).
The central positions of the two radio detections are in great agreement with the SDSS optical central positions.
 The integrated fluxes at $1.4$\,GHz given by the FIRST survey are $12.24\pm0.13$\,mJy and $2.57\pm0.13$\,mJy, corresponding to rest frame $P_{\rm 1.4 GHz}$ of $(6.77 \pm 0.07) \times 10^{22}$\,W\,Hz$^{-1}$ and $(1.45 \pm 0.07) \times 10^{22}$\,W\,Hz$^{-1}$, for J$0955+3954$EN and J$0955+3954$WS, respectively.
 For J$1338+4816$EN and J$1338+4816$WS, the integrated fluxes are, respectively, $47.20\pm0.14$\,mJy and $55.05\pm0.14$\,mJy, corresponding to  rest frame $P_{\rm 1.4 GHz}$ of  $(8.22 \pm 0.02) \times 10^{22}$\,W\,Hz$^{-1}$ and $(9.51 \pm 0.02) \times 10^{22}$\,W\,Hz$^{-1}$.
When deriving the rest frame 1.4\,GHz luminosity, a mean spectral index of $-0.7$ is assumed (e.g., \citealt{Pushkarev2012}, \citealt{Hovatta2014}).

To show the radio power from AGN activities, the $P_{\rm 1.4 GHz}$ versus $L_{\rm{H}\alpha}$ (good indicator of star formation rate) diagram of the two systems is shown in Fig.\,8.
 Clearly, both cores of the two systems have radio powers 10 times larger than those expected from the H$\alpha$-traced SFR, indicating significant radio-excess powers from AGN activities.
The above results provide unambiguous evidences of the AGN nature of the two dual AGNs.

\section{Conclusions}
\label{Conclusions}

\begin{enumerate}

\item 
 We have carried a systematic search for dual AGNs from 61 merging galaxies  by using the SDSS fiber spectra.
 20 dual AGNs are finally identified by the careful analysis of their emission lines.
 The successful efficiency is about 33\%, similar to that of our Paper II.
 For the 20 dual AGNs, 15 are discovered for the first time and the remaining 5 systems have been discovered in our Paper II. 

\item  
Interestingly, 2 of the identified dual AGNs show clear two radio cores in FIRST 1.4\,GHz images.
Compared to the H$\alpha$ luminosities of the two systems, both cores of the two dual AGNs are the so-called radio-excess AGN.

\item  
With the efforts from Paper II and the current work, we have constructed a large sample of 31 dual AGNs .
This sample, together with more candidates from future observations, will provide important constraints on understanding AGN physics and the coevolution between the supermassive black holes and their host galaxies.

\end{enumerate}

\section*{Acknowledgements}

We acknowledge the support of the staff of the Lijiang 2.4-m telescope. 
Funding for the telescope has been provided by CAS and the People's Government of Yunnan Province. 
The work of J. M. Bai is supported by the NSFC (grants 11133006, 11361140347) and the Strategic Priority Research Program ``The Emergence of Cosmological Structures'' of the Chinese Academy of Sciences (grant No. XDB09000000).
Y. Huang and X.-W. Liu acknowledge support by the National Key Basic Research Program of China 2014CB845700.
We thank XX for help. 

{\it Facilities:}  \facility{SDSS (SDSS spectrograph)} , \facility{LJT (Lijiang 2.4-m telescope)}, \facility{LAMOST (Large Sky Area Multi-Object Fiber Spectroscopy Telescope)}, \facility{FIRST (Faint Images of the Radio Sky at Twenty-Centimeters)}

\newpage

\bibliographystyle{aj}

\newpage
\begin{appendices} 

\appendix
\section{Notes For Individual Dual AGN}
\label{appendix}

\begin{figure*}[ht]
\centering
  \includegraphics[width=5.20cm,height=5.0cm]{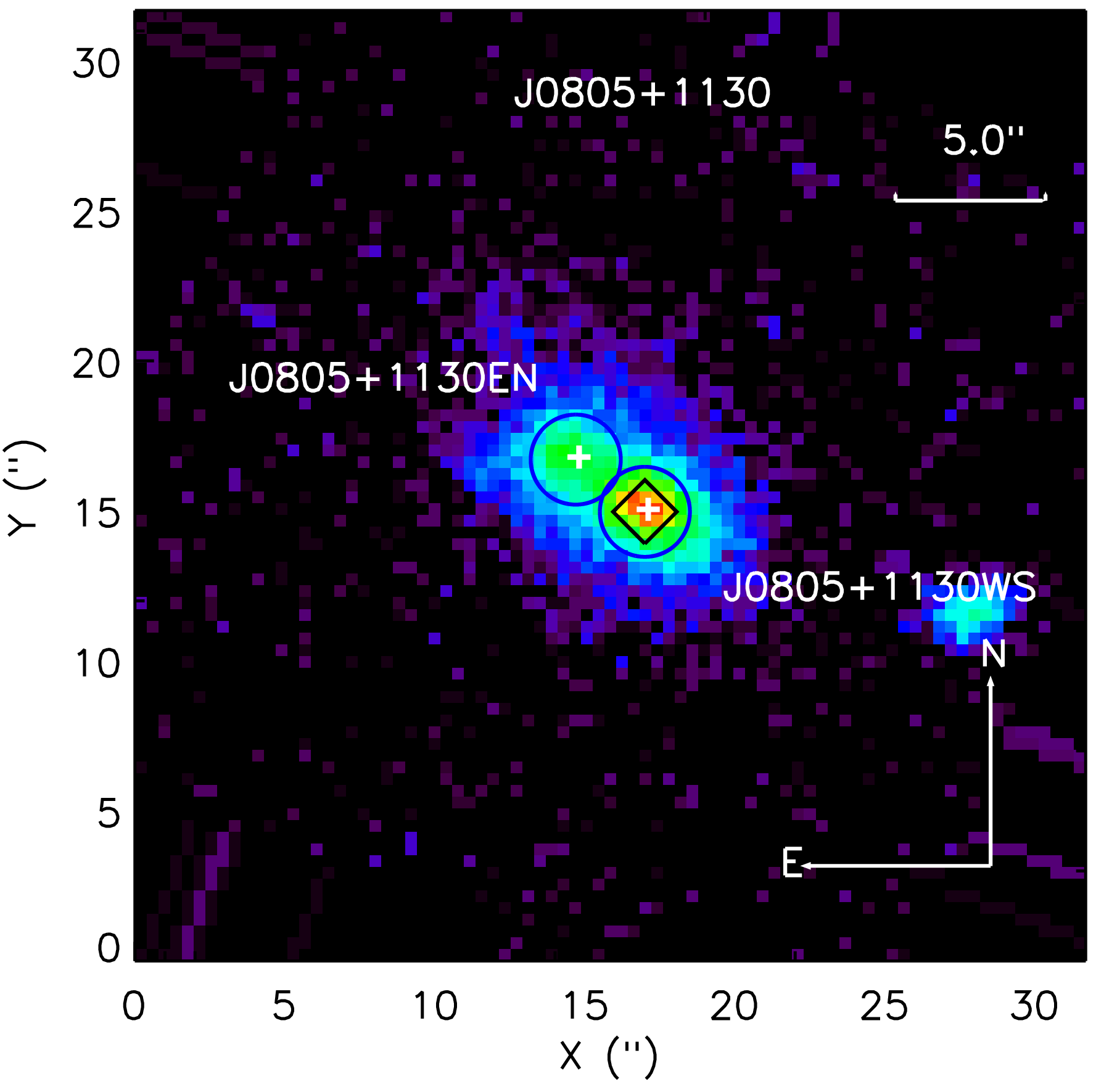}
  \includegraphics[width=12.2cm,height=5.2cm]{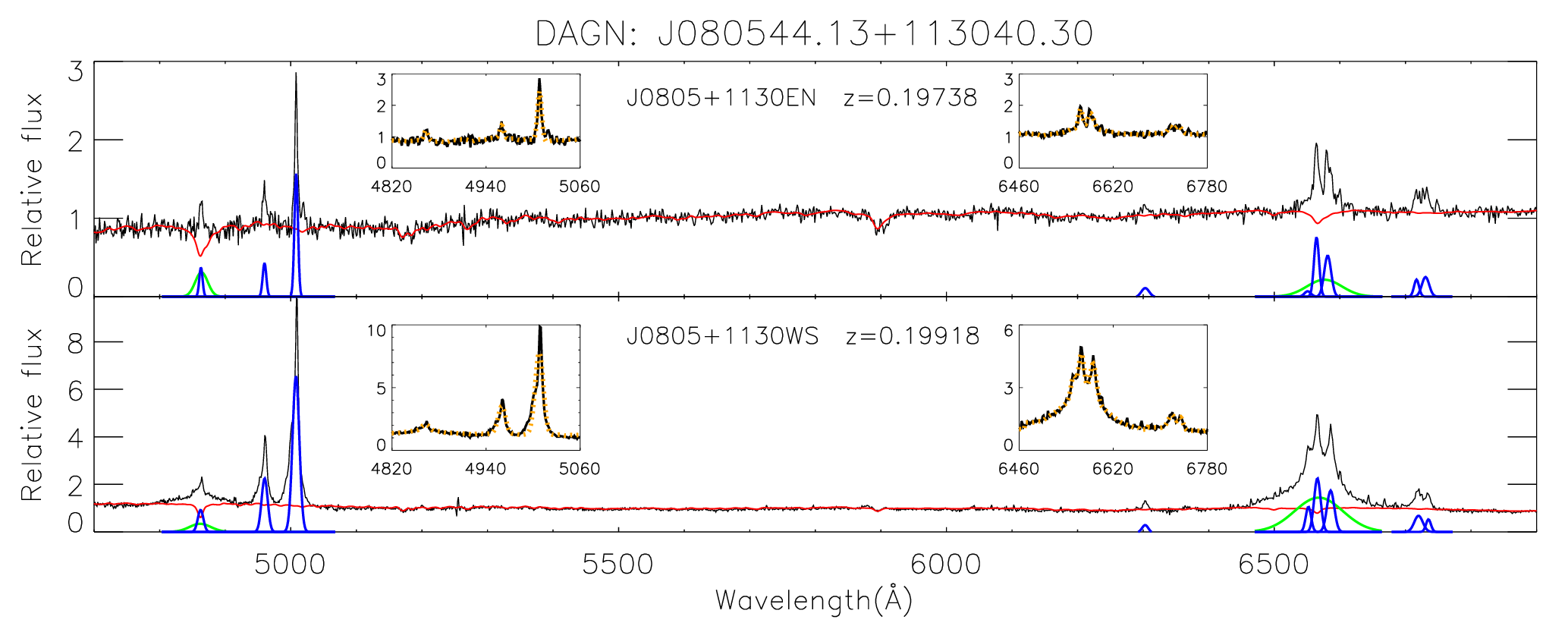}
 
\caption{Same as Fig.\,\ref{The spectra fitting of J1338+4816} but for dual AGN J080544.13+113040.30.} 
\label{The spectra fitting of J0805+1130}
\end{figure*}

\begin{figure*}[ht]
\centering
  \includegraphics[width=5.20cm,height=5.0cm]{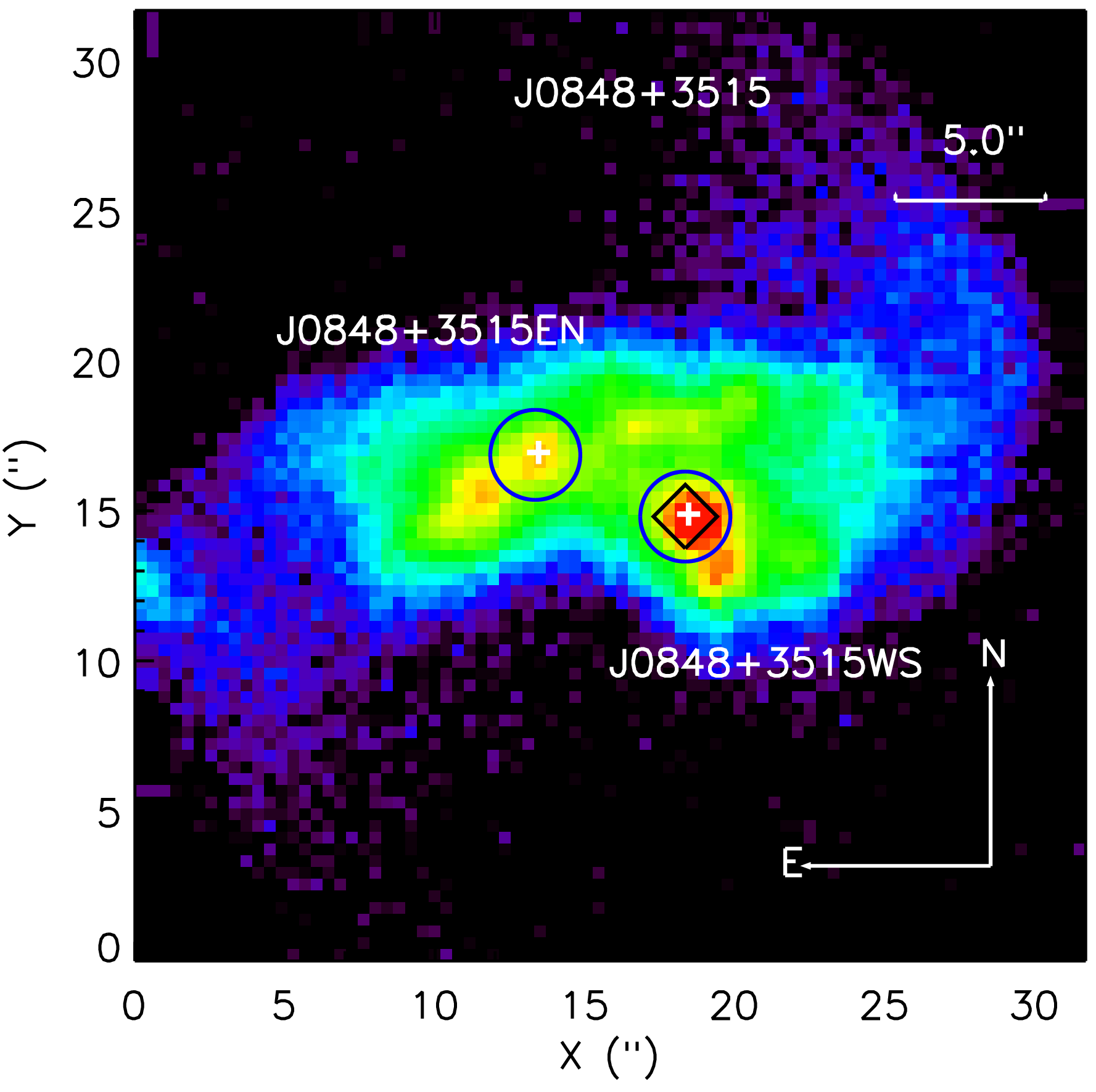}
  \includegraphics[width=12.2cm,height=5.2cm]{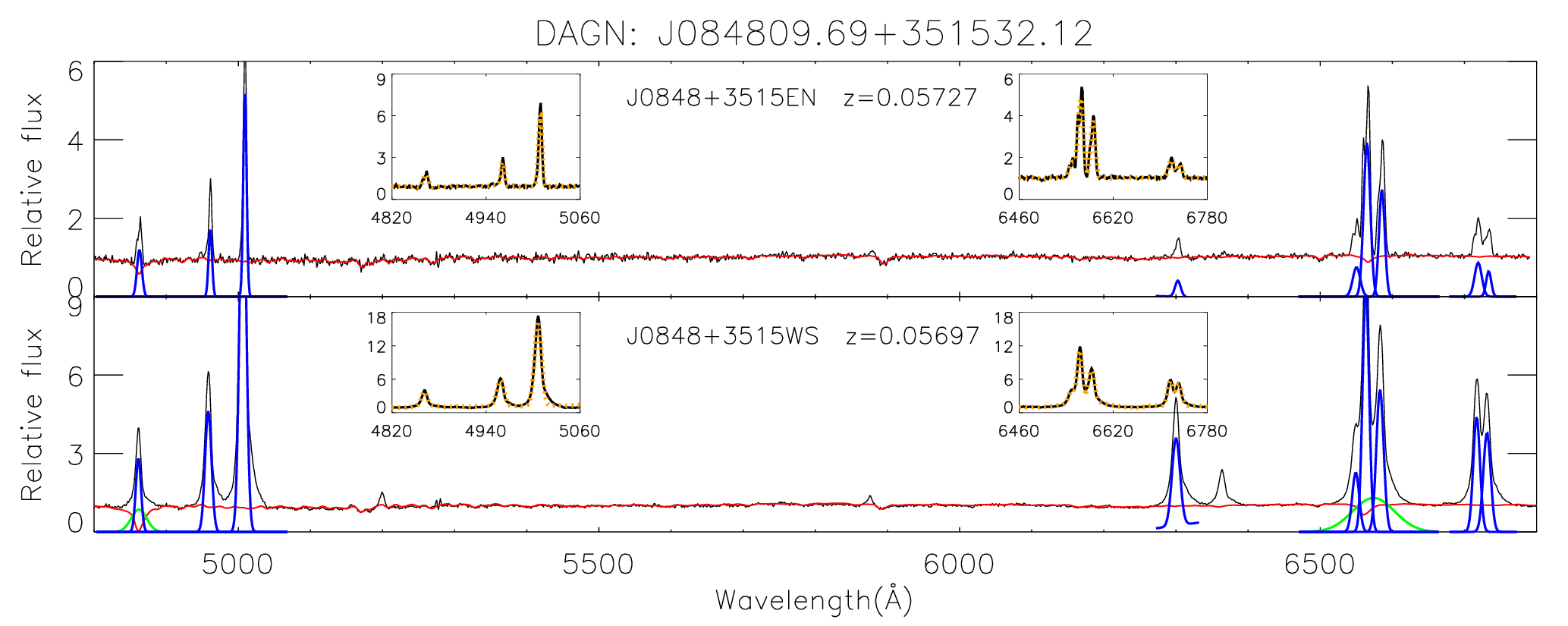}
\caption{Same as Fig.\,\ref{The spectra fitting of J1338+4816} but for dual AGN J084809.69+351532.12.} 
\label{The spectra fitting of J0848+3515}
\end{figure*}

\begin{figure*}[ht]
\centering
  \includegraphics[width=5.20cm,height=5.0cm]{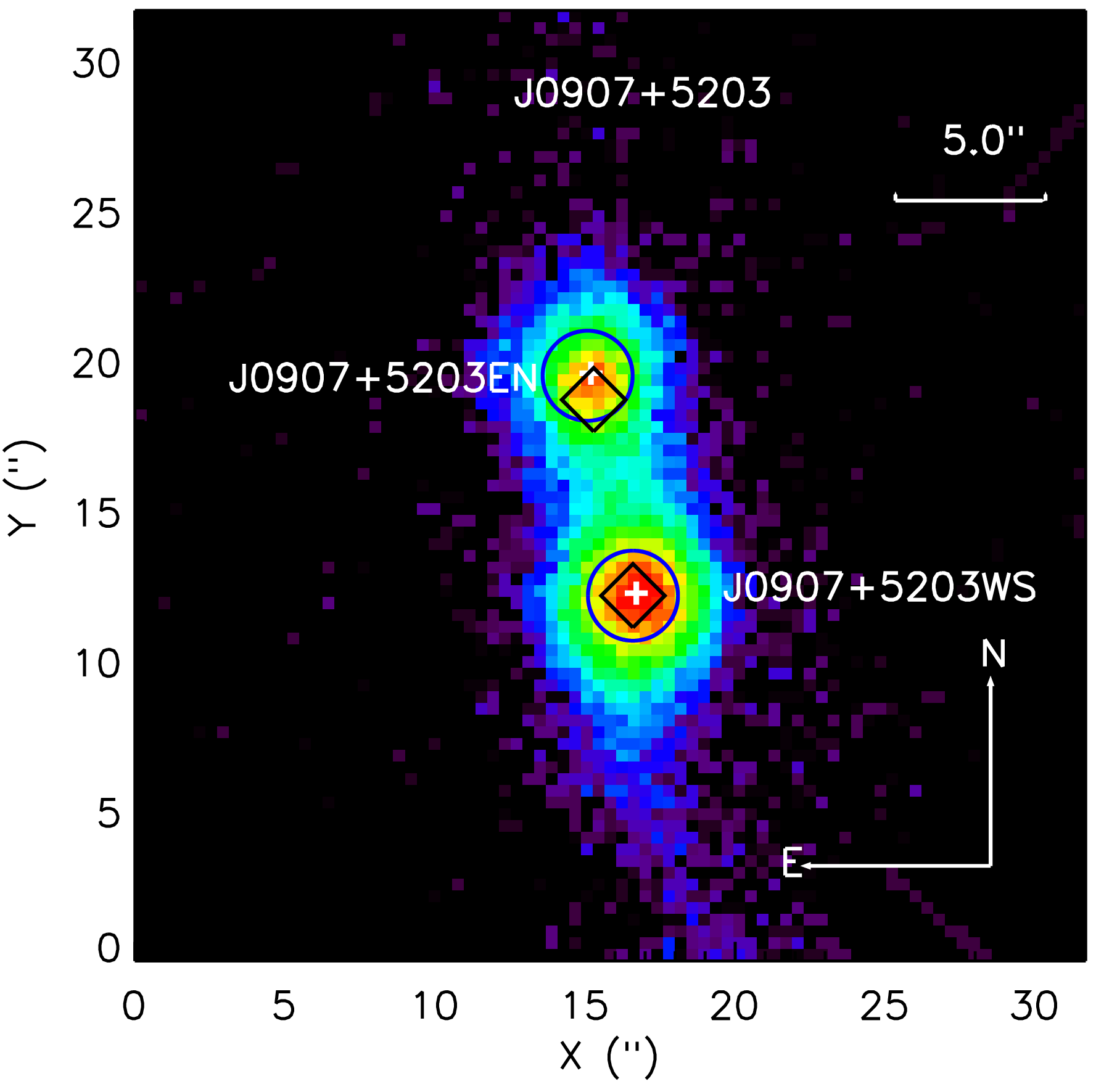}
  \includegraphics[width=12.2cm,height=5.2cm]{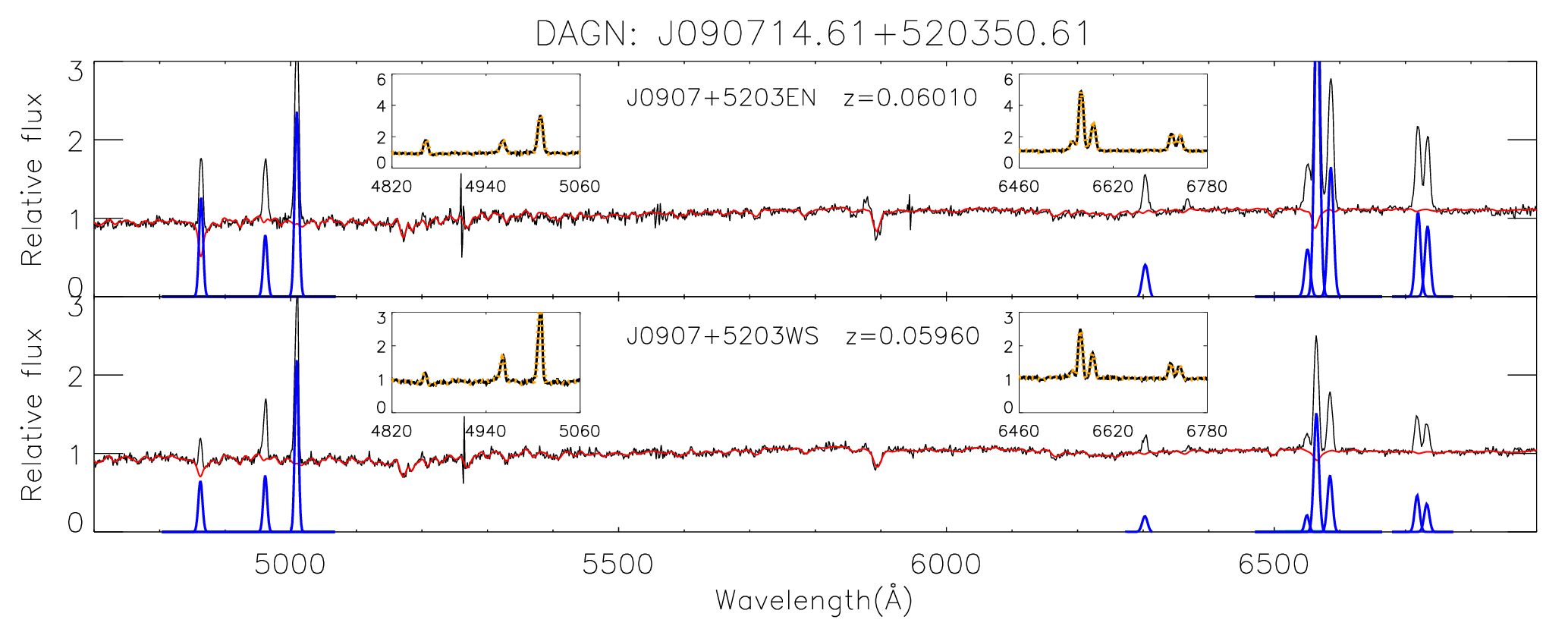}
\caption{Same as Fig.\,\ref{The spectra fitting of J1338+4816} but for dual AGN J090714.61+520350.61.} 
\label{The spectra fitting of J0907+5203}
\end{figure*}

\begin{figure*}[ht]
\centering
  \includegraphics[width=5.20cm,height=5.0cm]{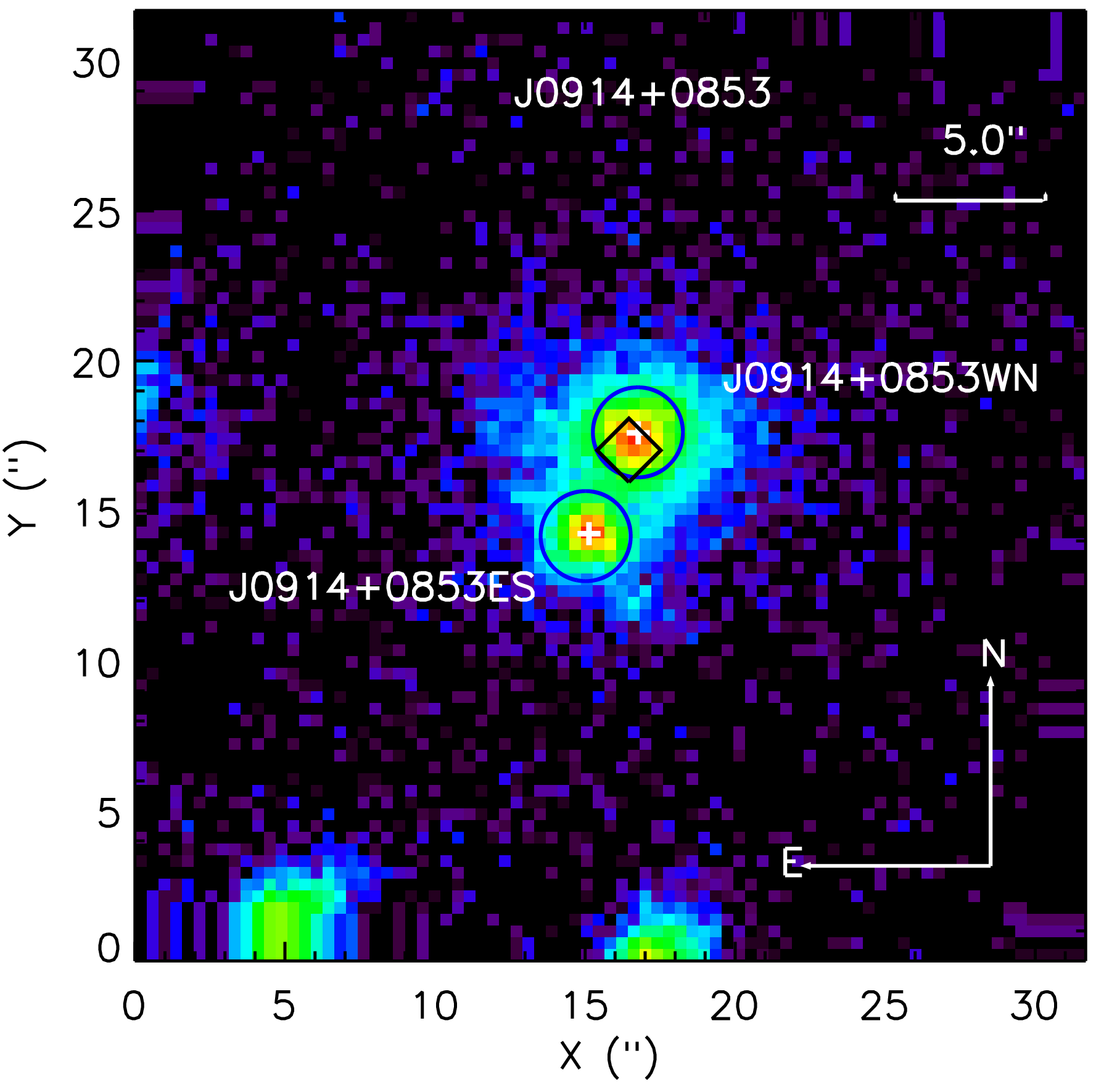}
  \includegraphics[width=12.2cm,height=5.2cm]{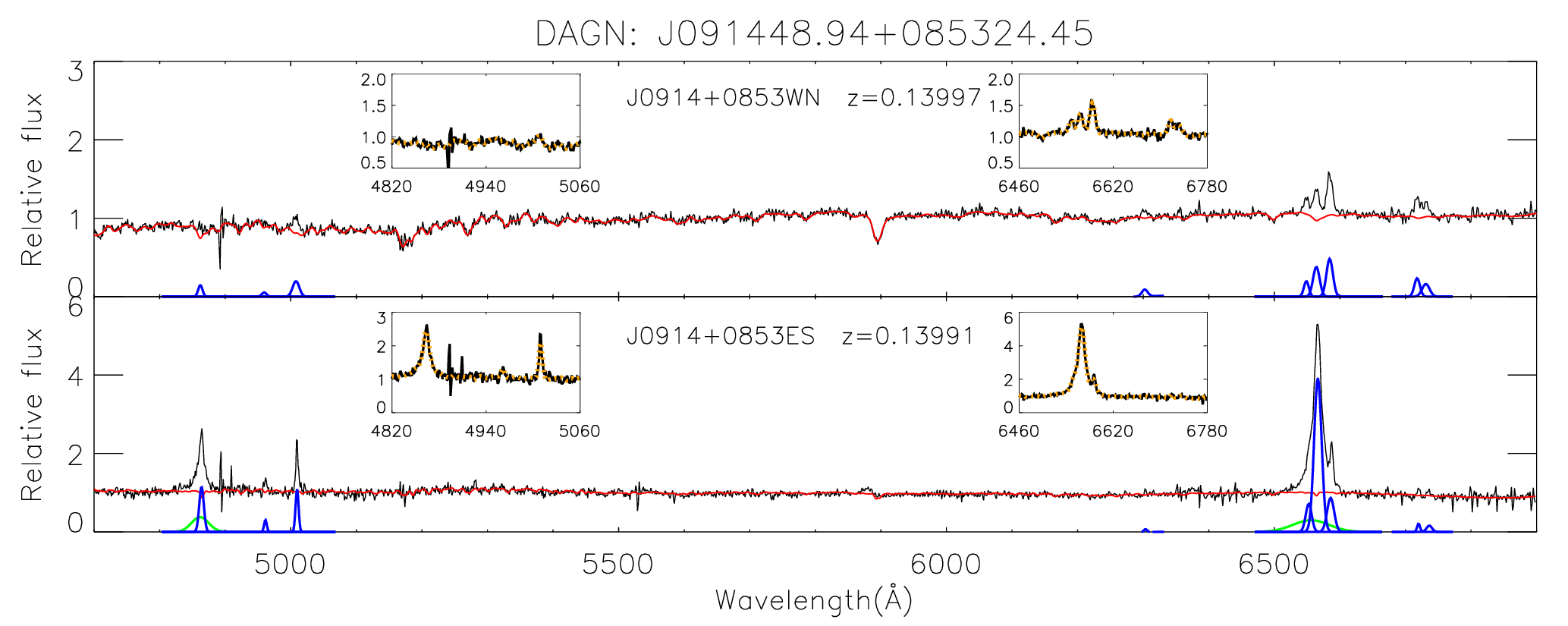}
 
\caption{Same as Fig.\,\ref{The spectra fitting of J1338+4816} but for dual AGN J091449.05+085321.10.} 
\label{The spectra fitting of J0914+0853}
\end{figure*}

\begin{figure*}[ht]
\centering
  \includegraphics[width=5.20cm,height=5.0cm]{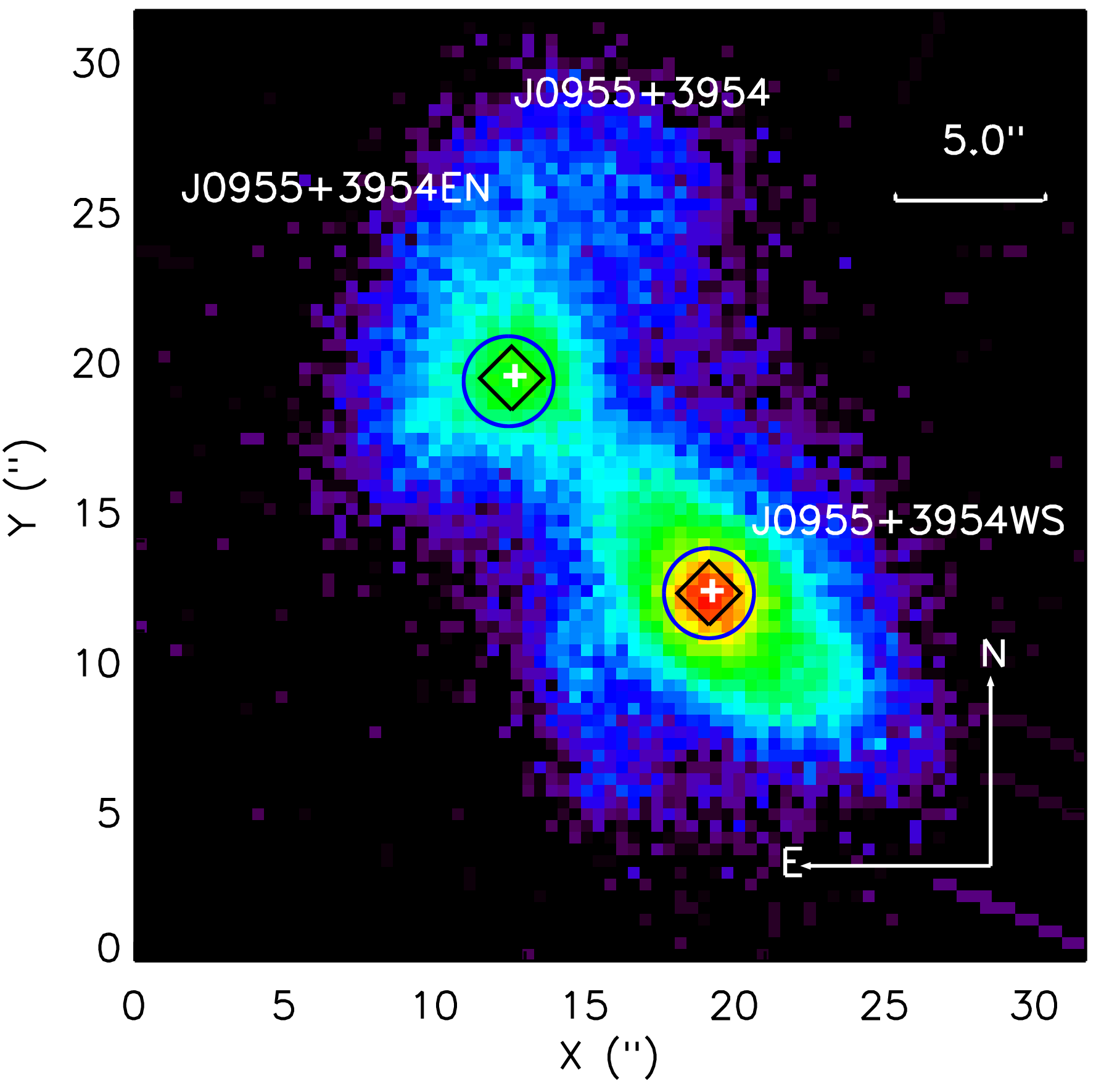}
  \includegraphics[width=12.2cm,height=5.2cm]{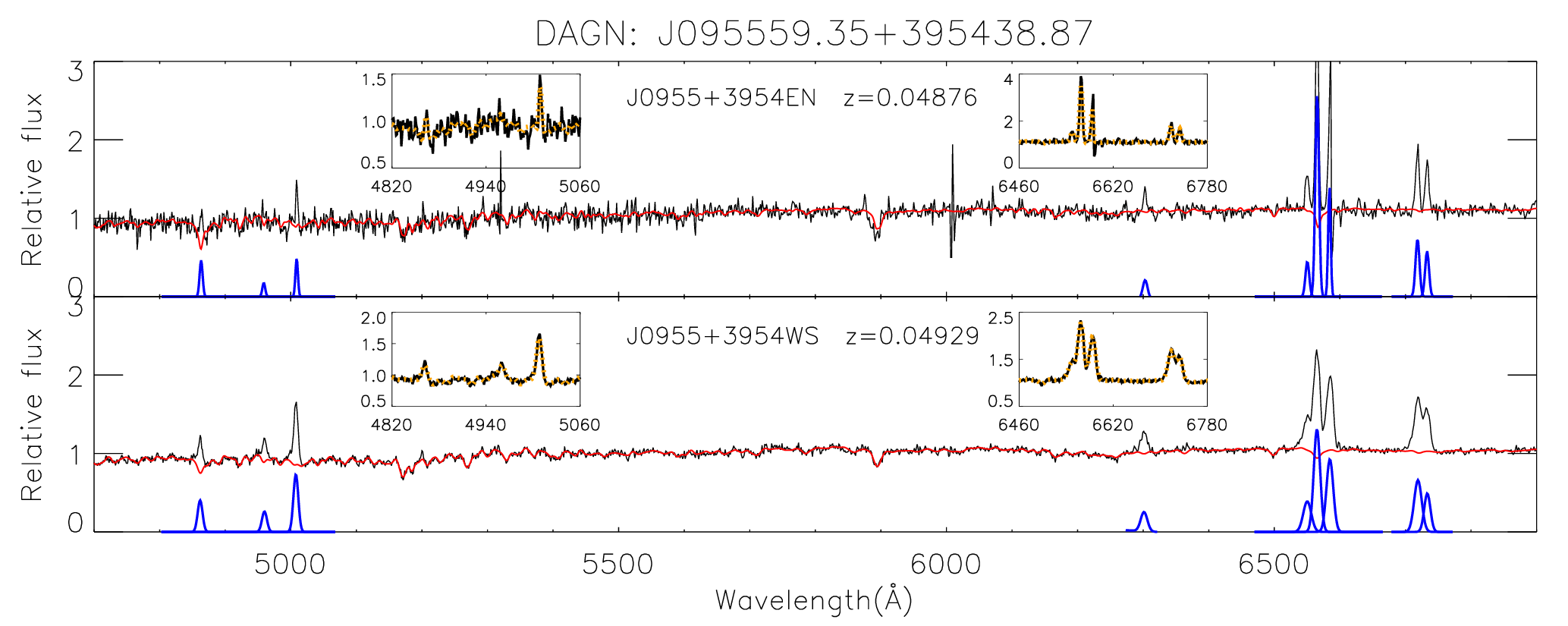}

\caption{Same as Fig.\,\ref{The spectra fitting of J1338+4816} but for dual AGN J095559.35+395438.87.}
\label{The spectra fitting of J0955+3954}
\end{figure*}

\begin{figure*}[ht]
\centering
  \includegraphics[width=5.20cm,height=5.0cm]{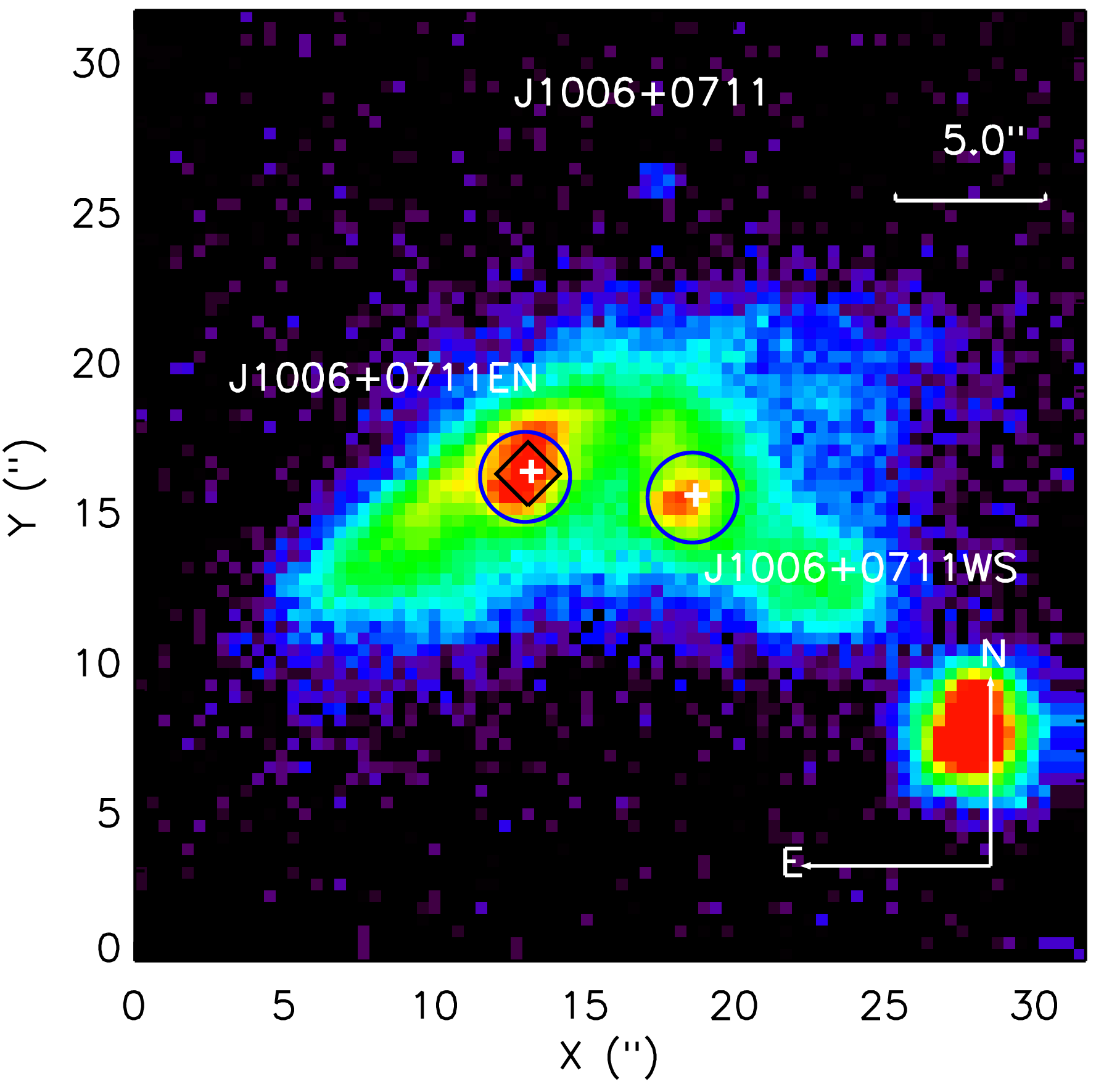}
  \includegraphics[width=12.2cm,height=5.2cm]{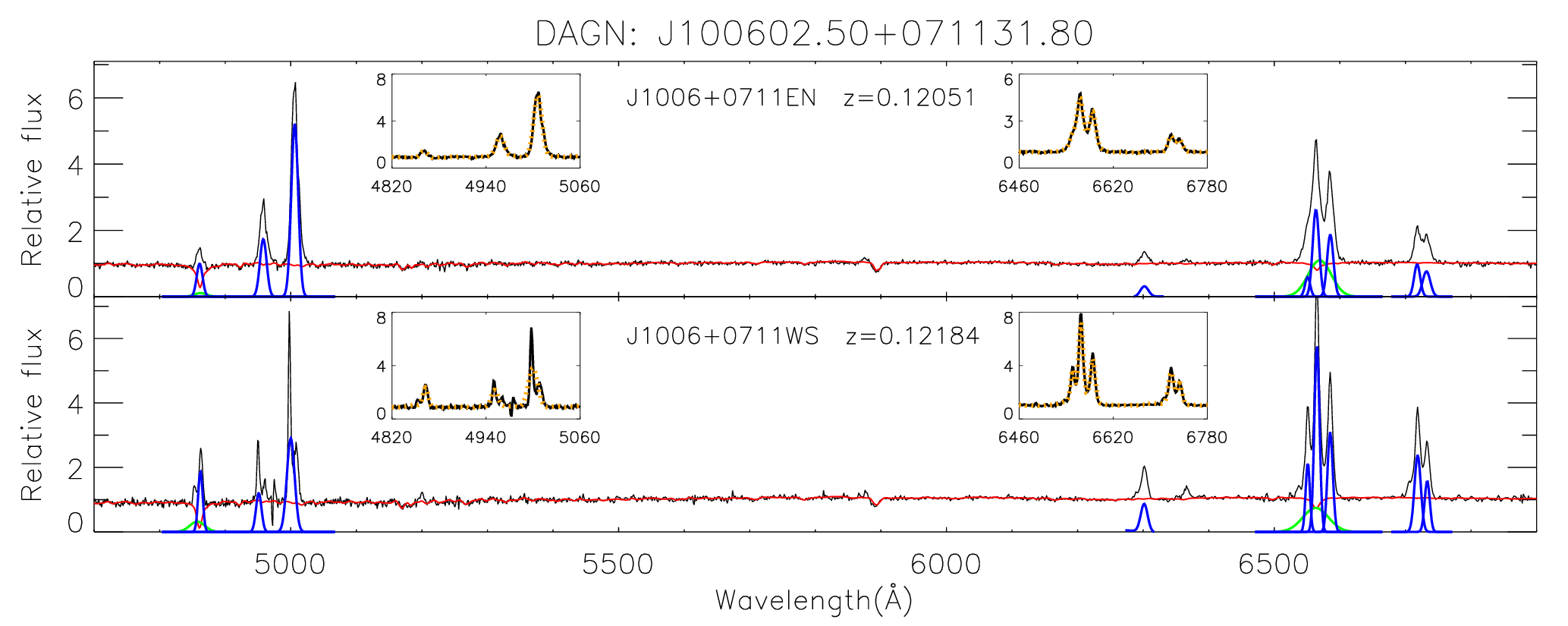}

\caption{Same as Fig.\,\ref{The spectra fitting of J1338+4816} but for dual AGN J100602.50+071131.80.}
\label{The spectra fitting of J1006+0711}
\end{figure*}

\begin{figure*}[ht]
\centering
  \includegraphics[width=5.20cm,height=5.0cm]{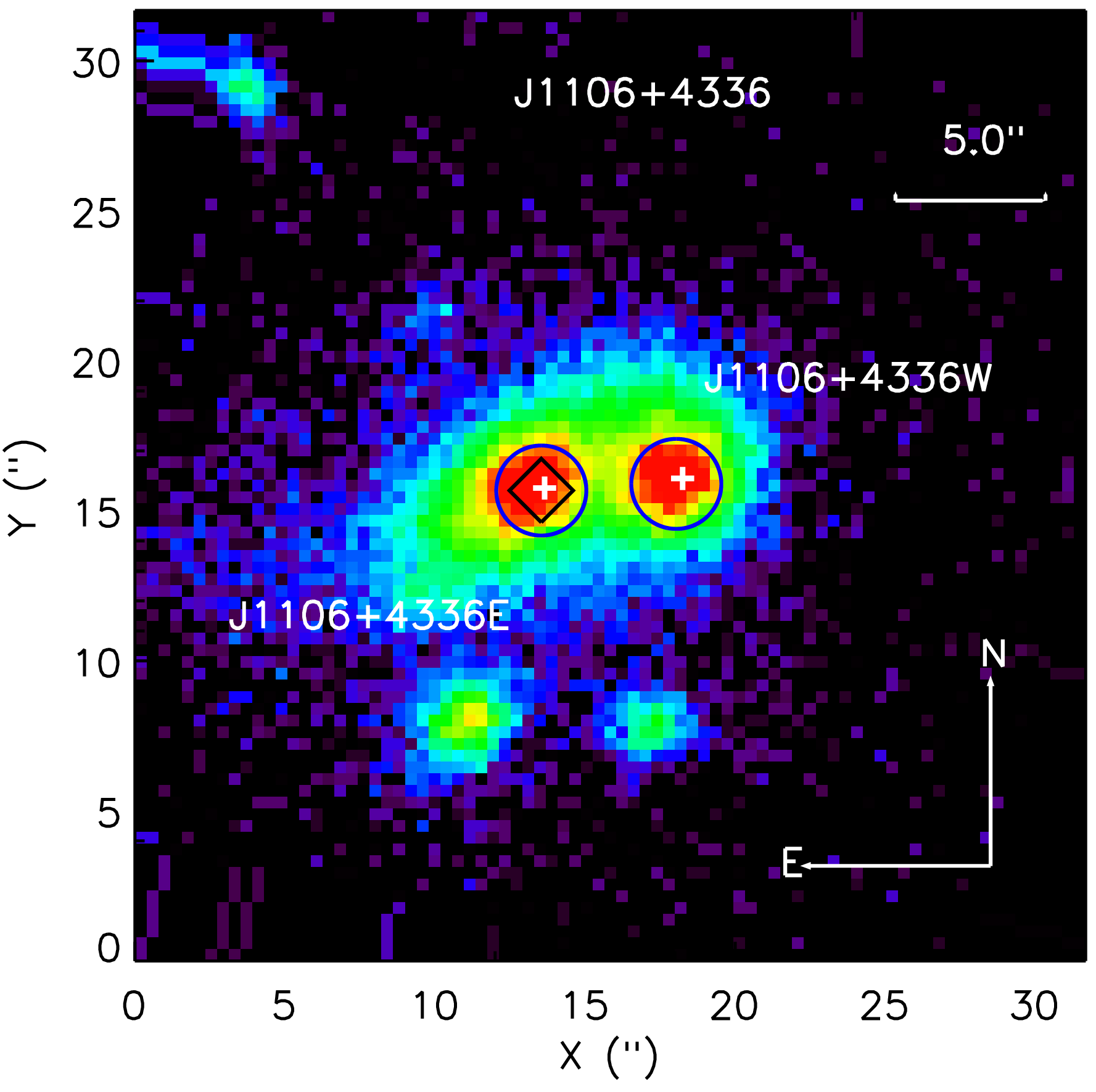}
  \includegraphics[width=12.2cm,height=5.2cm]{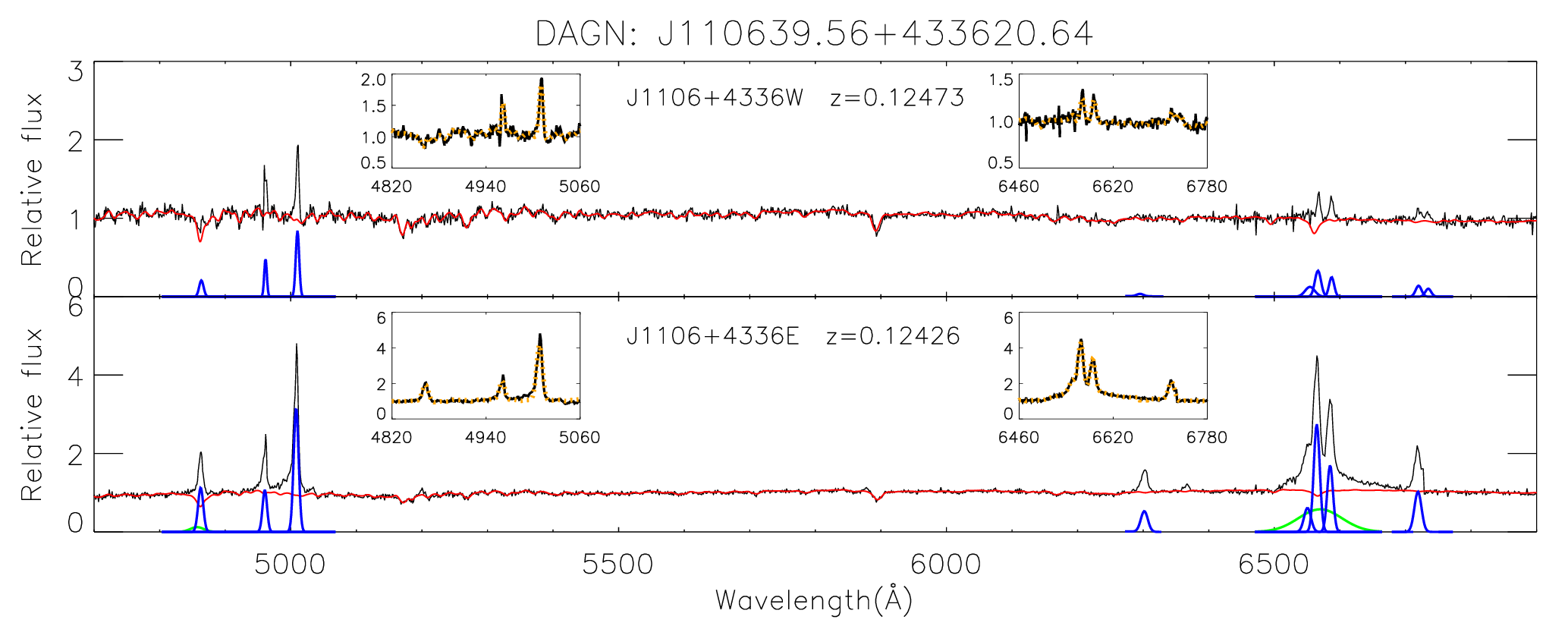}

\caption{Same as Fig.\,\ref{The spectra fitting of J1338+4816} but for dual AGN J110639.56+433620.64.}
\label{The spectra fitting of J1106+4336}
\end{figure*}

\begin{figure*}[ht]
\centering
  \includegraphics[width=5.20cm,height=5.0cm]{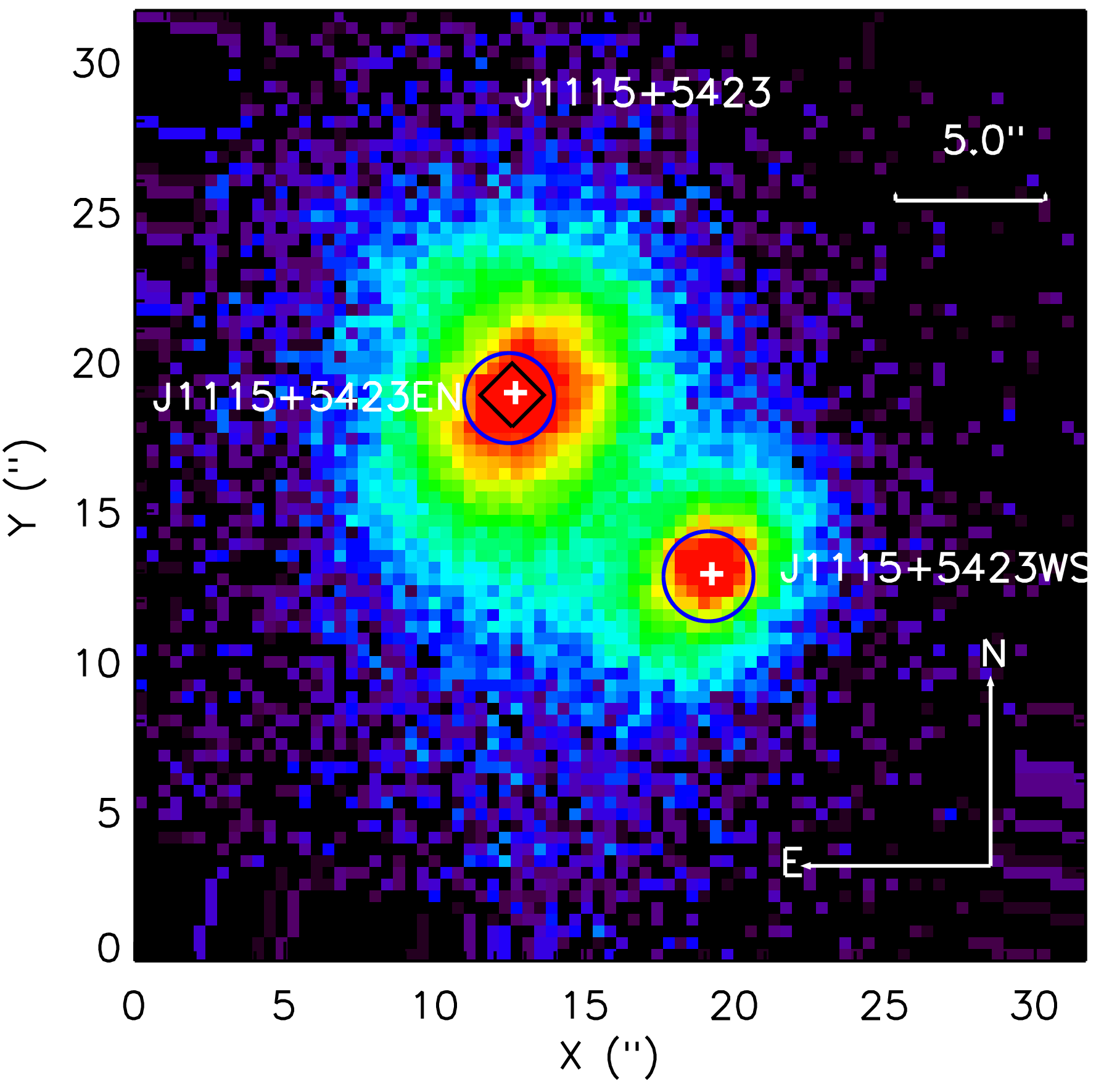}
  \includegraphics[width=12.2cm,height=5.2cm]{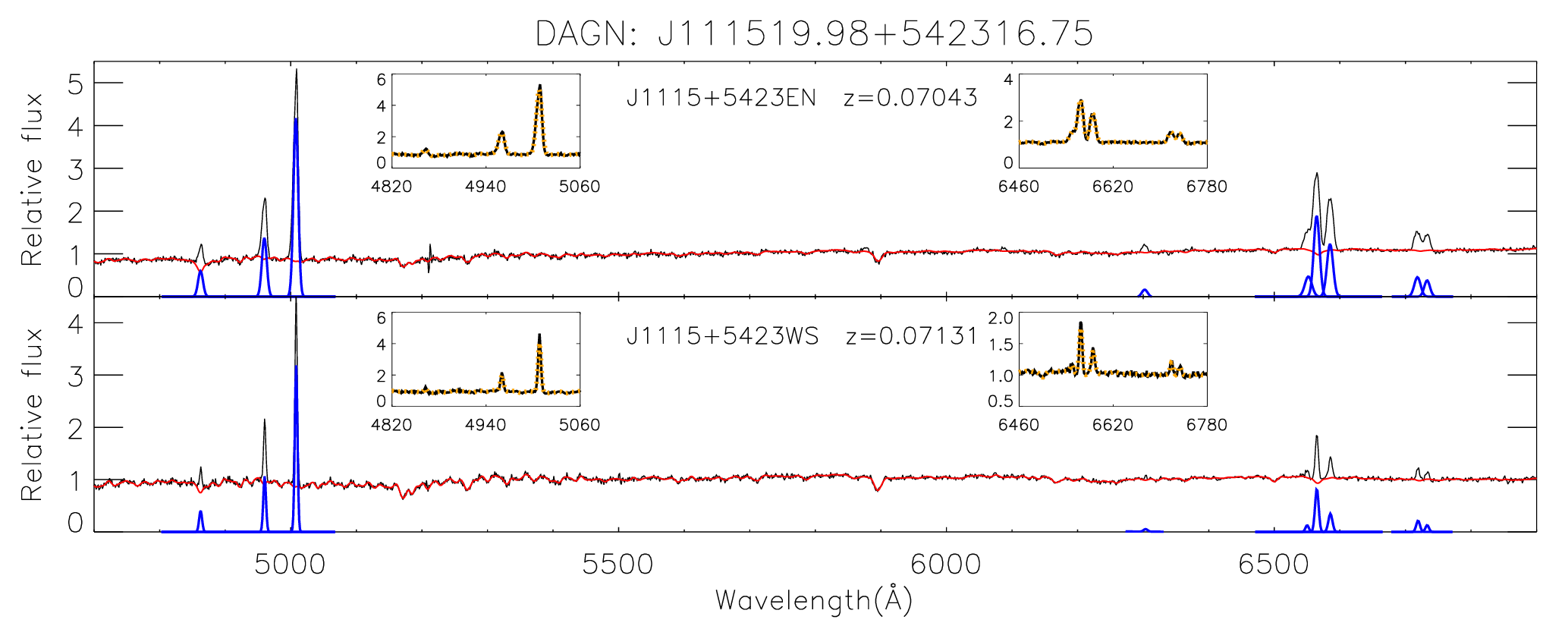}

\caption{Same as Fig.\,\ref{The spectra fitting of J1338+4816} but for dual AGN J111519.98+542316.75.}
\label{The spectra fitting of J1115+5423}
\end{figure*}

\begin{figure*}[ht]
\centering
  \includegraphics[width=5.20cm,height=5.0cm]{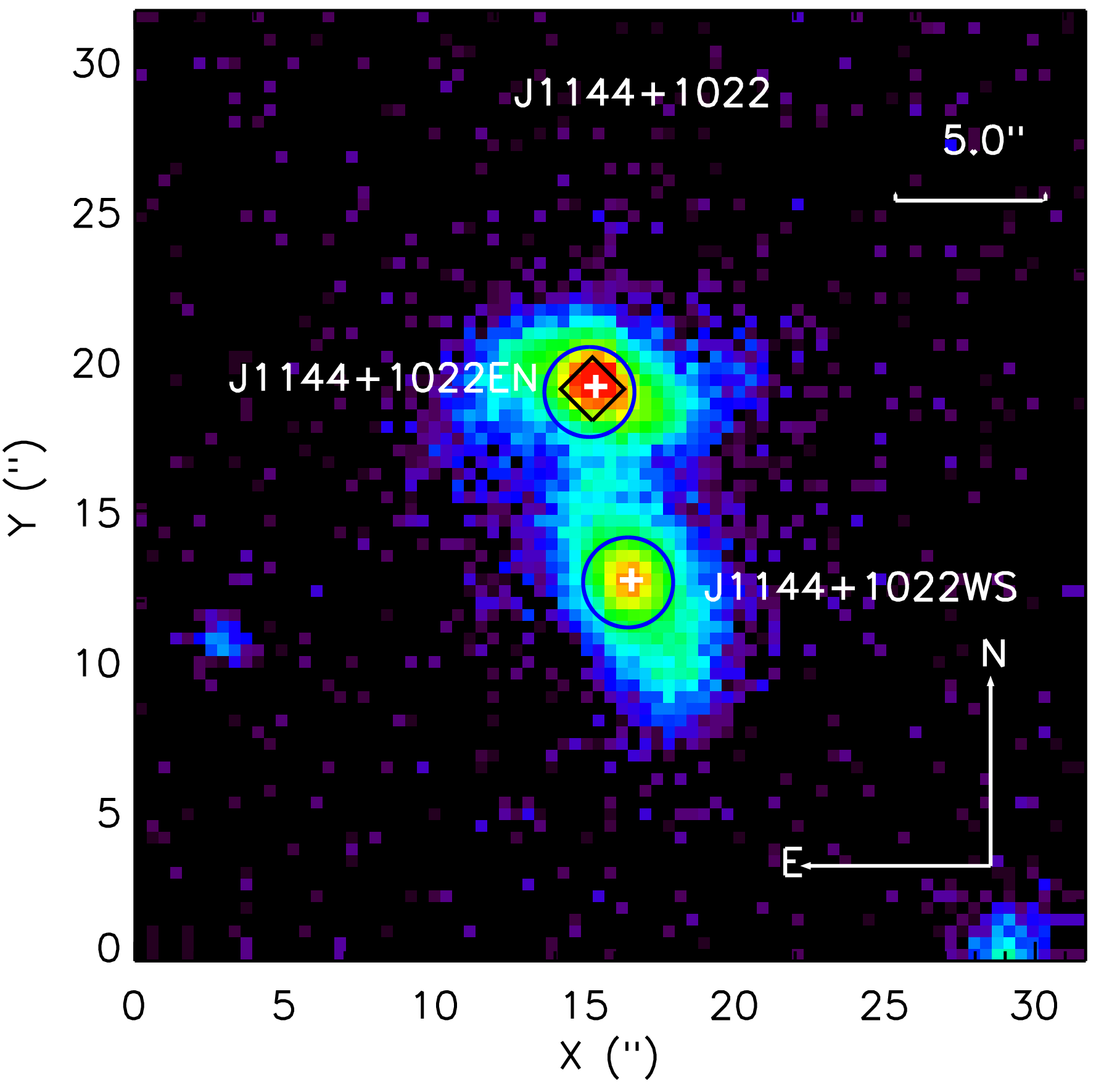}
  \includegraphics[width=12.2cm,height=5.2cm]{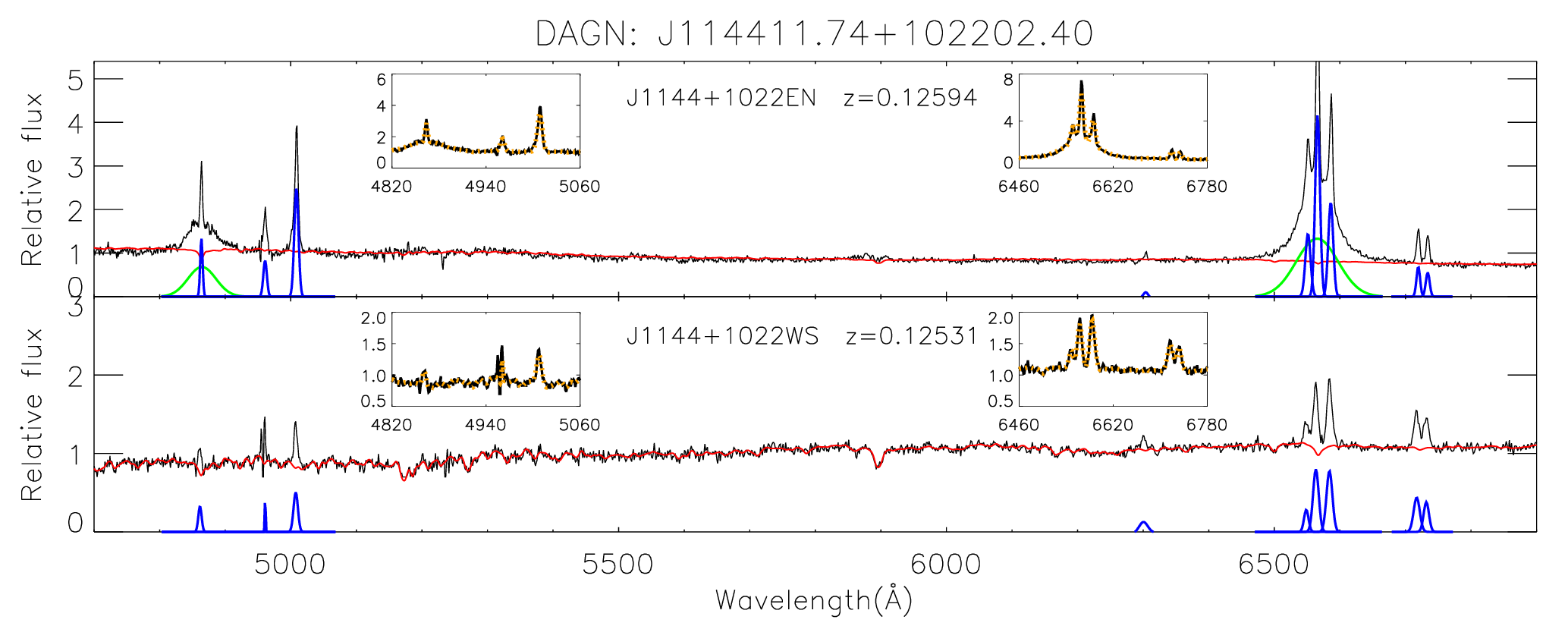}

\caption{Same as Fig.\,\ref{The spectra fitting of J1338+4816} but for dual AGN J114411.74+102202.40.}
\label{The spectra fitting of J1144+1022}
\end{figure*}

\begin{figure*}[ht]
\centering
  \includegraphics[width=5.20cm,height=5.0cm]{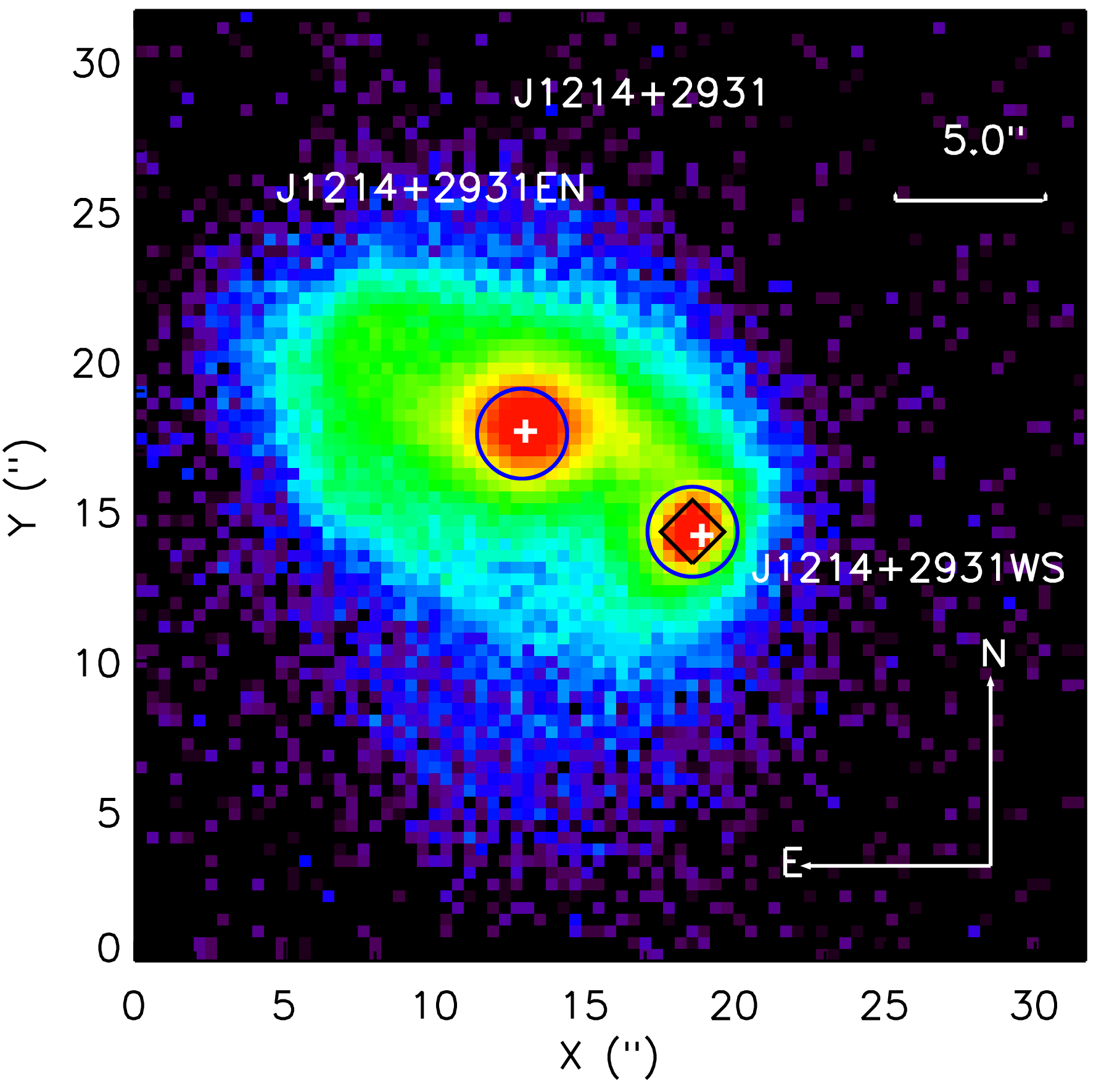}
  \includegraphics[width=12.2cm,height=5.2cm]{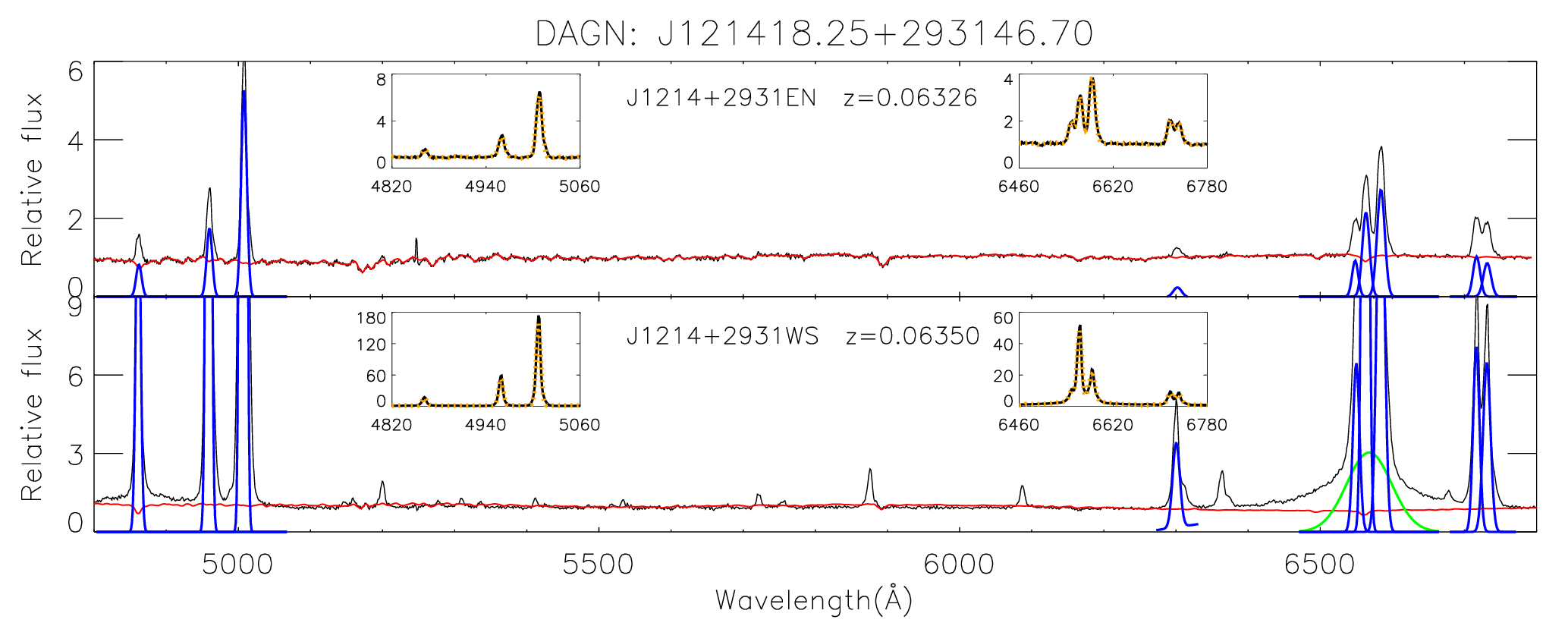}

\caption{Same as Fig.\,\ref{The spectra fitting of J1338+4816} but for dual AGN J121418.25+293146.70.}
\label{The spectra fitting of J1214+2931}
\end{figure*}

\begin{figure*}[ht]
\centering
  \includegraphics[width=5.20cm,height=5.0cm]{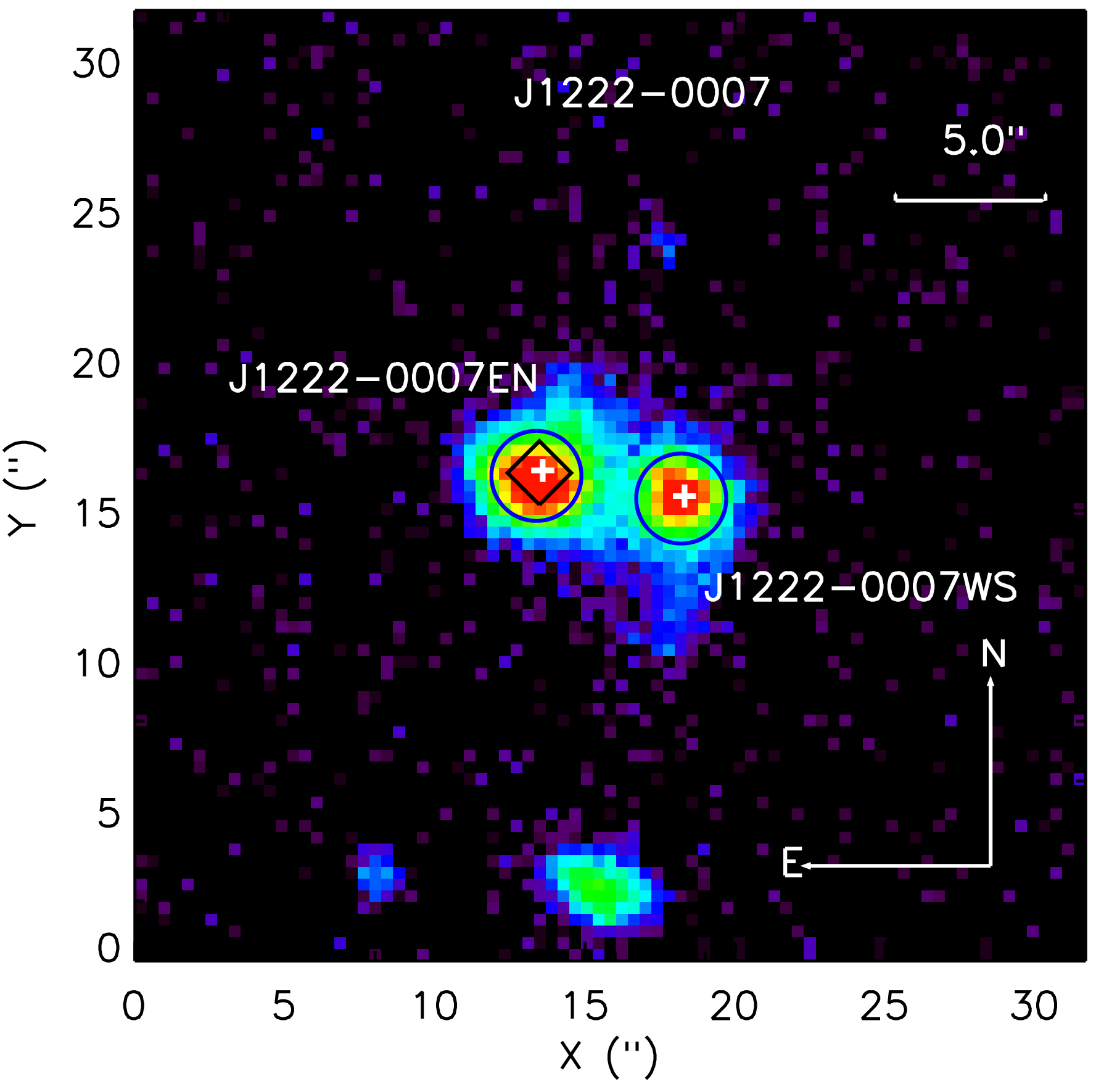}
  \includegraphics[width=12.2cm,height=5.2cm]{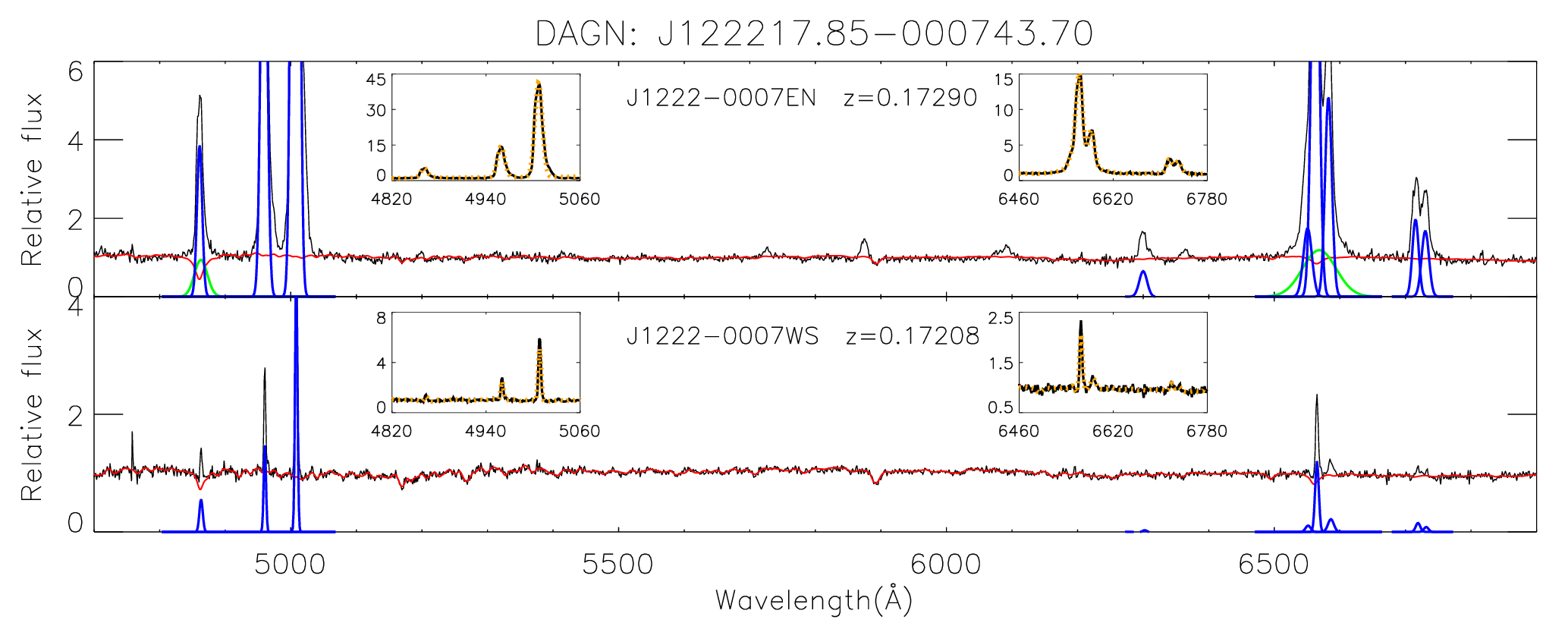}
  
\caption{Same as Fig.\,\ref{The spectra fitting of J1338+4816} but for dual AGN J122217.85-000743.70.}
\label{The spectra fitting of J1222-0007}
\end{figure*}

\begin{figure*}[ht]
\centering
  \includegraphics[width=5.20cm,height=5.0cm]{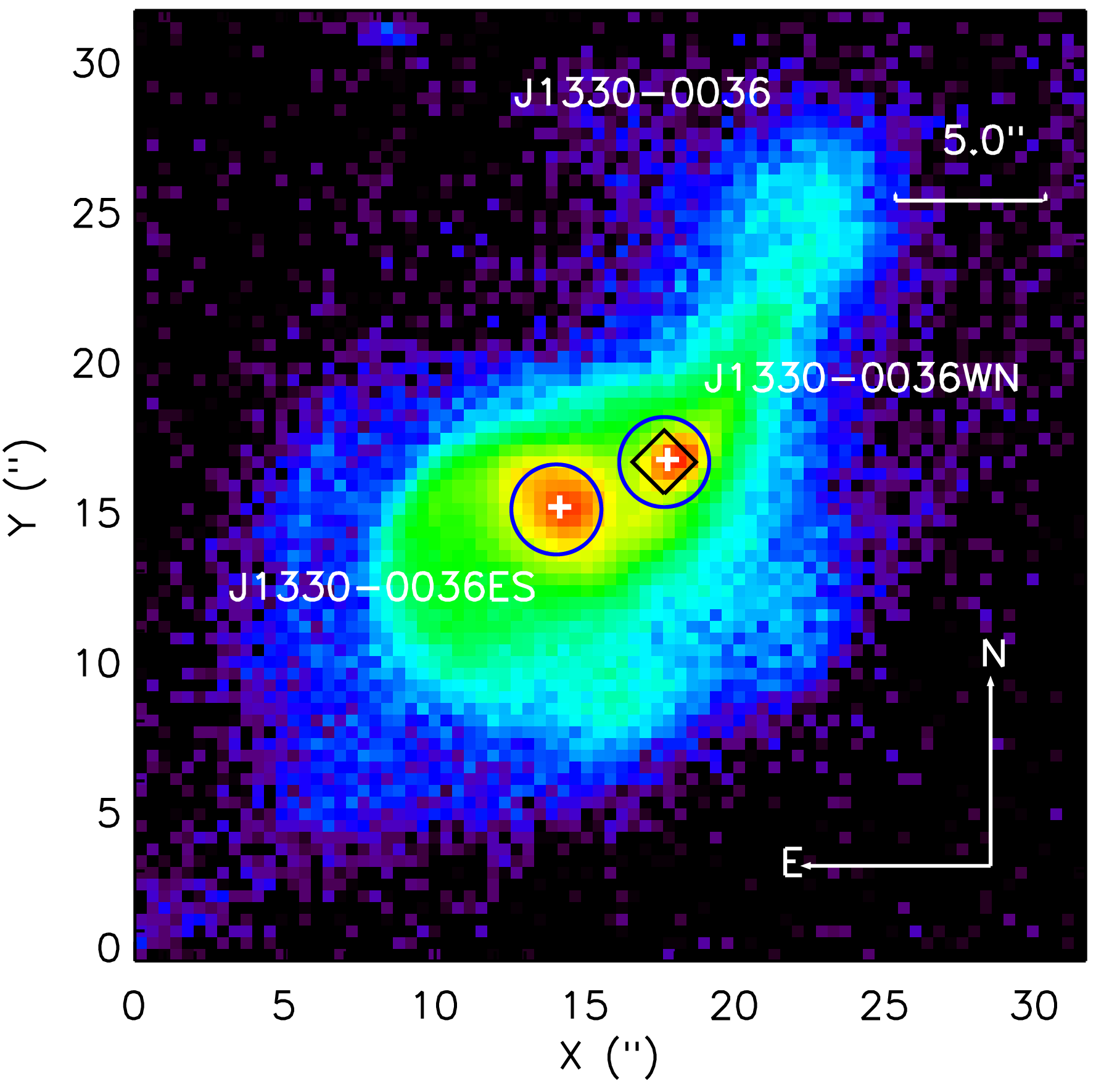}
  \includegraphics[width=12.2cm,height=5.2cm]{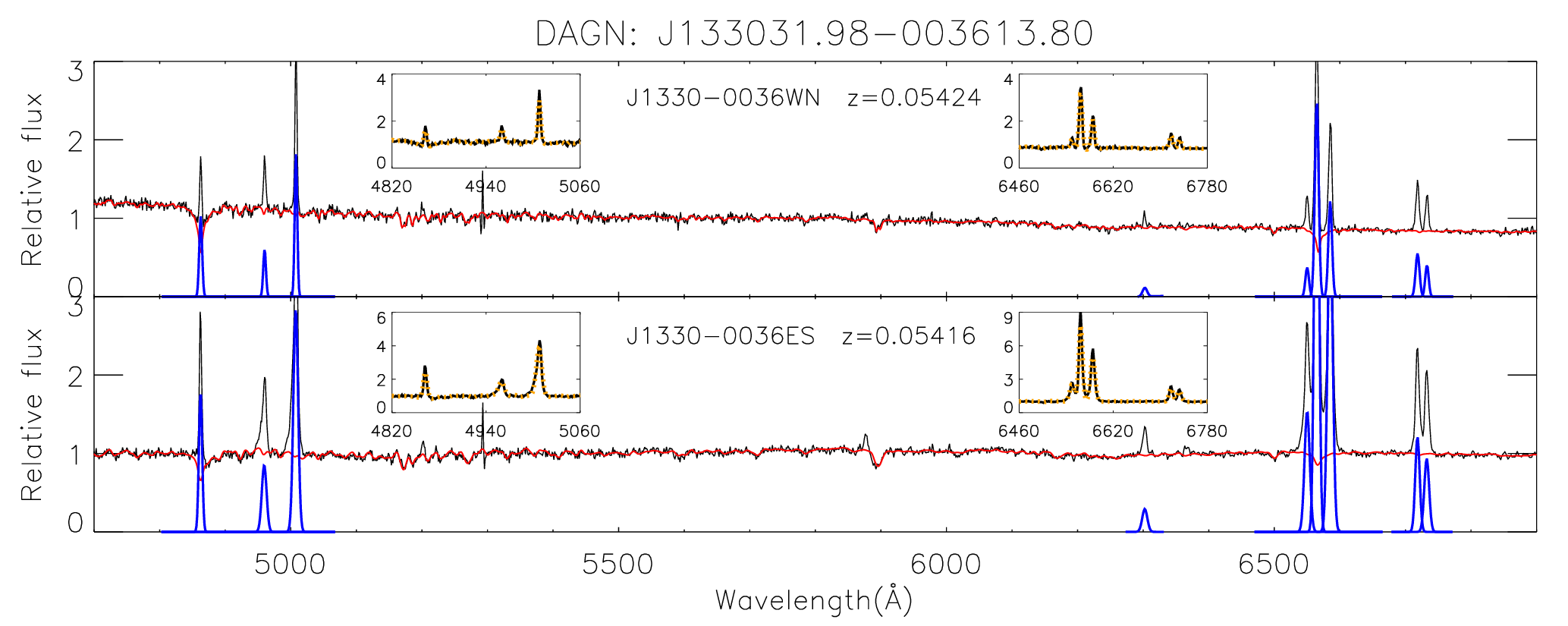}
  
\caption{Same as Fig.\,\ref{The spectra fitting of J1338+4816} but for dual AGN J133031.98-003613.80.}
\label{The spectra fitting of J1330-0036}
\end{figure*}

\begin{figure*}[ht]
\centering
  \includegraphics[width=5.20cm,height=5.0cm]{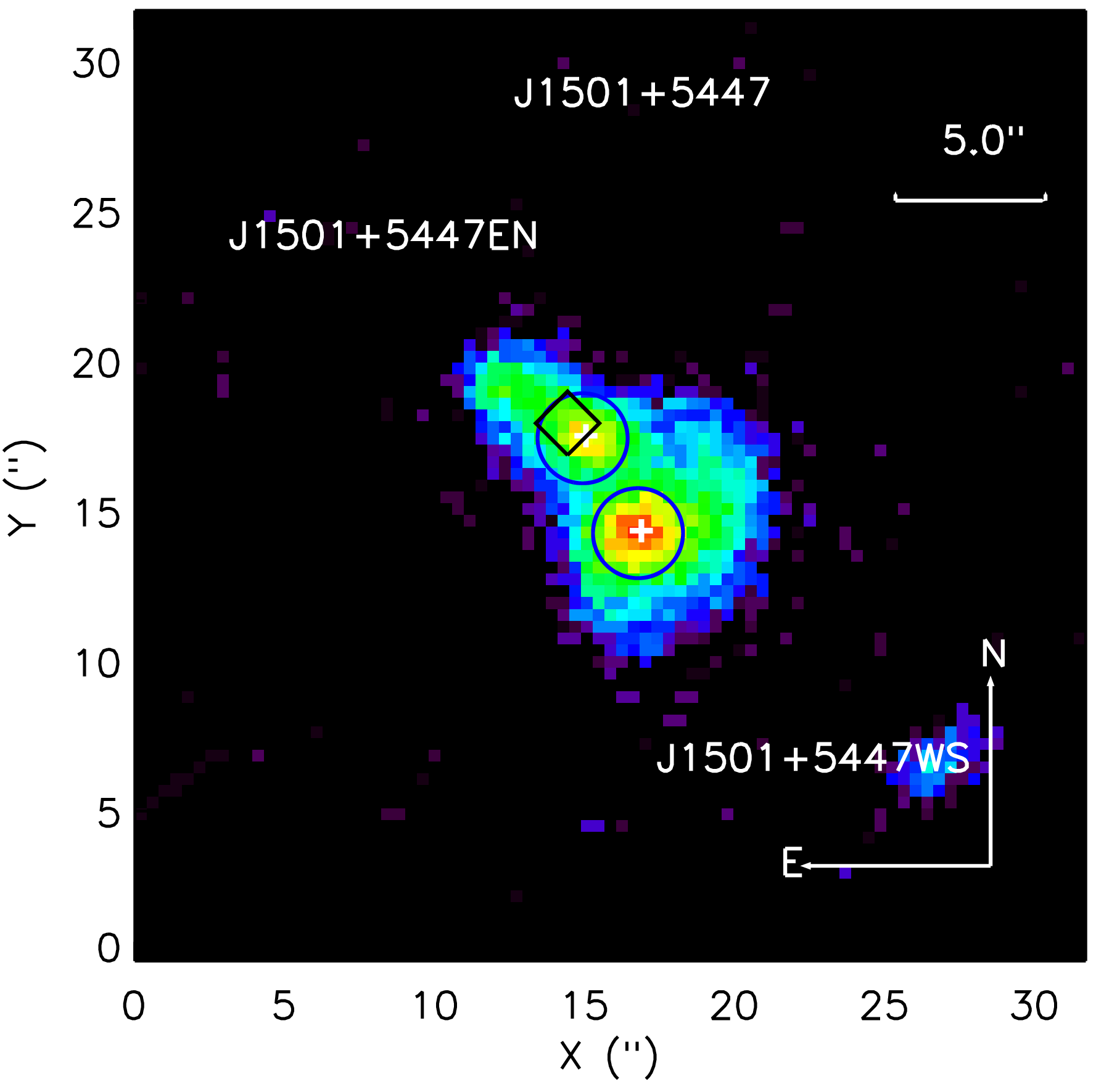}
  \includegraphics[width=12.2cm,height=5.2cm]{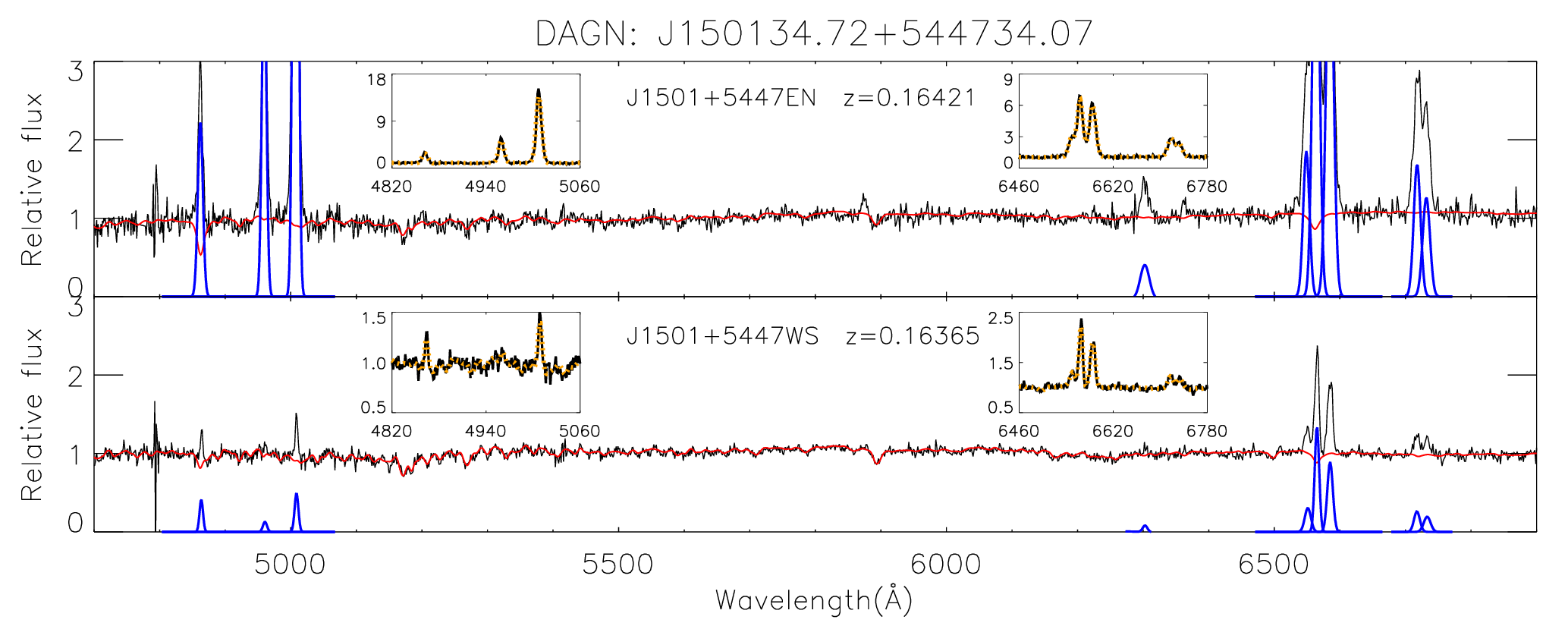}
  
\caption{Same as Fig.\,\ref{The spectra fitting of J1338+4816} but for dual AGN J150134.72+544734.07.}
\label{The spectra fitting of J1501+5447}
\end{figure*}

\begin{figure*}[ht]
\centering
  \includegraphics[width=5.20cm,height=5.0cm]{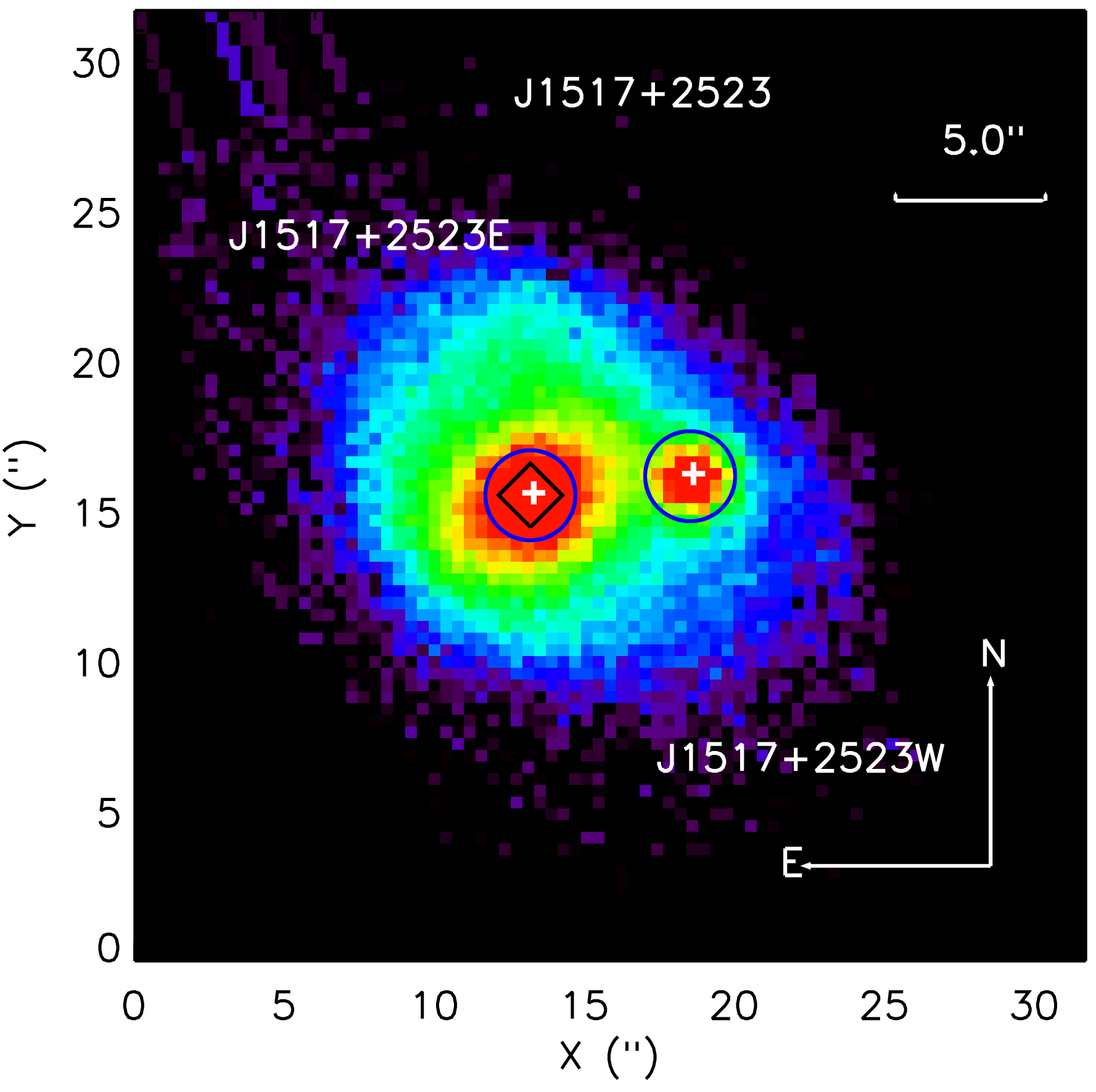}
  \includegraphics[width=12.2cm,height=5.2cm]{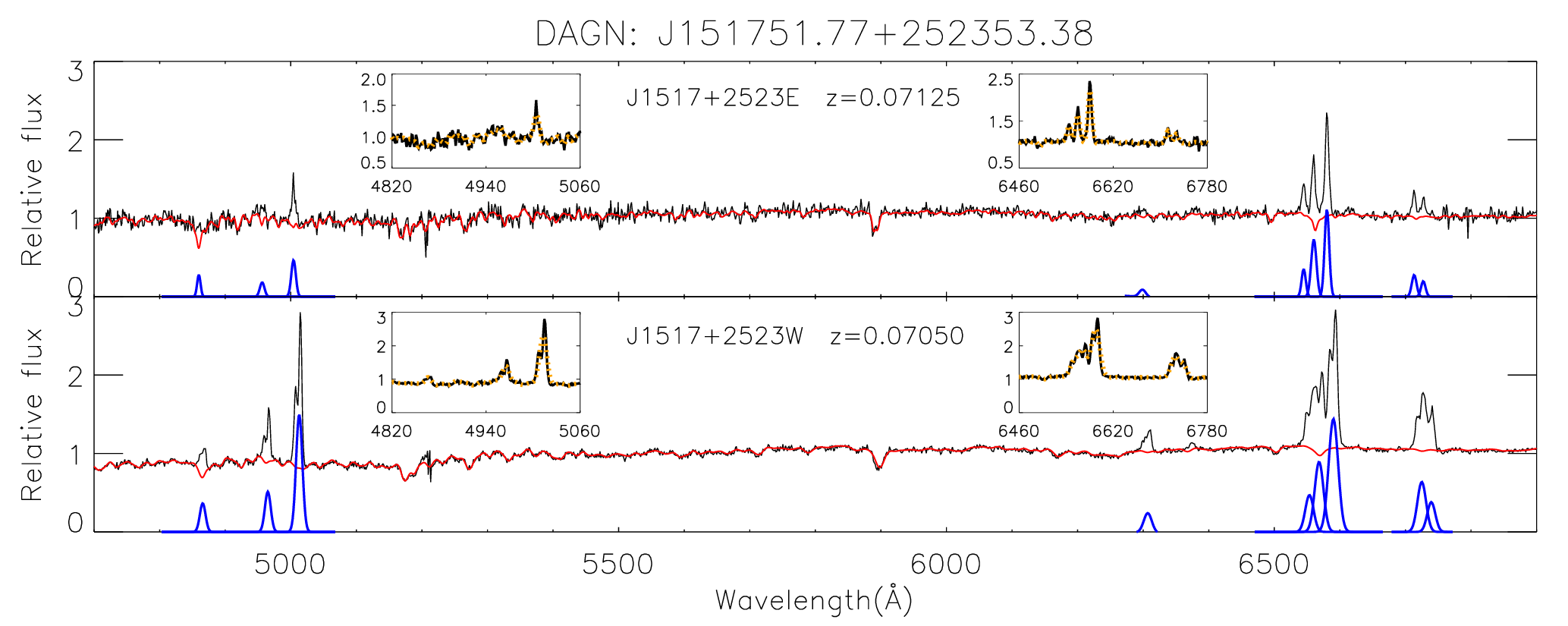}
  
\caption{Same as Fig.\,\ref{The spectra fitting of J1338+4816} but for dual AGN J151751.77+252353.38.}
\label{The spectra fitting of J1517+2523}
\end{figure*}

\begin{figure*}[ht]
\centering
  \includegraphics[width=5.20cm,height=5.0cm]{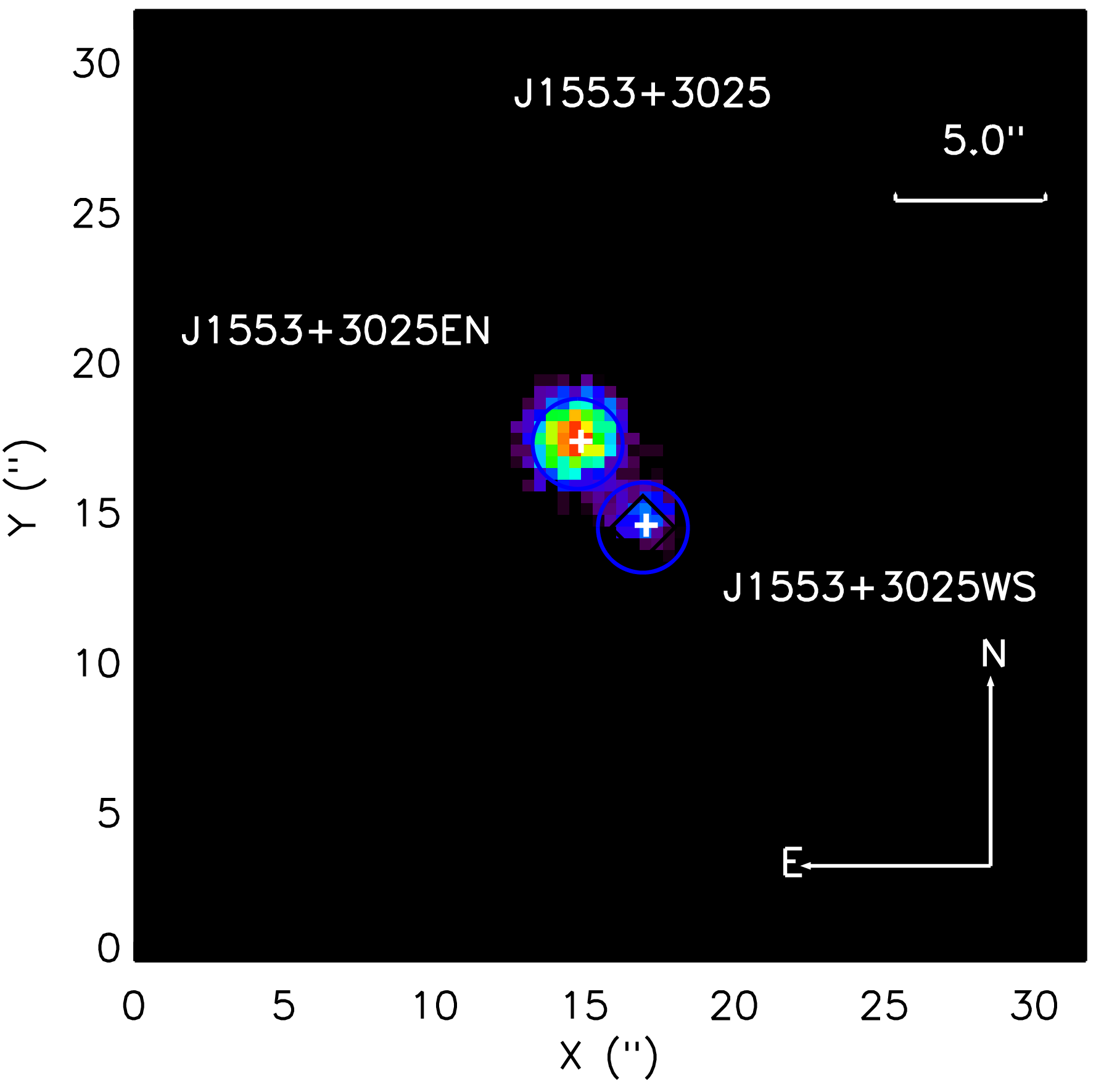}
  \includegraphics[width=12.2cm,height=5.2cm]{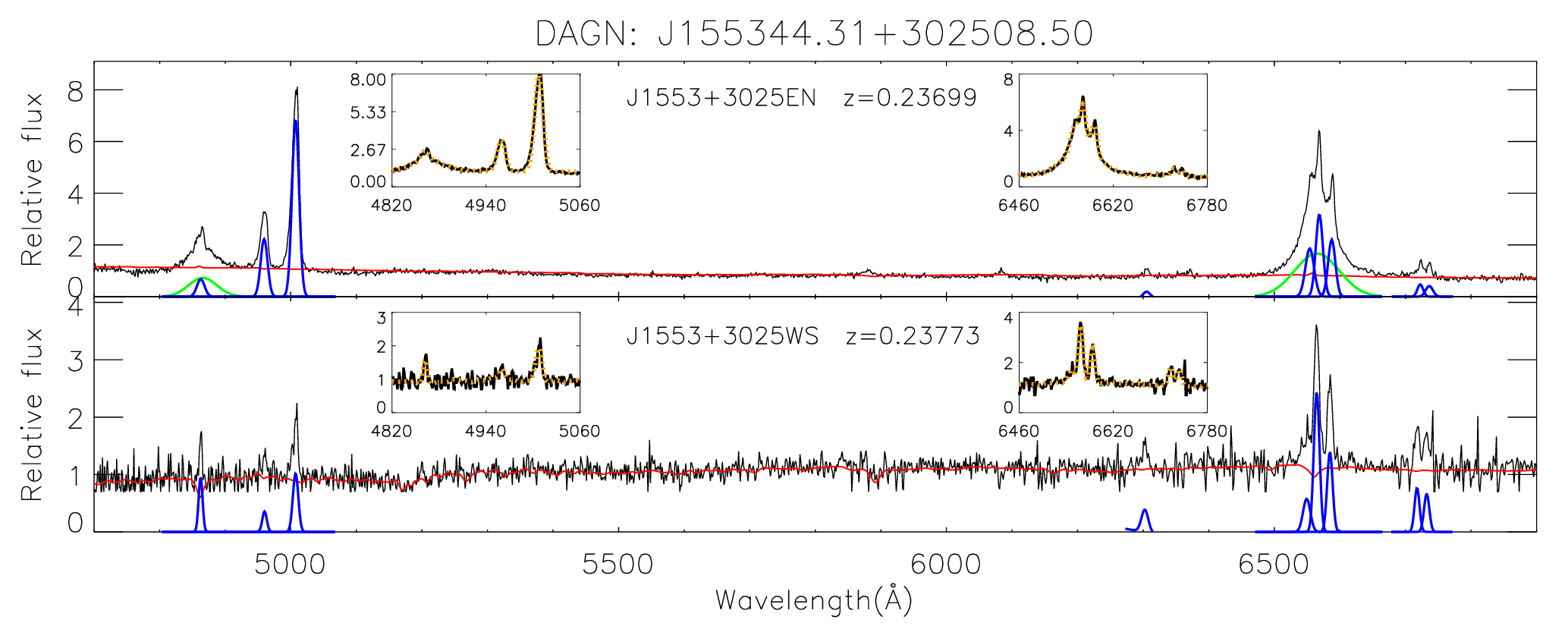}
  
\caption{Same as Fig.\,\ref{The spectra fitting of J1338+4816} but for dual AGN J155344.31+302508.50.}
\label{The spectra fitting of J1553+3025}
\end{figure*}

\begin{figure*}[ht]
\centering
  \includegraphics[width=5.20cm,height=5.0cm]{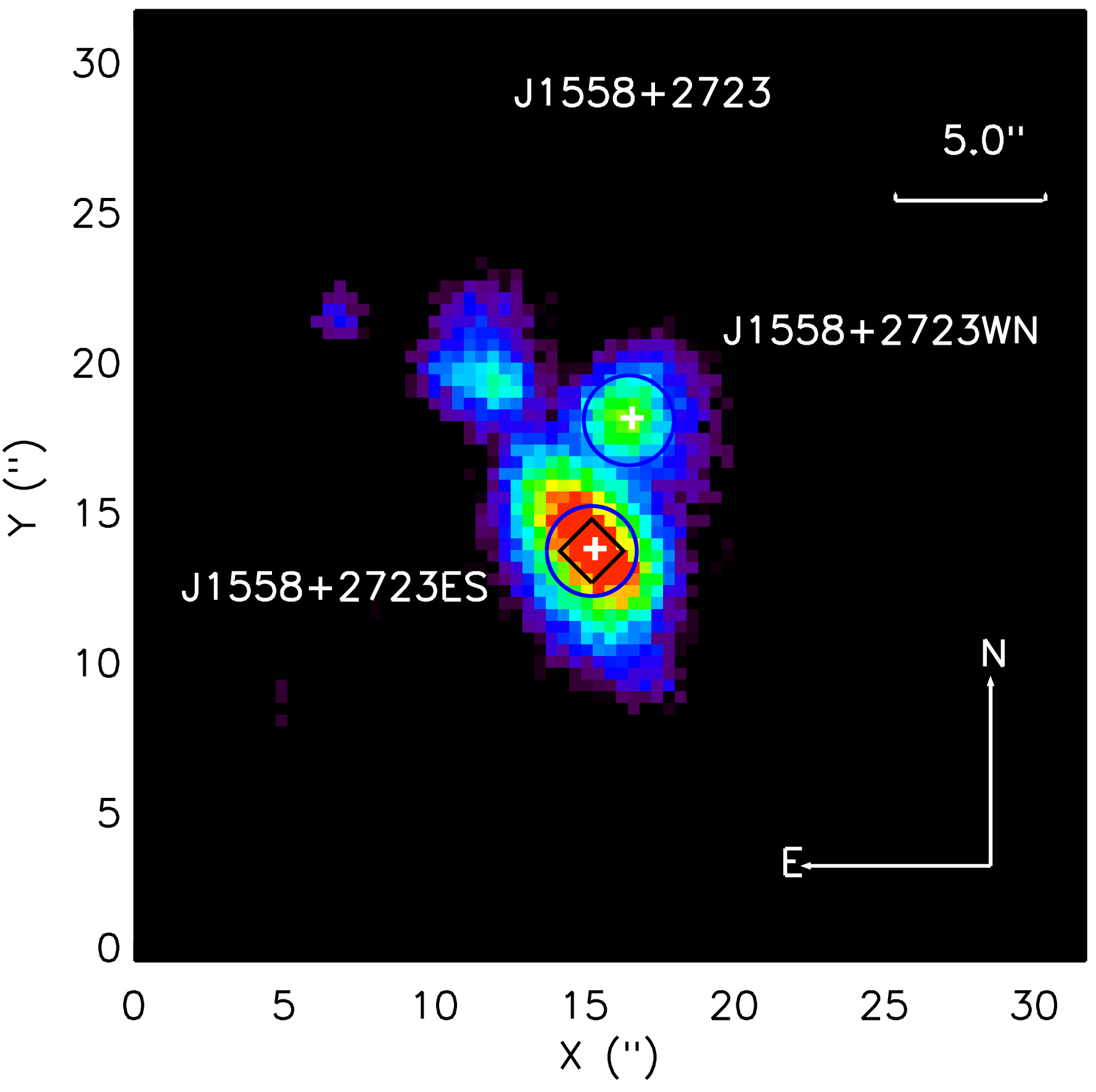}
  \includegraphics[width=12.2cm,height=5.2cm]{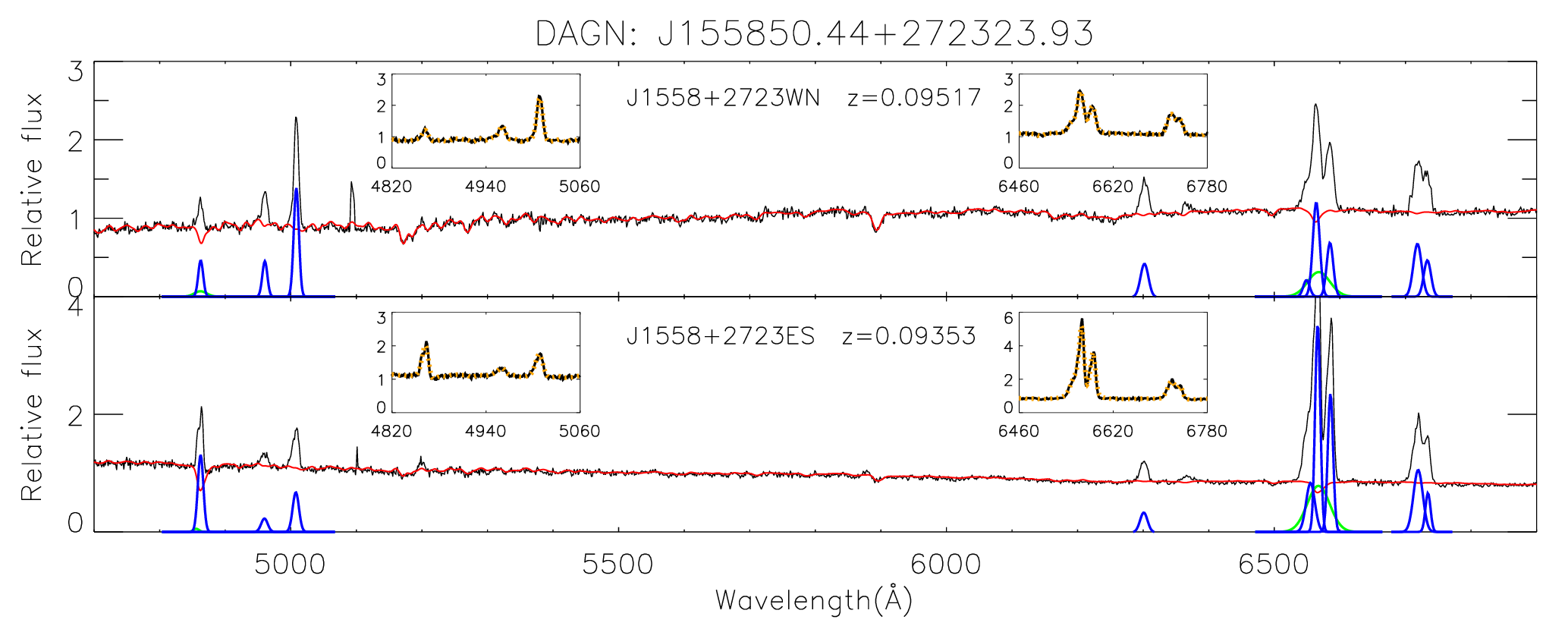}
  
\caption{Same as Fig.\,\ref{The spectra fitting of J1338+4816} but for dual AGN J155850.44+272323.93.}
\label{The spectra fitting of J1558+2723}
\end{figure*}

\begin{figure*}[ht]
\centering
  \includegraphics[width=5.20cm,height=5.0cm]{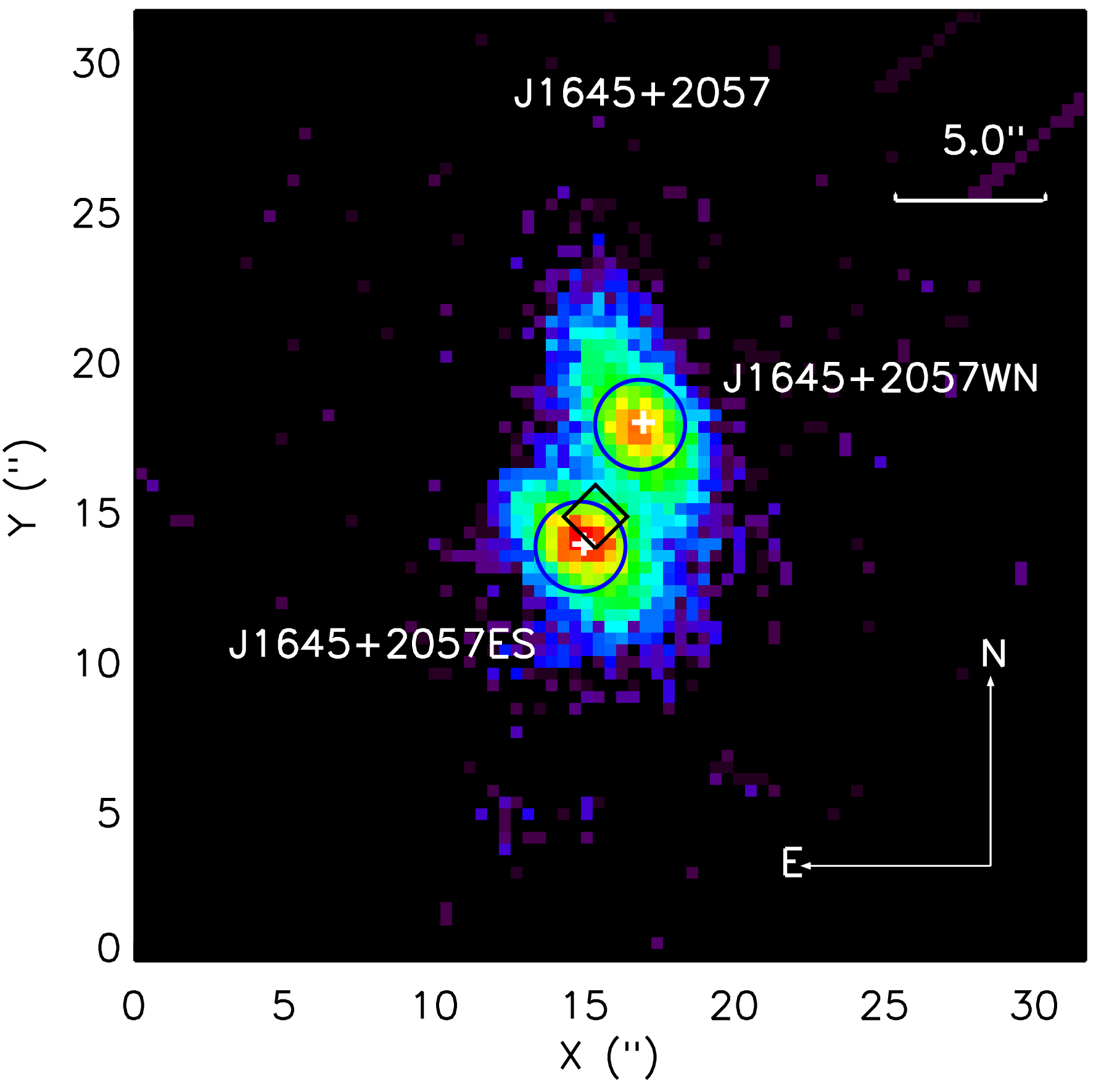}
  \includegraphics[width=12.2cm,height=5.2cm]{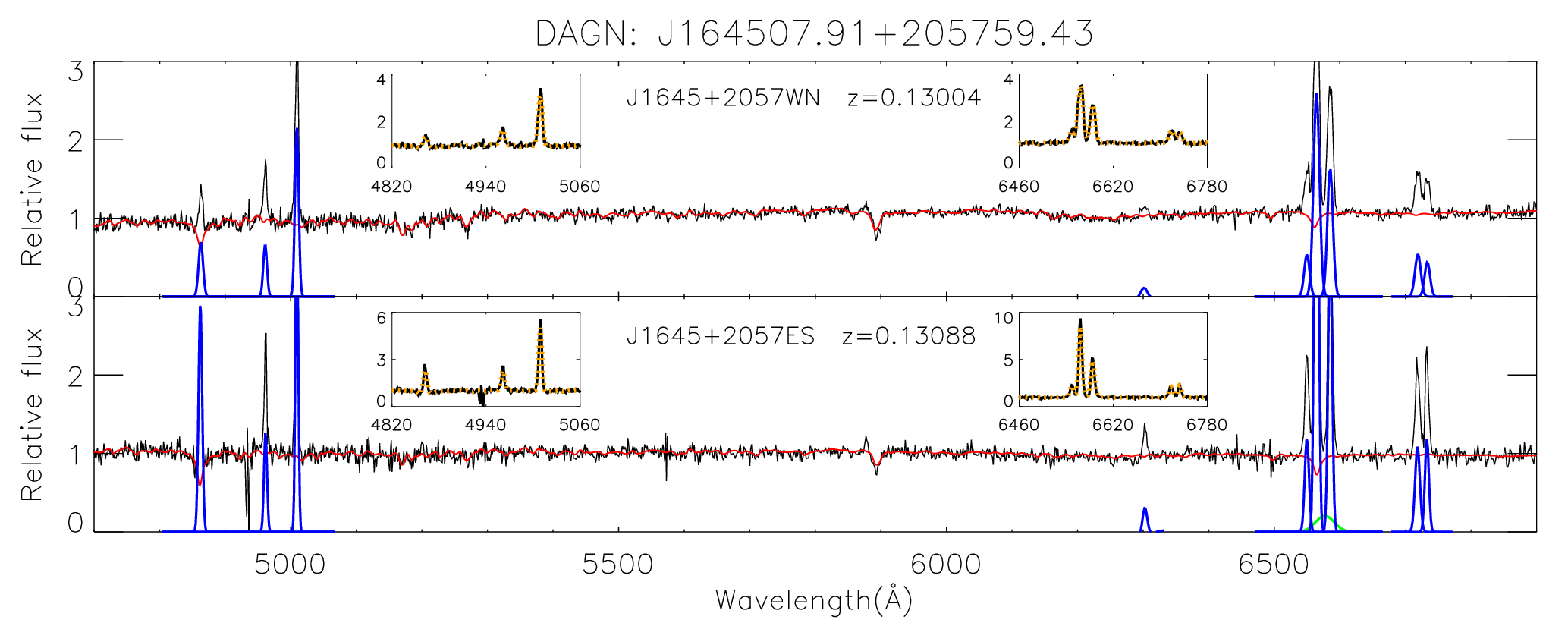}
  
\caption{Same as Fig.\,\ref{The spectra fitting of J1338+4816} but for dual AGN J164507.91+205759.43.}
\label{The spectra fitting of J1645+2057}
\end{figure*}

\begin{figure*}[ht]
\centering
  \includegraphics[width=5.20cm,height=5.0cm]{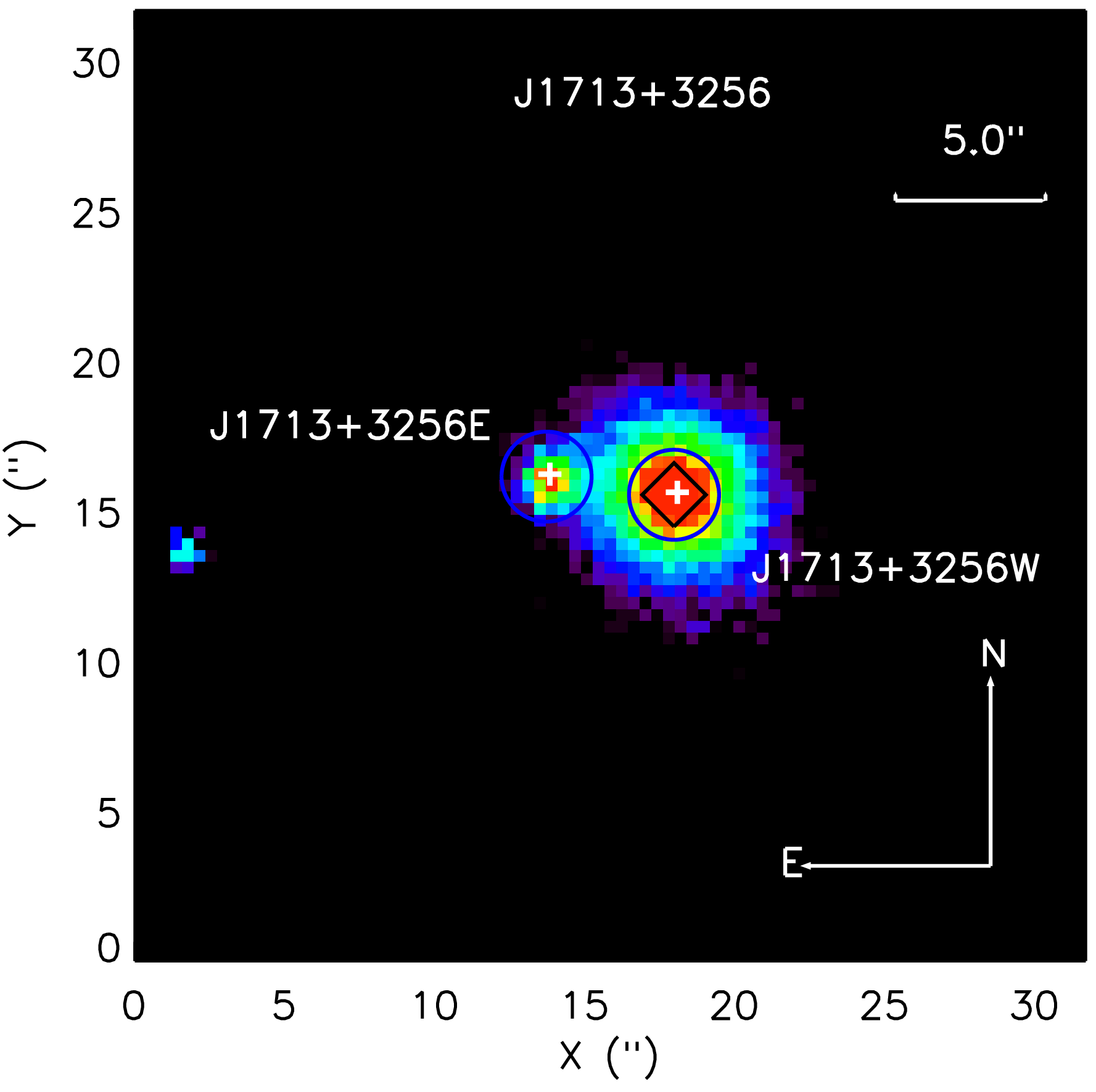}
  \includegraphics[width=12.2cm,height=5.2cm]{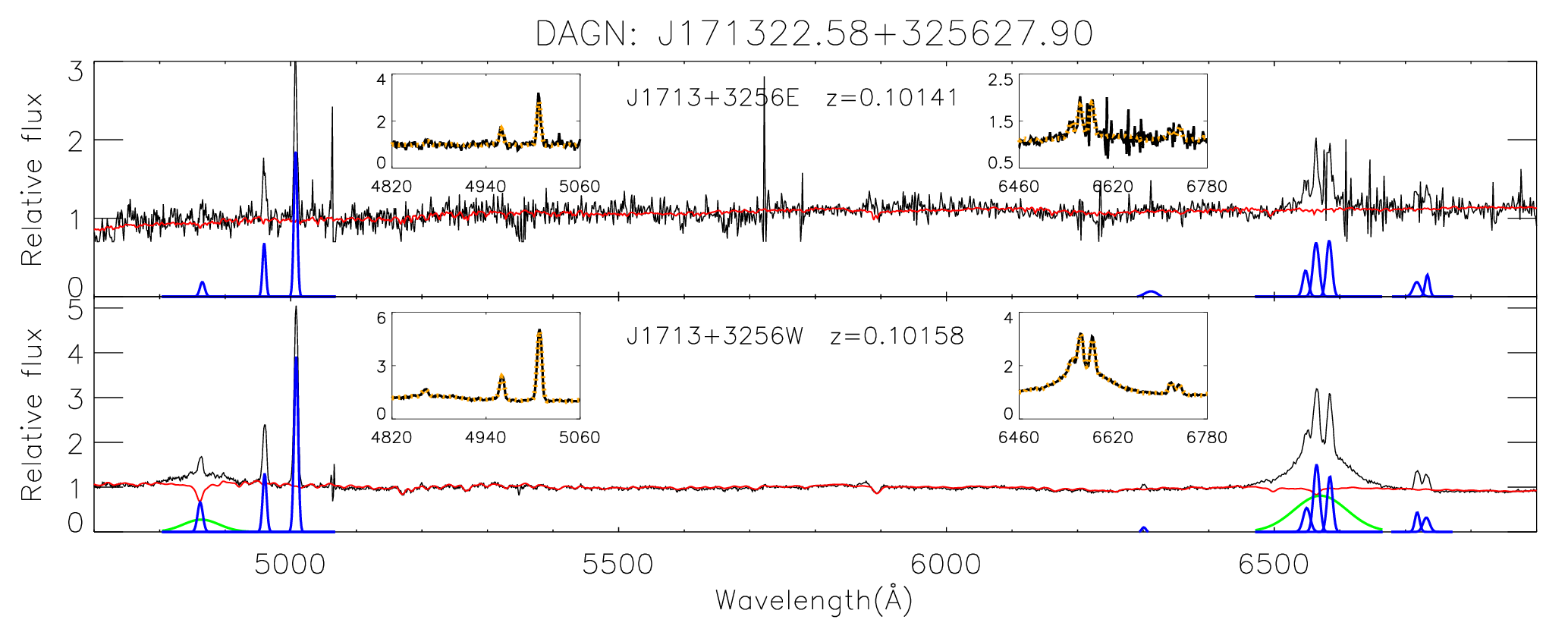}
  
\caption{Same as Fig.\,\ref{The spectra fitting of J1338+4816} but for dual AGN J171322.58+325627.90.}
\label{The spectra fitting of J1713+3256}
\end{figure*}

\begin{figure*}[ht]
\centering
  \includegraphics[width=5.20cm,height=5.0cm]{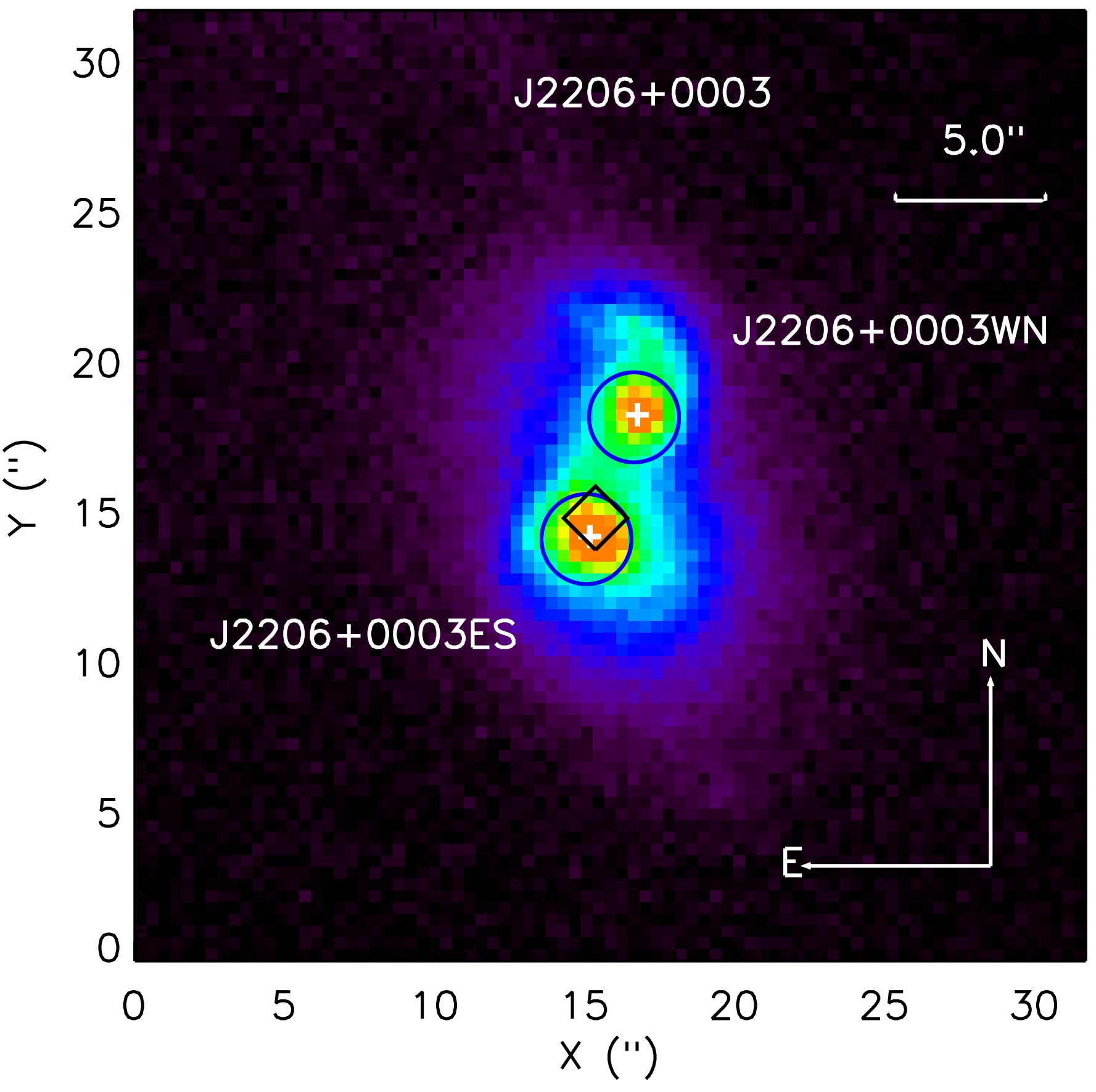}
  \includegraphics[width=12.2cm,height=5.2cm]{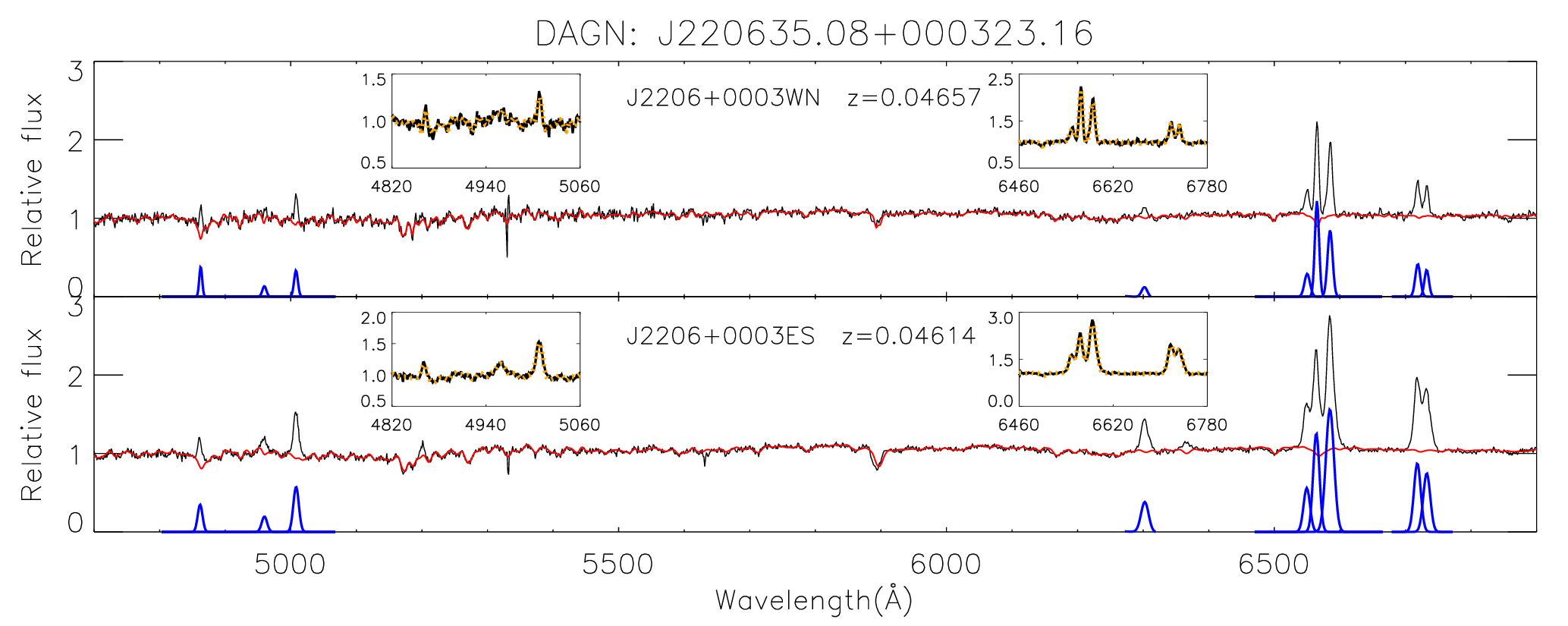}
   
\caption{Same as Fig.\,\ref{The spectra fitting of J1338+4816} but for dual AGN J220635.08+000323.16.}
\label{The spectra fitting of J2206+0003}
\end{figure*}

\end{appendices} 
\label{lastpage}
\end{document}